%% file: KS.tex
\documentclass[12pt,epsf,epsfig,psfig]{article}
\usepackage{graphicx}
\usepackage{epsfig}
\usepackage{cite}
\usepackage{multirow}
\def\B{\boldmath}

\begin{document}

\input{KScont.tex}

\newpage

\input{Sa.tex} 

\input{NNkl.tex}

\input{out.tex}

\input{ref.tex}

\end{document}

%% file: KScont.tex
\oddsidemargin 15pt
\topmargin 0pt
\headheight 00pt
\headsep 00pt
\textheight 235mm
\textwidth 161mm
\hoffset=-0.5cm
\parindent=0pt
\thispagestyle{empty}

\def\J{$J/\psi$}



~~\vskip 2cm

\centerline{\Large \bf  Color Deconfinement and Charmonium Production}

\vskip0.5cm

\centerline{\Large \bf in Nuclear Collisions}

\vskip1cm

\centerline{\large \bf Louis Kluberg$^1$ and Helmut Satz$^2$} 

\vskip0.5cm

\centerline{$1$ LLR, Ecole Polytechnique, CNRS-IN2P3, Palaiseau, France}

\centerline{$2$ Fakult\"at f\"ur Physik, Universit\"at Bielefeld,
Bielefeld, Germany}

\vskip 2cm

\centerline{\bf Abstract:}

\vskip0.5cm

In statistical QCD, color deconfinement and the properties of the  
quark-gluon plasma determine the in-medium behavior of heavy quark bound 
states. In high energy nuclear collisions, charmonia probe 
the partonic medium produced in the early stages of the interaction. We 
survey the present theoretical status and provide a critical evaluation
of the charmonium production measurements in experiments at the CERN-SPS
and the BNL-RHIC.

\vfill

\newpage

~~~~~~~~

\newpage

\centerline{\Large \bf Contents}

\vskip 1truecm

{\bf \large Introduction}

\vskip20pt

{\large \bf I.\ Theory}

\begin{enumerate}

\item{{\bf Heavy Quarks and Quarkonia}

\vskip 10pt

\item{\bf Quarkonium Binding and Dissociation}

\vskip 10pt

\item{{\bf Thermal Quark Dissociation}
\vskip 5pt \hskip 15pt
3.1 Interaction Range and Color Screening
\vskip 1pt \hskip 15pt
3.2 Potential Models
\vskip 1pt \hskip 15pt
3.3 Lattice Studies of Charmonium Correlators} 

\vskip 10pt

\item{{\bf Charmonium Production in Hadronic Collisions}
\vskip 5pt \hskip 15pt
4.1 Elementary Collisions
\vskip 1pt \hskip 15pt
4.2 p-A Collisions
\vskip 1pt \hskip 15pt
4.3 Nuclear Collisions
\vskip 1pt \hskip 15pt
4.4 Transverse Momentum Behavior}

\vskip 10pt

\item {\bf Conclusions}}

\end{enumerate}

\vskip20pt

{\large \bf II.\ Experiment}

\medskip

\begin{enumerate}

\item{\bf Charmonium Experiments at the CERN SPS}
\vskip 5pt \hskip 15pt
1.1 The Nuclear Dependence of Charmonium Production
\vskip 1pt \hskip 15pt
1.2 Normal Charmonium Production
\vskip 1pt \hskip 15pt
1.3 The First Hints of an Anomaly in Pb-Pb Collisions
\vskip 1pt \hskip 15pt
1.4 Anomalous \J~suppression in Pb-Pb Collisions

\vskip 10pt

\item{\bf Features of \B$\psi'$ Suppression at SPS energies}  

\vskip 10pt

\item{\bf More Results from SPS and RHIC}
\vskip 5pt \hskip 15pt
3.1 \J~Suppression in In-In Collisions at 158 GeV 
\vskip 1pt \hskip 15pt
3.2 \J~Suppression in A-A Collisions at $\sqrt s=$ 200 GeV

\vskip 10pt

\item{\bf Discussion and Evaluation}

\vskip 10pt

\item{\bf Summary of the Experimental Status}

\end{enumerate}

\vskip20pt

{\large \bf  Outlook}

\vskip20pt

{\large \bf  References}

%% file: Sa.tex
\oddsidemargin 15pt
\topmargin 0pt
\headheight 00pt
\headsep 00pt
\textheight 235mm
\textwidth 160mm
\voffset=0.5cm
\hoffset=-0.5cm
\parindent=0pt

\def\J{$J/\psi$}
\def\j{J/\psi}
\def\X{$\chi_c$}
\def\x{\chi}
\def\P{$\psi'$}
\def\p{\psi'}
\def\U{$\Upsilon$}
\def\u{\Upsilon}
\def\C{c{\bar c}}
\def\B{b{\bar b}}
\def\cg{c{\bar c}\!-\!g}
\def\bg{b{\bar b}\!-\!g}
\def\b{b{\bar b}}
\def\q{q{\bar q}}
\def\Q{Q{\bar Q}}
\def\e{\epsilon}
\def\t{\tau}
\def\l{\Lambda_{\rm QCD}}
\def\A{$A_{\rm cl}$}
\def\a{\alpha}
\def\N{$n_{\rm cl}$}
\def\n{n_{\rm cl}}
\def\R{R_{\C}}
\def\s{s_{\rm cl}}
\def\bb{\bar \beta}
\def\chiral{\psi {\bar \psi}}
\def\CMP{{ Comm.\ Math.\ Phys.\ }}
\def\NP{{ Nucl.\ Phys.\ }}
\def\PL{{ Phys.\ Lett.\ }}
\def\PR{{ Phys.\ Rev.\ }}
\def\PRep{{ Phys.\ Rep.\ }}
\def\PRL{{ Phys.\ Rev.\ Lett.\ }}
\def\RMP{{ Rev.\ Mod.\ Phys.\ }}
\def\ZP{{ Z.\ Phys.\ }}
\def\EPJ{{Eur.\ Phys.\ J.\ }}
\def\B{\boldmath}

\def\be{\begin{equation}}
\def\ee{\end{equation}}
\def\lsim{\raise0.3ex\hbox{$<$\kern-0.75em\raise-1.1ex\hbox{$\sim$}}}
\def\gsim{\raise0.3ex\hbox{$>$\kern-0.75em\raise-1.1ex\hbox{$\sim$}}}


{\LARGE \bf Introduction}

\vskip0.5cm

Statistical QCD predicts that strongly interacting matter undergoes
a deconfining transition to a new state, the quark-gluon plasma, in 
which the colored quarks and gluons are no longer bound to colorless
hadrons. How can this state be investigated - which phenomena provide 
information about its thermal properties? The main probes considered 
so far are electromagnetic signals (real or virtual photons), heavy 
flavor mesons ($Q\bar q$/$\bar Q q$ states),
quarkonia ($\Q$ pairs), and jets (energetic partons).
In theory, the ultimate aim is to carry out {\sl ab initio} calculations
of the in-medium behavior of these probes in finite temperature QCD.
In high energy nuclear collisions, we want to study the deconfinement
transition and the QGP in the laboratory, to compare experimental results 
to QCD predictions. In our report, we want to consider as specific case 
the spectral analysis of quarkonia in a hot QGP and its application to 
nuclear collisions.

\medskip

\begin{figure}[htb]
\centerline{\epsfig{file=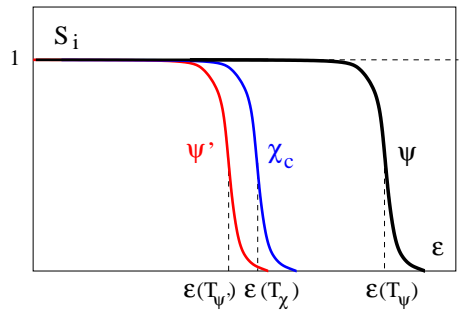,width=6cm}}
\vskip-0.2cm
\caption{Charmonium survival probabilities vs.\ energy density} 
\label{survival}
\end{figure}

The theoretical basis for this analysis is:
\begin{itemize}
\item{The QGP consists of deconfined color charges, so that the binding
of a $\Q$ pair in such a medium is subject to the effects of color screening.}
\vspace*{-0.2cm}
\item{The color screening radius $r_D(T)$ decreases with temperature $T$.}
\vspace*{-0.2cm}
\item{When $r_D(T)$ falls below the binding radius $r_i$ of a $\Q$ state 
$i$, the $Q$ and the $\bar Q$ can no longer bind, so that quarkonium $i$ 
cannot exist.}
\vspace*{-0.2cm}
\item{The quarkonium dissociation points $T_i$, specified through 
$r_D(T_i)\simeq r_i$, determine energy density and temperature of the QGP, 
as schematically illustrated in Fig. \ref{survival}.} 
\end{itemize}

\vskip-0.2cm

\medskip

Clearly the essential aspect here is a precise determination of the
dissociation points of the different quarkonium states, in 
temperature and/or energy density. This will therefore be the primary
task of theoretical studies of the in-medium behavior of quarkonia.

\medskip

Quarkonium production in high energy collisions
provides an excellent experimental tool to address a variety of
challenging questions.
\begin{itemize}
\vspace*{-0.2cm}
\item{In $pA$ collisions, quarkonium production can be studied in cold
nuclear matter, in order to provide a reference to 
determine which features of $AA$ collisions
are ``anomalous'', i.e., due to a newly produced medium.} 
\vspace*{-0.2cm}
\item{In $AA$ collisions, quarkonium production can be measured as function 
of collision energy, centrality, transverse momentum, and $A$, providing
a variety of different conditions of the produced medium.}
\vspace*{-0.2cm}
\item{The onset of anomalous suppression for the different quarkonium 
states can be determined and correlated to thermodynamic variables, such 
as the temperature or the energy density.}
\vspace*{-0.2cm}
\item{The resulting thresholds in the survival probabilities $S_i$ of 
states $i$ can then be compared to the relevant QCD predictions of
Fig.\ \ref{survival}.}
\end{itemize}

In this way we could, at least in principle, obtain a direct comparison
between experimental results and quantitative predictions from finite
temperature QCD.

\medskip

Starting its operation at CERN in 1986, immediately after the theoretical
prediction \cite{M-S} of \J~suppression as deconfinement signal, the 
experiment NA38 pioneered
the systematic study of \J~production and suppression. It was 
followed, some years later, by experiments NA50 and NA51,
with adapted and upgraded versions of the same detector. 
The first of these experiments, NA38, had submitted a Letter of Intent 
in 1984 aiming at the study of thermal muon pair production. 
It was on the floor in 1986, but it had changed
its main physics goal, as stated in its original proposal, to the 
study of \J~suppression. 
The three experiments made use of an exceptionally performant 
muon pair spectrometer with good measurement resolution  
and high luminosity operating capability; both features were 
inherited from the previous NA10 experiment, which was designed for 
and devoted to the study of high mass Drell-Yan muon pairs. 
It took almost 20 years of efforts to improve the 
detectors and beams, refine the analyses, overcome various 
difficulties and collect useful data under various experimental
conditions. This allowed double, and sometimes triple checks of
the outcoming results. In the meantime,
new results are presently being provided by much more 
recent experiments, PHENIX at the BNL-RHIC collider 
and NA60 at the CERN-SPS.    

\medskip

The main issue to be addressed is whether an {\em anomalous} 
\J~suppression has or has not been discovered, as claimed in
past publications. In order to answer this question, independently 
from any theoretical model, a mandatory prerequisite is to 
unambiguously define and precisely establish what is or should be 
considered  {\em normal \J~behavior}. 
We will see that the investment required to fill this part of the 
paradigm is always huge and, moreover, specific to each set of 
experimental conditions. It is, nevertheless, a {\em sine qua non} 
condition to avoid an insurmountable potential dead end to the 
problem. 

\vskip1.5cm

{\bf \LARGE I  Theory}

\medskip

\section{Heavy Quarks and Quarkonia}

Heavy quarks first showed up in the discovery of 
the \J ~meson \cite{Ting}, of mass of 3.1 GeV; it is a bound state of a 
charm quark ($c$) and its antiquark ($\bar c$), each having a mass 
of some 1.2--1.5 GeV. On the next level there is the \U~meson \cite{Leder}, 
with a mass of about 9.5 GeV, made up of a bottom or  
quark-antiquark pair ($b \bar b$), with each quark here having a mass 
around 4.5 - 4.8 GeV. Both charm and bottom quarks can of course also 
bind with normal light quarks, giving rise to open charm ($D$) and open 
beauty ($B$) mesons. The lightest of these `light-heavy' mesons have 
masses of about 1.9 GeV and 5.3 GeV, respectively. 

\medskip

The bound states of a heavy quark $Q$ and its antiquark $\bar Q$ are 
generally referred to as quarkonia. Besides the initially discovered 
vector ground states \J~and \U, both the $c \bar c$ and the $b \bar b$ 
systems give rise to a number of other {\it stable} bound states of 
different quantum numbers. They are stable in the sense that their mass 
is less than that of two light-heavy mesons, so that strong decays 
into open charm or beauty are forbidden. The measured stable charmonium 
spectrum contains the $1S$ 
scalar $\eta_c$ and vector \J, three $1P$ states \X~(scalar, vector 
and tensor), and the $2S$ vector state \P, whose mass is just below 
the open charm threshold. There are further charmonium states above the 
\P; these can decay into $D \bar D$ pairs, and we shall here restrict 
our considerations only to quarkonia stable under strong interactions.
These are summarized in tables 1 and 2 for charmonia and bottomonia,
respectively. The binding energies $\Delta E$ listed there are the 
differences between the quarkonium masses and the open charm or beauty 
threshold, respectively.

\vskip0.5cm

\centerline{
\renewcommand{\arraystretch}{1.4}
\begin{tabular}{|c|c|c|c|c|c|c|}
\hline
{\rm state}& $\eta_c$ & $J/\psi$ & $\chi_{c0}$ & 
 $\chi_{c1}$ &  $\chi_{c2}$ & $\psi'$ \\
\hline
{\rm mass~[GeV]}&
2.98&
3.10&
3.42&
3.51&
3.56&
3.69 \cr
\hline
$\Delta E$ {\rm[GeV]}&
0.75&
0.64&
0.32&
0.22&
0.18&
0.05\cr
\hline
\end{tabular}}

\vskip0.5cm

\centerline{~~Table 1: Charmonium states and binding energies}

\vskip0.8cm

\centerline{
\renewcommand{\arraystretch}{1.4}
\begin{tabular}{|c|c|c|c|c|c|c|c|c|c|}
\hline
{\rm state}
& $\Upsilon$ 
& $\chi_{b0}$
& $\chi_{b1}$ 
& $\chi_{b2}$ 
& $\Upsilon'$
& $\chi'_{b0}$ 
& $\chi'_{b1}$ 
& $\chi'_{b2}$ 
& $\Upsilon''$ \\
\hline
{\rm mass~[GeV]}&
9.46&
9.86&
9.89&
9.91&
10.02&
10.23&
10.26&
10.27&
10.36\cr
\hline
$\Delta E$ {\rm[GeV]}&
1.10&
0.70&
0.67&
0.64&
0.53&
0.34&
0.30&
0.29&
0.20\cr
\hline
\end{tabular}}

\bigskip

\centerline{~~Table 2: Bottomonium states and binding energies}

\vskip0.5cm

Quarkonia are rather unusual hadrons. The masses of the light hadrons,
in particular those of the non-strange mesons and baryons, arise almost
entirely from the interaction energy of their nearly massless quark 
constituents. In contrast, the quarkonium masses are largely determined
by the bare charm and bottom quark masses. These large quark masses allow 
a very straightforward calculation of many basic quarkonium properties, 
using non-relativistic potential theory. It is found that
the ground states and the lower excitation levels of quarkonia are very
much smaller than the normal hadrons, and that they are very tightly bound.
Now deconfinement is a matter of scales: when the separation between normal 
hadrons becomes much less than the size of these hadrons, they melt to form 
the quark-gluon plasma. What happens at this point to the much smaller 
quarkonia?  When do they become dissociated? As already indicated,
the disappearance of quarkonia can signal the presence of a 
deconfined medium \cite{M-S}, and the melting of specific quarkonium
states can determine the temperature of this medium \cite{K-M-S}. 
Thus the study of the quarkonium spectrum in a given medium 
is akin to the spectral analysis of stars \cite{K-S}, where the
presence or absence of specific excitation or absorption lines allows 
a determination of the temperature of the stellar matter.

\medskip

After these introductory comments, we shall review the basic properties of
quarkonia and the dynamics of their dissociation, and then survey the
different approaches to quarkonium binding in QCD thermodynamics. 
Following this, we consider the main theoretical features of quarkonium 
production in elementary as well as $p-\!A$ and nucleus-nucleus collisions.
With the theoretical basis provided, we then address the present experimental 
results.

\section{Quarkonium Binding and Dissociation}

We had defined quarkonia as bound states of heavy quarks which are 
stable under strong decay into open heavy flavor, i.e., $M_{c\bar c} 
\leq 2 M_D$ for charmonia and $M_{b\bar b} \leq 2 M_B$ for bottomonia. 
Since the quarks are so heavy, quarkonium spectroscopy can be studied 
quite well in non-relativistic potential theory \cite{Schrodinger}.
The simplest (``Cornell'') confining
potential \cite{Cornell} for a $\Q$ at separation distance $r$ has the form
\be
V(r) = \sigma ~r - {\alpha \over r}
\label{cornell}
\ee
with a string tension $\sigma \simeq 0.2$ GeV$^2$ and a Coulomb-like term
with a gauge coupling $\alpha \simeq \pi/12$. The corresponding 
Schr\"odinger equation
\be
\left\{2m_c -{1\over m_c}\nabla^2 + V(r)\right\} \Phi_i(r) = M_i \Phi_i(r)
\label{schroedinger}
\ee
then determines the bound state masses $M_i$ and the wave functions 
$\Phi_i(r)$, and with
\be
\langle r_i^2 \rangle = \int d^3r~ r^2 |\Phi_i(r)|^2 / 
\int d^3r~|\Phi_i(r)|^2 .
\label{radii}
\ee
the latter in turn provide the (squared) average bound state diameters.

\medskip

The solution of eq.\ (\ref{schroedinger}) gives in fact a very good 
account of the full (spin-averaged) quarkonium spectroscopy, as seen
in Table 3. The line labelled $\Delta M$ shows the differences 
between the experimental and the calculated values; they are in all cases 
less than 1 \%. Here $r_0$ gives the $\Q$ separation for the state in 
question, and the input parameters are $m_c=1.25$ GeV, $m_b=4.65$ GeV, 
$\sqrt \sigma = 0.445$ GeV, $\alpha=\pi/12$ \cite{HSjpg}.

\vskip0.5cm

\hskip1.5cm
\renewcommand{\arraystretch}{1.4}
\begin{tabular}{|c|c|c|c|c|c|c|c|c|}
\hline
{\rm state}& $J/\psi$ & $\chi_c$ & $\psi'$  & $\Upsilon$
 & $\chi_b$ & 
$\Upsilon'$ & $\chi_b'$ & $\Upsilon''$ \\
\hline
{\rm mass~[GeV]}&
3.10&
3.53&
3.68&
9.46&
9.99&
10.02&
10.26&
10.36 \\
\hline
$\Delta E$ {\rm[GeV]}&0.64&0.20&0.05 &1.10&
0.67&0.54&0.31&0.20 \cr
\hline
$\Delta M$ {\rm[GeV]}&0.02&-0.03&0.03 & 0.06&
-0.06&-0.06&-0.08&-0.07 \cr
\hline
{$r_0$ \rm [fm]}&0.50&0.72&0.90&
0.28&0.44& 0.56& 0.68 &0.78 \cr
\hline
\end{tabular}

\bigskip

\centerline{Table 3:
Quarkonium Spectroscopy from Non-Relativistic Potential Theory \cite{HSjpg}}

\bigskip

We thus see that in particular the \J~ and the lower-lying bottomonium
states are very tightly bound ($ 2M_{D,B} - M_0 \gg \l)$ and of very
small spatial size ($r_0 \ll r_h \simeq 1.5 - 2$ fm); here $r_h$ denotes
the typical hadron diameter. Through what kind of
interaction dynamics can they then be dissociated? 

\medskip

\begin{figure}[htb]
\centerline{\epsfig{file=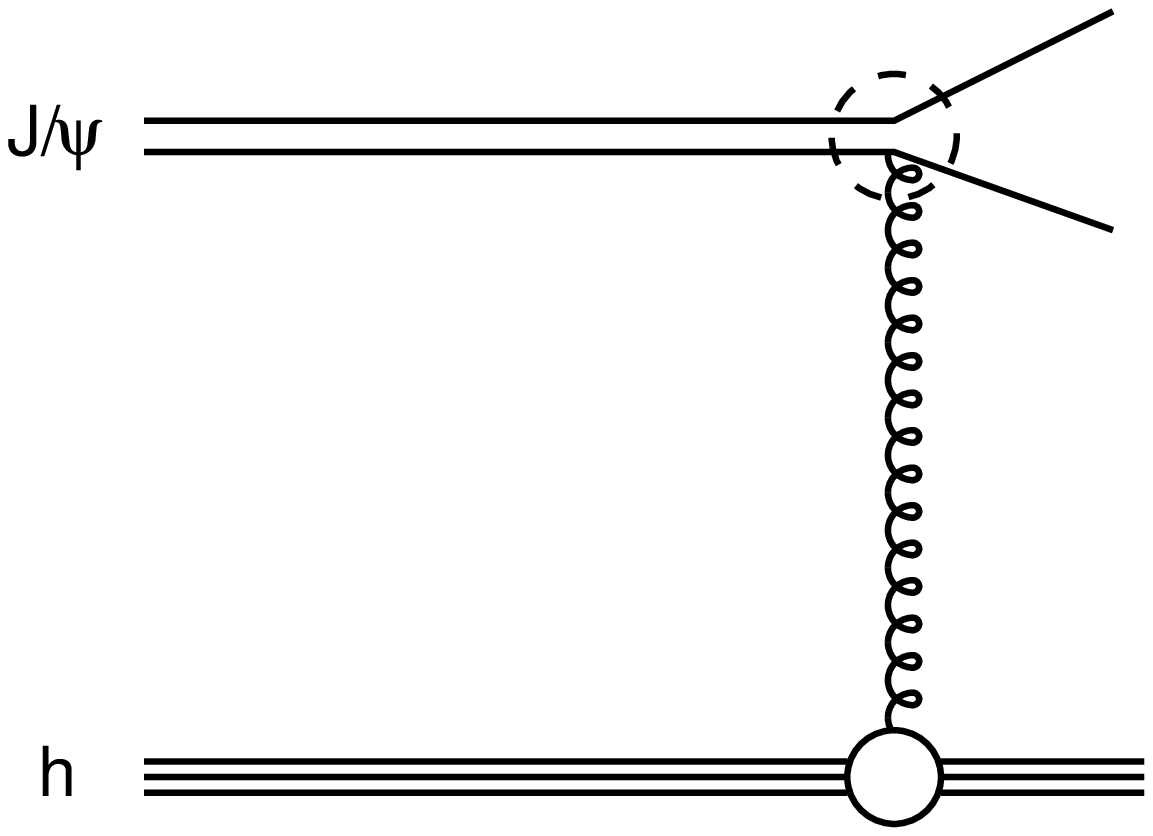,width=5cm}}
\caption{\J-hadron interaction}
\label{hadrodis}
\end{figure}

As illustration, we consider the collision dissociation of a \J. 
It is very small ($r_{\j} \sim 0.5$ fm) and hence can only be resolved
by a sufficiently hard probe \cite{K-S-OP}. It is moreover tightly bound
($2M_D-M_{\j} \sim 0.6$ GeV), so that only a sufficiently energetic
projectile can break the binding. In the collision with a normal hadron, 
the \J~can thus only be dissociated through the interaction with a hard 
gluonic contituent of the hadron, not with the hadron as a whole (see Fig.\ 
\ref{hadrodis}).  

\medskip

The gluon momentum distribution $g(x)$ in a hadron is determined in deep
inelastic lepton-hadron scattering; with $k_h$ denoting the gluon momentum,
$x = k_h/p_h$ specifies the fraction of the incident hadron momentum $p_h$
carried by the gluon. For mesons, one finds for large momenta
\be
g(x)  \sim (1-x)^3,
\label{gluon-pdf} 
\ee
so that the average momentum of a hadronic gluon is
\be
\langle k \rangle_h = {1\over 5} \langle p_h \rangle.
\label{gluon-mom}
\ee
For thermal hadrons in confined matter, 
$\langle p_h \rangle \sim 3T$, with $T < 175$ MeV,
so that with
\be
\langle k \rangle_h = {3\over 5} T \leq 0.1~{\rm GeV} 
\ll \Delta E \simeq 0.6~{\rm GeV}
\label{gluonT}
\ee
the gluon momentum is far too low to allow a dissociation of the 
\J. 

\medskip

On the other hand, the average momentum of a deconfined thermal
gluon in a quark-gluon plasma will be
\be
\langle k_g \rangle \simeq 3~T 
\label{gluon-dec}
\ee
and for $T~\! \gsim~\! 1.2~ T_c$, this provides enough energy to overcome
the \J~binding.  We thus expect that a hot deconfined medium can
lead to \J~dissociation, while the gluons available in a confined
medium are too soft to allow this. 

\medskip

\begin{figure}[htb]
\centerline{\epsfig{file=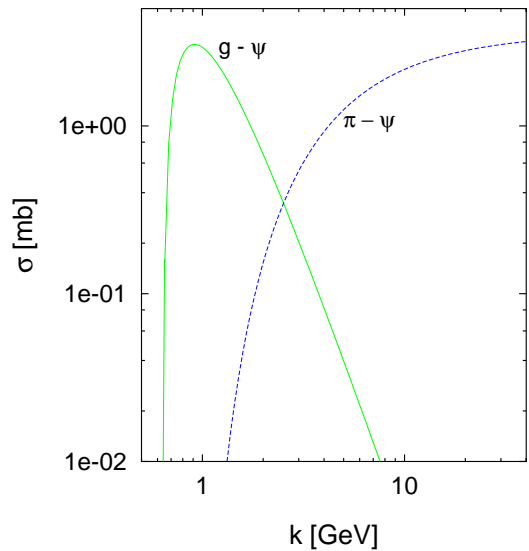,width=7.5cm,angle=-90}}
\caption{Gluon and hadron \J~dissociation cross-sections 
\protect\cite{K-S-OP}}
\label{g-h-dis}
\end{figure}

To make these considerations quantitative, one first has to calculate
the cross-section for the gluon-dissociation of a \J, a QCD analogue
of the photo-effect. This can be carried out using the operator product
expansion \cite{K-S-OP,Bhanot}, and the result is
\be
\sigma_{g-\j} \sim {1\over m_c^2}~{(k/\Delta E_{\psi}-1)^{3/2}\over
 (k/\Delta E_{\psi})^{5}}
\label{gluo-effect}
\ee
with $\Delta E_{\j} = 2M_D - M_{\j}$. The corresponding cross-section
for the hadron dissociation is then obtained by convoluting the
gluon-dissociation cross-section (\ref{gluo-effect}) with the
hadronic gluon distribution function $g(x)$, which for \J-meson
interactions leads to 
\be
\sigma_{h-\j} \simeq \sigma_{\rm geom} (1 - \lambda_0/\lambda)^{5.5}
\label{hadro-J}
\ee
with $\lambda \simeq (s-M_{\psi}^2)/M_{\psi}$ and 
$\!\lambda_0 \simeq (M_h + \Delta E_{\psi}$). Here $\sigma_{\rm geom}
\simeq \pi (r_{\j}/2)^2 \simeq 2$ mb is the geometric \J~cross-section and
$M_h$ denotes the mass of the incident meson. In Fig.\ \ref{g-h-dis},
we compare the two dissociation cross-sections (\ref{gluo-effect})
and (\ref{hadro-J}) as function of the incident projectile momentum.


\medskip

This result confirms the qualitative arguments given at the beginning
of this section: typical thermal gluon momenta near 1 GeV produce a large 
dissociation cross-section, whereas hadron momenta in a thermal range (up 
to 2 - 3 GeV) still lead to a vanishingly small cross-section. In other 
words, the \J~should survive in confining media, but become dissociated 
in a hot quark-gluon plasma \cite{K-S-OP}. 

\medskip

The calculations leading to eq.\ (\ref{gluo-effect}) are exact in the
large quark mass limit $m_Q \to \infty$. It is not clear if the charm 
quark mass really satisfies this condition, and so the dissociation cross 
section in \J-hadron collisions has been discussed in other approaches, 
some of which lead to much larger values \cite{J-dis1,J-dis2}. The basic 
question is apparently whether or not the charmonium wave function
has enough overlap with that of the usual hadrons to lead to a sizeable
effect. It may well be that this question is resolvable only
experimentally, by testing if slow charmonia in normal nuclear matter
suffer significant dissociation. Such experiments are definitely
possible \cite{inv-kine}. 
    
\section{Thermal Quarkonium Dissociation} 

We shall here address the first of our ``basic'' questions: 
Is it possible to specify the state of strongly interacting matter 
in terms of observable quantities calculated in QCD?

\subsection{Interaction Range and Color Screening}

\begin{figure}[htb]
\hspace*{0.1cm}
\centerline{\epsfig{file=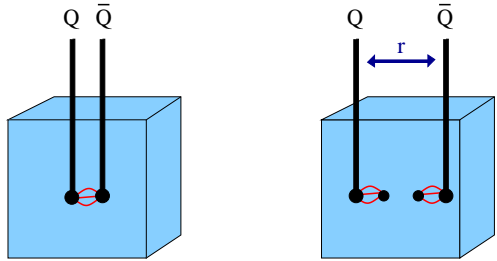,width=6.5cm}}
\caption{String breaking for a $\Q$ system}
\label{string-b}
\end{figure}

Consider a color-singlet bound state of a heavy quark $Q$ and its 
antiquark $\bar Q$, put into the medium in such a way that we can measure 
the energy of the system as function of the $\Q$ separation $r$ (see Fig.\ 
\ref{string-b}). The quarks are assumed to be heavy so that they are static 
and any energy changes indicate changes in the binding energy. We consider 
first the case of vanishing baryon density; at $T=0$, the box is therefore 
empty. 

\medskip

\medskip

In vacuum, i.e., at $T=0$, the free energy of the $\Q$ pair is
assumed to have the string form \cite{Cornell} 
\be
F(r) \sim \sigma r
\label{string}
\ee
where $\sigma \simeq 0.16$ GeV$^2$ is the string tension as 
determined in the spectroscopy of heavy quark resonances (charmonium
and bottomonium states). Thus $F(r)$ increases with separation
distance; but when it reaches the value of a pair of dressed
light quarks (about the mass of a $\rho$ meson), it becomes energetically
favorable to produce a $\q$ pair from the vacuum, break the string
and form two light-heavy mesons ($Q\bar q$) and ($\bar Q q$). These
can now be separated arbitrarily far without changing the energy of the
system (Fig.\ \ref{string-b}). 

\medskip

The string breaking energy for charm quarks is found to be
\be
F_0 = 2(M_D - m_c) \simeq 1.2~{\rm GeV};
\label{charm-break}
\ee
for bottom quarks, one obtains the same value,
\be
 F_0 = 2(M_B - m_b) \simeq 1.2~{\rm GeV},
\label{bottom-break}
\ee
using in both cases the quark mass values obtained in the solution
leading to Table 3. Hence the onset of string breaking is evidently 
a property of the vacuum as a medium. It occurs when the two heavy 
quarks are separated by a distance
\be
r_0 \simeq 1.2~{\rm GeV}/\sigma \simeq 1.5~{\rm fm},
\label{stringbreak}
\ee
independent of the mass of the (heavy) quarks connected by the string.

\medskip

\begin{figure}[h]
\hspace*{0.1cm}
\centerline{\epsfig{file=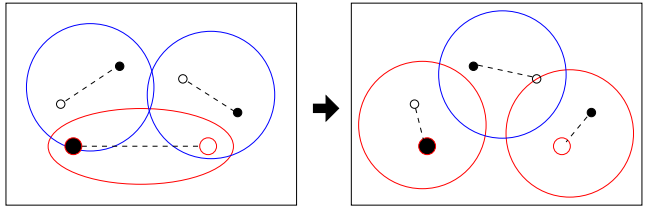,width=8.5cm}}
\caption{In-medium string breaking through recoupling}
\label{flip}
\end{figure}

If we heat the system to get $T>0$, the medium begins to contain light mesons, 
and the large distance $\Q$ free energy $F(\infty,T)$ decreases, since 
we can use these light hadrons to achieve an earlier string breaking through 
a kind of flip-flop recoupling of quark constituents \cite{miya}, resulting 
in an effective screening of the interquark force (see Fig.\ \ref{flip}).
Near the deconfinement point, the hadron density increases rapidly, and
hence the recoupling dissociation becomes much more effective, causing
a considerable decrease of $F(\infty,T)$.

\medskip

A further increase of $T$ will eventually bring the medium to the 
deconfinement point $T_c$, at which chiral symmetry restoration causes 
a rather abrupt drop of the light quark dressing (equivalently, of the 
constituent quark mass), increasing strongly the density of constituents. 
As a consequence, $F(\infty,T)$ now continues to drop sharply.   
Above $T_c$, light quarks and gluons become deconfined color charges, and 
this quark-gluon plasma leads to a color screening, which limits the range
of the strong interaction. The color screening radius $r_D$, which determines 
this range, is inversely proportional to the density of charges, so that
it decreases with increasing temperature. As a result, the $\Q$ interaction
becomes more and more short-ranged. 

\medskip

\begin{figure}[htb]
\hspace*{-0.3cm}
\centerline{\epsfig{file=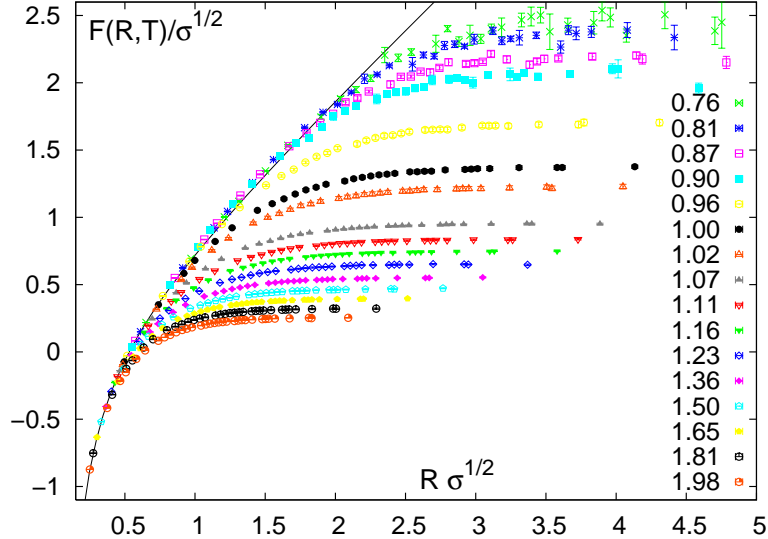,width=11cm}}
\caption{The color singlet $\Q$ free energy $F(r,T)$ vs.\ $r$ at 
different $T$ \cite{K-Z}}
\label{OKfree}
\end{figure}

In summary, starting from $T=0$, the $\Q$ probe first tests vacuum string 
breaking, then a screening-like
dissociation through recoupling of constituent quarks, and 
finally genuine color screening. In Fig.\ \ref{OKfree}, we show the 
behavior obtained in full two-flavor QCD for the color-singlet
$\Q$ free energy as a function of $r$ for different $T$ \cite{K-Z}. 

\medskip

\begin{figure}[htb]
\mbox{
\hskip0.5cm
\epsfig{file=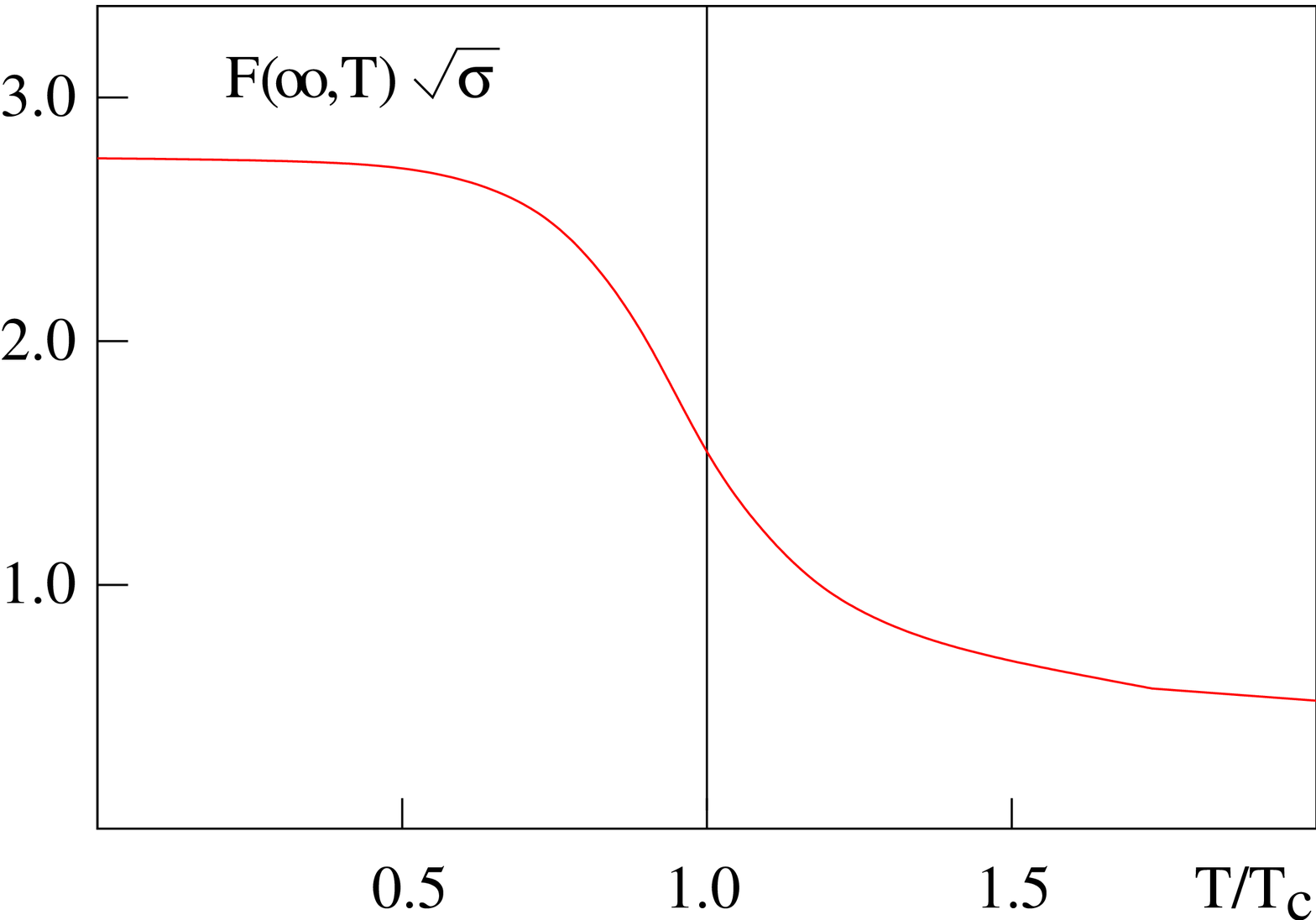,width=6.5cm}
\hskip1.5cm
\epsfig{file=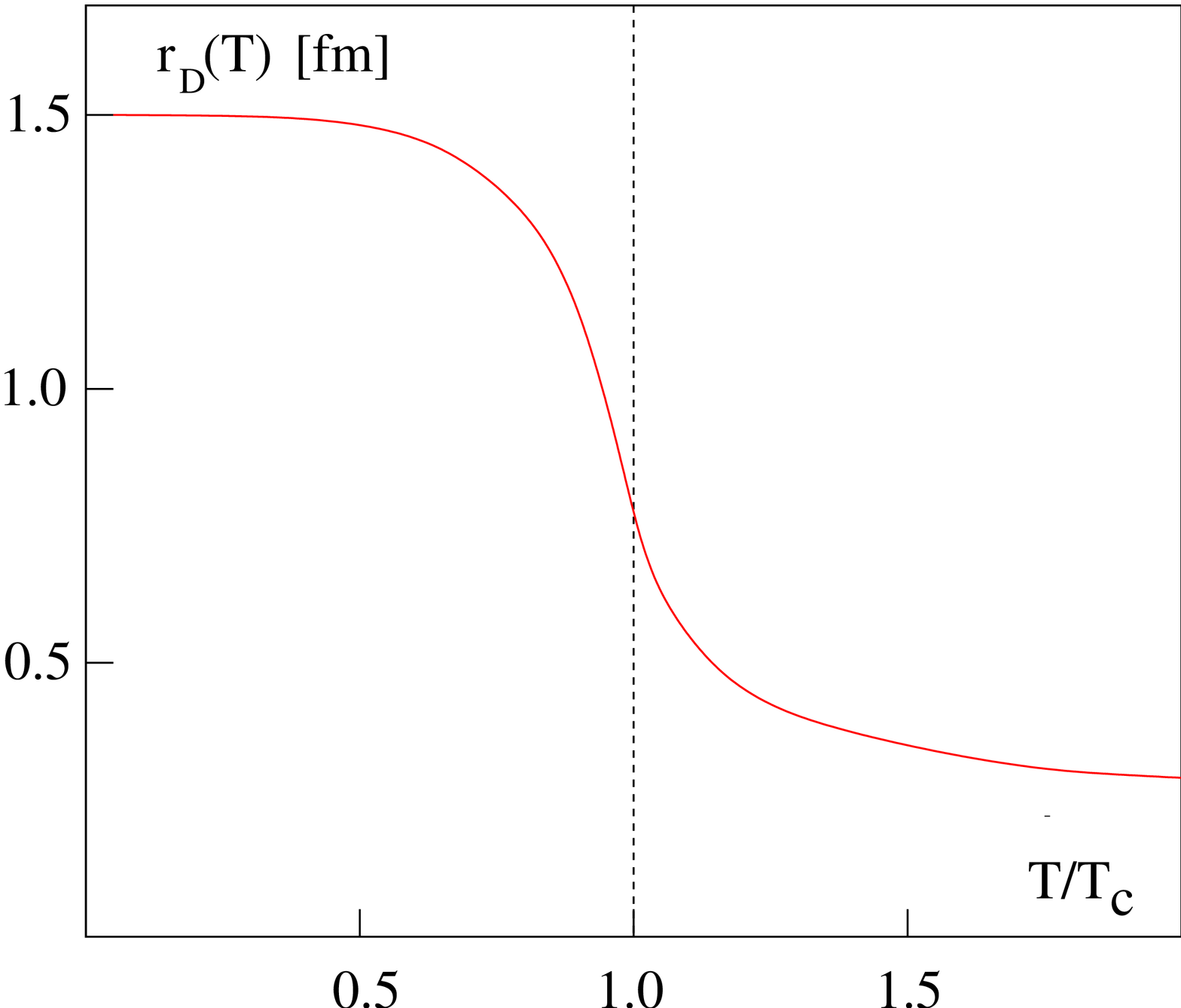,width=6.5cm,height=4.5cm}}
\caption{Asymptotic free energy and interaction range at different 
temperatures}
\label{free-T}
\end{figure}

It is evident in Fig.\ \ref{OKfree} that the asymptotic value $F(\infty,T)$, 
i.e., the energy needed to separate the $\Q$ pair, decreases with increasing 
temperature, as does the separation distance at which ``the string breaks''.
Here we consider the latter to be defined by the point beyond which 
the free energy remains constant within errors. The behavior of both 
quantities is shown in Fig.\ \ref{free-T}. Deconfinement is thus reflected 
very clearly in the temperature behavior of the free energy of a 
heavy quark potential: both the string breaking energy 
and the interaction range drop sharply around $T_c$. The latter decreases 
from hadronic size in the confinement region to much smaller values in the 
deconfined medium, where color screening is operative. 

\medskip

The in-medium behavior of heavy quark bound states thus serves 
quite well as probe of the state of matter in QCD thermodynamics. We
had so far just considered $\Q$ bound states in general. Let us now
turn to a specific state such as the \J. What happens when the range 
of the binding force becomes smaller than the radius of the state?
Since the $c$ and the $\bar c$ can now no longer see each other, the
\J~ must dissociate for temperatures above this point. Hence the
dissociation points of the different quarkonium states provide a 
way to measure the temperature of the medium. The effect had already been
illustrated schematically in Fig. \ref{survival}, showing how with increasing 
temperature the different charmonium states disappear sequentially
as function of their binding strength; the most loosely bound state 
disappears first, the ground state last.

\medskip

\begin{figure}[htb]
\hspace*{-0.3cm}
\centerline{\epsfig{file=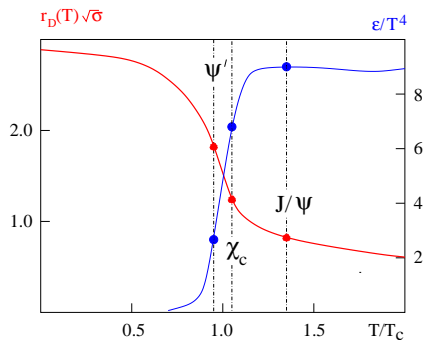,width=5.5cm}}
\caption{Charmonium dissociation vs.\ temperature and energy density}
\label{melt-e}
\end{figure}

Moreover, since finite temperature lattice QCD also provides the
temperature dependence of the energy density, the melting of the
different charmonia or bottomonia can be specified as well in terms
of $\e$. In Fig\ \ref{melt-e}, we illustrate this, combining the 
the energy density calculated in lattice QCD \cite{K-L-P,schlad}
and the force radii from Fig.\ \ref{free-T}. It is evident that 
although \P~and \X~are expected to melt around $T_c$, the corresponding 
dissociation energy densities will presumably be quite different.

\medskip

\vspace*{-0.3cm}

\medskip

To make these considerations quantitative, we thus have to find a way
to determine the in-medium melting points of the different quarkonium
states. This problem has been addressed in two different approaches:
\vskip-0.2cm
\begin{itemize}
\vskip-0.2cm
\item{Solve the Schr\"odinger equation (\ref{schroedinger}) with a
temperature-dependent potential $V(r,T)$, obtained either through
model considerations or from heavy quark lattice studies.}
\vskip-0.5cm
\item{Calculate the quarkonium spectrum directly in finite temperature
lattice QCD.}
\end{itemize}
Clearly the last is the only model-independent way, and it will
in the long run provide the decisive determination. However, the direct
lattice study of charmonium spectra has become possible only quite
recently, and corresponding studies for bottomonia are still more difficult.
Hence much of what is known so far is based on Schr\"odinger equation
studies with different model inputs. To illustrate the model-dependence
of the dissociation parameters, we first cite early work modelling
the temperature dependence of $V(r,T)$, then some more recent 
studies based on lattice results for $F(r,T)$, and finally summarize
the present state of direct lattice calculations of charmonia in finite
temperature media.

\subsection{Potential Model Studies}

The first quantitative studies of finite temperature charmonium dissociation
\cite{K-M-S,K-S} were based on screening in the form obtained in 
one-dimensional QED, the so-called Schwinger model. The confining part of the 
Cornell
potential (\ref{cornell}), $V(r) \sim \sigma r$, is the solution of the 
Laplace equation in one space dimension. In this case, Debye-screening leads 
to \cite{Dixit}
\be
V(r,T) \sim \sigma r \left\{ {1-e^{-\mu r} \over \mu r} \right\}, 
\label{schwinger-e}
\ee
where $\mu(T)$ denotes the screening mass (inverse Debye radius) for the
medium at temperature $T$. This form reproduces at least qualitatively 
the convergence to a finite large distance value $V(\infty,T) = 
\sigma/\mu(T)$, and since $\mu(T)$ increases with $T$, it also gives
the expected decrease of the potential with increasing temperature.
Combining this with the usual Debye screening for the $1/r$ part of
eq.\ (\ref{cornell}) then leads to  
 \be
V(r,T) \sim \sigma r \left\{ {1-e^{-\mu r} \over \mu r} \right\} 
- {\alpha \over r} e^{-\mu r} = {\sigma \over \mu} 
\left\{ 1-e^{-\mu r} \right\} - {\alpha \over r} e^{-\mu r}
\label{schwinger}
\ee
for the screened Cornell potential. In \cite{K-M-S}, the screening mass 
was assumed to have the form $\mu(T) \simeq 4~T$, as obtained in first
lattice estimates of screening in high temperature $SU(N)$ gauge theory.  
Solving the Schr\"odinger equation with these inputs, one found that
both the \P~and the \X~become dissociated essentially at $T \simeq T_c$,
while the \J~persisted up to about $1.2~T_c$. Note that as function of
the energy density $\e \sim T^4$, this meant that the \J~really survives
up to much higher $\e$.

\medskip

This approach has two basic shortcomings:
\vspace*{-0.2cm}
\begin{itemize}
\item{The Schwinger form (\ref{schwinger-e}) corresponds to the screening
of $\sigma r$ in one space dimension; the correct result in three space
dimensions is different \cite{Dixit}.}
\vspace*{-0.2cm}
\item{The screening mass $\mu(T)$ is assumed in its high energy form;
lattice studies show today that its behavior near $T_c$ is quite 
different \cite{D-K-K-S}.}
\end{itemize}

While the overall behavior of this approach provides some first insight
into the problem, quantitative aspects require a more careful treatment.

\medskip

From thermodynamics it is known that $F$ can be decomposed into the
internal energy $U$ and the entropy $S$ of the system,
\be
F = U -TS.
\label{F}
\ee
Since
\be
F = -T \ln Z
\ee
in terms of the partition function $Z$, we have
\be
U= -T^2\left({\partial[F/T] \over \partial T}\right)
\label{int}
\ee
and 
\be
S= -\left({\partial F \over \partial T}\right).
\label{entr}
\ee
Given $F(r,T)$, eqns.\ (\ref{int}) and (\ref{entr}) determine $U(r,T)$ 
and $S(r,T)$. It should be emphasized that the $F(r,T)$ calculated in 
finite temperature lattice QCD specifies the difference in free energy 
between a system with and a system without a $\Q$ pair, so that also
$U$ and $S$ are such differences. Since $U(r,T)$ thus is the expectation
value of the the Hamiltonian with a heavy quark pair, minus that without
such a pair, it is the relevant quantity to determine the binding
potential. We shall see shortly, however, that other effects enter
as well. 

\medskip

\begin{figure}[htb]
\centerline{\epsfig{file=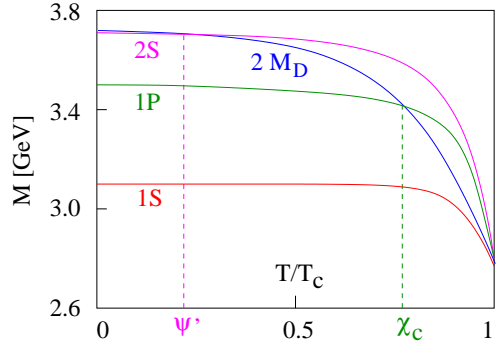,width=6.5cm}}
\caption{Temperature dependence of open and hidden charm masses \cite{D-P-S1}}
\label{masses}
\end{figure}

When the first lattice results for the color averaged value 
$F_{\rm av}(r,T)$ of the free energy became available, they were 
used to determine melting points for the different charmonium states
\cite{D-P-S1}. It was assumed that in the thermodynamic relation 
(\ref{int}) the entropy term $-T(\partial F /\partial T)$ could be 
neglected, thus equating binding potential and free energy,
\be
V(r,T) \simeq F_{\rm av}(r,T).
\label{DPS}
\ee
Using this potential in the Schr\"odinger equation (\ref{schroedinger})
specifies the temperature dependence of the different charmonium 
masses. On the other hand, the large distance limit of $V(r,T)$ 
determines the temperature variation of the open charm meson $D$,
\be
2 M_D(T) \simeq 2 m_c + V(\infty,T)
\label{D-mass}
\ee
In fig.\ \ref{masses}, we compare the resulting open and hidden charm 
masses as function of temperature. It is seen that the \P~mass falls
below $2~M_D$ around 0.2 $T_c$, that of the \X~at about 0.8 $T_c$;
hence these states disappear by strong decay at the quoted temperatures.
Only the ground state \J~survives up to $T_c$ and perhaps
slightly above; the lattice data available at the time did not extend 
above $T_c$, so further predictions were not possible. 

\medskip
 
The main shortcoming of this approach is also quite evident. 
The neglect of the entropy term in the potential reduces $V(r,T)$ 
and hence the binding. As a result, the $D$ mass drops faster with
temperature than that of the charmonium states, and it is this effect
which leads to the early charmonium dissociation. Moreover, 
in the lattice studies used here, only the color averaged free energy
was calculated, which leads to a further reduction of the binding
force.

\medskip

\begin{figure}[htb]
\hskip1cm{\epsfig{file=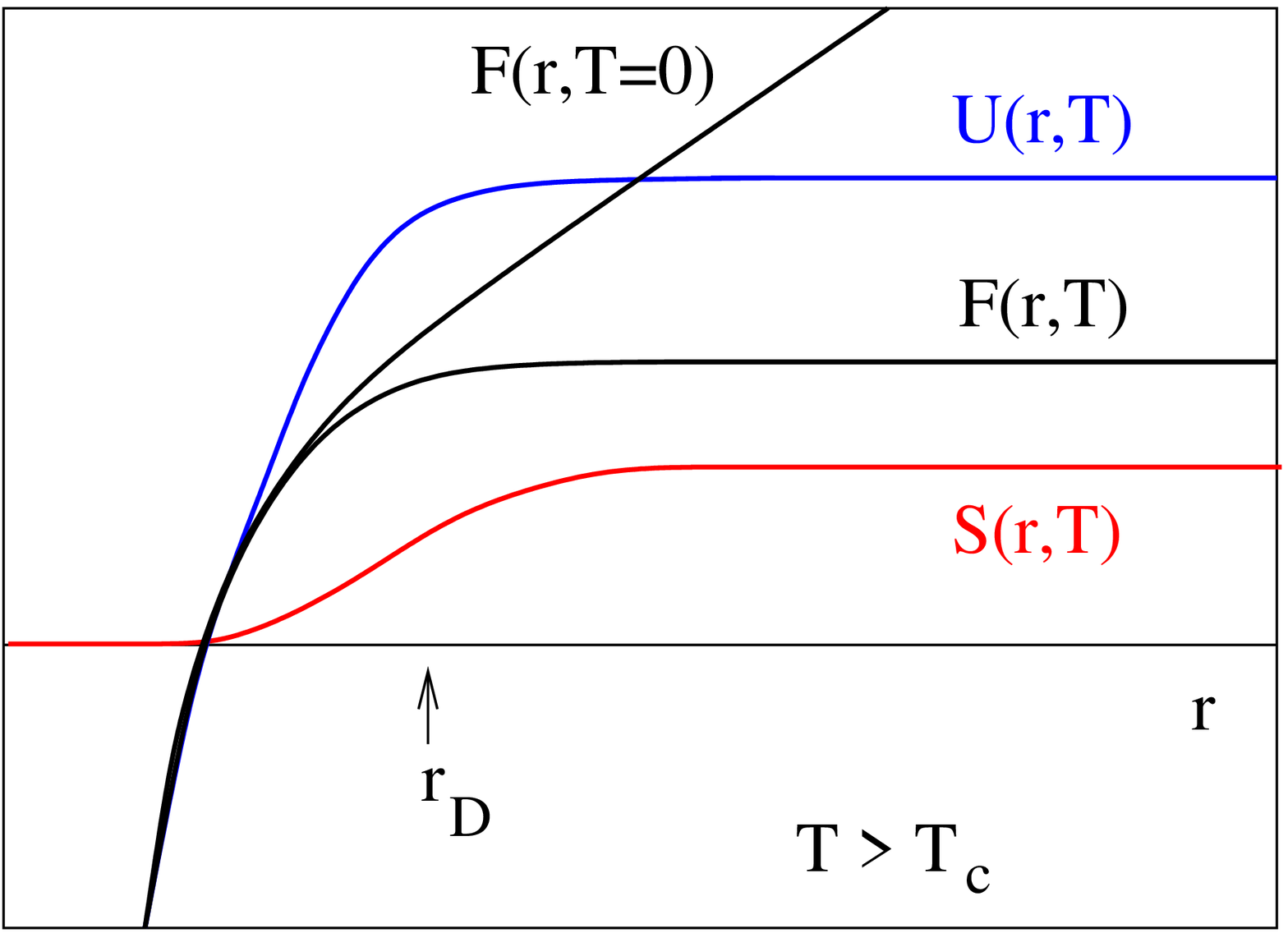,width=6cm,height=4cm}}
\hfill{\epsfig{file=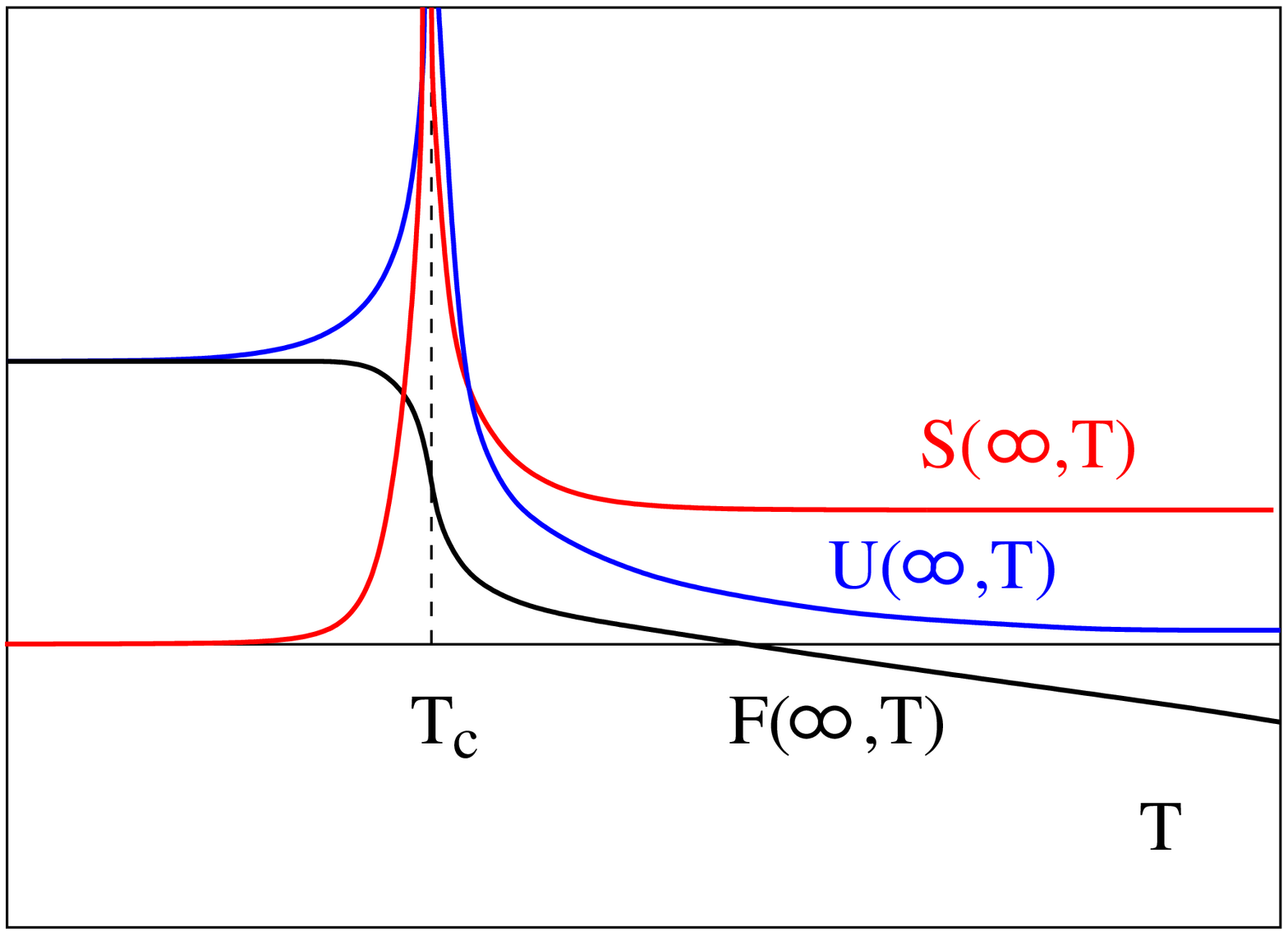,width=6cm,height=4cm}\hskip1cm}
\vskip0.6cm
\caption{Thermodynamic potentials as function of $r$ and $T$}
\label{pot}
\end{figure}

Today, finite temperature lattice QCD with two light quark flavors 
\cite{K-Z,P-P} provides the {\sl color singlet} free energy $F(r,T)$ of a 
static $\Q$ pair in a strongly interacting medium at temperature $T$.
Here $F(r,T)$ is again the difference between a system with and a system 
without a $\Q$ pair; it was shown in Fig.\ \ref{OKfree} for several 
temperatures below and above the deconfinement point $T_c$. For very 
small separation $r\ll T^{-1}$, the $\Q$ pair as a color neutral 
entity does not ``see'' the medium. Hence the results for different 
temperatures must coincide in the short distance limit and fall on 
the $T=0$ curve $F(r,T=0) = \sigma r - \alpha/r$;
this is used to normalize the results for $T>T_c$. 
Given $F(r,T)$, we can now extract the corresponding color singlet
behavior of the internal energy and the entropy; in Fig.\ \ref{pot}, 
we show schematically the resulting behavior of the different 
thermodynamic potentials as function of $r$ at some fixed $T > T_c$ 
and as function of $T$ for $r \to\infty$ \cite{OK}. At very short 
$\Q$ separation distance and finite T, for $r\ll T^{-1}$, the $\Q$ feels 
no medium and the medium does not see the pair; hence $U(r,T)=F(r,T)=F(r,0)$ 
and $S(r,T)=0$. The onset of the entropy starts around $r \simeq r_D(T)$, 
i.e, when the $\Q$ separation distance reaches the value of the screening 
length of the medium. 

\medskip

\medskip

For a quantitative potential theory study, we now have to extract
the in-medium heavy quark potential $V(r,T)$ from the color singlet 
internal energy obtained in lattice QCD. It appears that so far this 
task is not solved in an unambiguous way, in spite of a number of 
attempts. We therefore summarize these and their results, indicating 
the remaining problems.

\medskip

The internal energy $U$ is the expectation of the Hamiltonian $\cal H$, 
in our case the difference in interaction energy between a system with 
and one without the heavy quark pair at rest in the medium. Hence we have
\be
U(r,T) =  {P^2\over m} + V(r,T) \rangle, 
\label{intern}
\ee
where the first term gives the relative kinetic energy and $V(r,T)$ the 
interaction potential. The lattice results we want to use are obtained 
for a static $\Q$ pair, so that there is no relative kinetic energy,
and hence $U(r,T)$ is our effective potential. However, at very
large separation, the $Q$ and the $\bar Q$ do not see each other
any more and hence cannot interact.  Nevertheless, they will 
still polarize the medium in their vicinity and thus make $U(\infty,T)$
different from zero. This polarization cloud can be interpreted as an 
effective heavy quark thermal mass $m(T) \geq m_c$. Near $T_c$, the
gluonic correlation length becomes very large (or even diverges), 
thus increasing the size of the clouds. It is this effect which causes
the dramatic increase in $U(\infty,T)$ as $T$ approaches $T_c$ (see
the right part of Fig.\ \ref{pot}). As the $Q$ and the 
$\bar Q$ get closer to each other, the polarization clouds begin to 
overlap and hence interact. The interaction enhances the binding
potential between the two heavy quark components, and this cloud-cloud
interaction is what leads $U(r,T)$ near $r_D$ to overshoot the $T=0$
form. The free energy does not contain this component and hence 
approaches the Cornell potential. For a further discussion, see
\cite{SQM08}.

\medskip

The relevant Schr\"odinger equation thus becomes
\be
\left\{2m_c -{\nabla^2\over m_c} + 
U(r,T)\right\} \Phi_i(r,T) = M_i(T) \Phi_i(r,T),
\label{Schr3}
\ee
and the resulting dissociation temperatures are listed in Table 4.
As we shall see shortly, they agree quite well with the values
presently quoted in lattice studies. However, the inclusion of the 
cloud-cloud interaction has been put to question, and so the 
Schr\"odinger equation has also been solved with potentials of an
intermediate form $V(r,T;x) = x~\!U(r,T) + (1-x)~\!F(r,T)$, with 
$0 \leq x \leq 1$. The results tend towards those of 
Table 4 for $x \to 1$ and give lower dissociation temperatures as 
$x\to 0$ \cite{SZ,Wong,Alberico}.
 
\medskip

\begin{center}
\renewcommand{\arraystretch}{1.4}
\begin{tabular}{|c||c|c|c||c|c|c|c|c|}
\hline
 state & J/$\psi(1S)$ & $\chi_c$(1P) & $\psi^\prime(2S)$&$\Upsilon(1S)$&
$\chi_b(1P)$&$\Upsilon(2S)$&$\chi_b(2P)$&$\Upsilon(3S)$\\
\hline
\hline
$T_d/T_c$ & 2.10  & 1.16 & 1.12 & $>4.0$ & 1.76 & 1.60& 1.19 & 1.17 \\
\hline
\end{tabular}\end{center}

\medskip

\centerline{Table 4: Quarkonium Dissociation Temperatures \cite{HSjpg}}

\medskip

\begin{figure}[htb]
\centerline{\epsfig{file=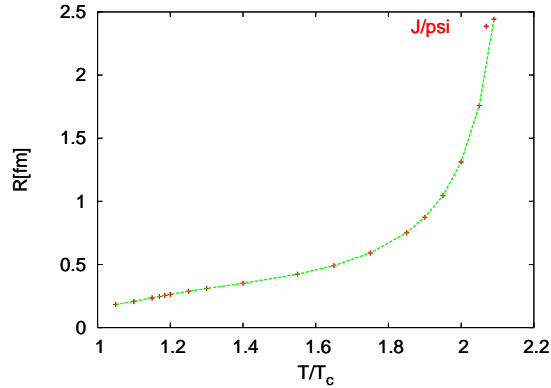,width=7.5cm}}
\caption{$T$-dependence of the bound state diameter for the 
\J~\protect\cite{HSjpg}}
\label{c-binding}
\end{figure}

We thus have to conclude that at this time, a final answer is still not
provided in the potential study approach. Besides the mentioned 
potential ambiguities, a further difficulty shared by all potential models 
is that the dissociation points are defined as those temperature
values for which the diameter $r_i$ of a given state $i$ diverges
(see Fig.\ \ref{c-binding}). 
As seen, this leads to ``bound state'' regions in which $r_i \gg T^{-1}$ 
and in fact surpasses the normal hadronic size $r_h$. One may expect 
\cite{K-ML-S} that under such conditions, thermal dissociation will break 
up any bound state, so that the actual bound state survival ends when its 
radius reaches the size of the screening radius \cite{D-P-S2}.

\subsection{Charmonium Correlators}

\begin{figure}[htb]
\centerline{\epsfig{file=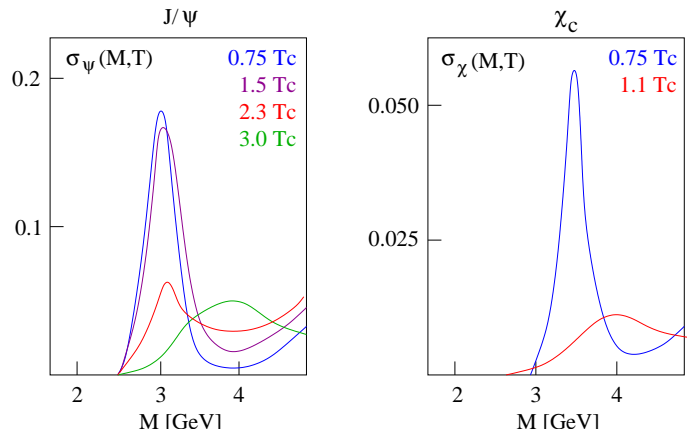,width=9cm}}
\caption{\J~and \X~spectral functions at different temperatures}
\label{spec}
\end{figure}

The direct spectral analysis of charmonia in finite temperature lattice 
has come within reach only in very recent years \cite{Umeda,Asakawa,
Datta,Iida,Jacovac,Skullerud}. It is possible now to evaluate the correlation 
functions $G_i(\t,T)$ for charmonium quantum number channels $i$ in terms 
of the Euclidean time $\t$ and the temperature $T$. These correlation 
functions are directly related to the corresponding spectral function 
$\sigma_i(M,T)$,
\be
G_i(\tau,T) = \int d\omega~ \sigma_i(\omega,T)~ K(\omega,\tau,T),
\label{cor1}
\ee
which describe the distribution in mass $M$ at temperature $T$ for
the channel in question. Here the kernel
\be
K(\omega,\tau,T) = {\cosh[\omega(\tau - (1/2T))] \over \sinh(\omega/2T)}
\label{cor2}
\ee
provides the relation betweeen the imaginary time $\tau$ and the
$\C$ energy $\omega$. The inversion of eq.\ (\ref{cor1}) with the
help of the maximum entropy method (MEM) provides the desired
spectrum \cite{MEM}. In Fig.\ \ref{spec}, schematic results at different
temperatures are shown for the \J~and the \X~channels. It is seen that
the spectrum for the ground state \J~ remains essentially unchanged
even at $1.5~T_c$. At $3~T_c$, however, it has disappeared; the
remaining spectrum is that of the $\C$ continuum of \J~quantum numbers
at that temperature. In contrast, the \X~is already absent at $1.1~T_c$, 
with only the corresponding continuum present. 

\medskip

These results are very promising: they indicate that in a foreseeable
future, the dissociation parameters of quarkonia can be determined 
{\sl ab initio} in lattice QCD. After first calculations performed in
quenched QCD \cite{Umeda,Asakawa,Datta,Iida,Jacovac}, i.e., without 
dynamical quark loops, there now are also results from two-flavor QCD 
\cite{Skullerud}, and these fully support the late dissociation of the \J. 
The widths of the
observed spectral signals are at present determined by the precision of
the lattice calculations; to study the actual physical widths, much
higher precision is needed. Moreover, one has so far only first signals
at a few selected points; a temperature scan also requires higher performance
computational facilities. Since the next generation of computers, in the
multi-Teraflops range, is presently going into operation, the next
years should bring the desired results. 

\medskip

So far, in view of the mentioned uncertainties in both approaches, the 
results from direct lattice studies and those from the potential model 
calculations of the previous section appear compatible. We note
at this point, however, that numerous attempts to obtain information 
on charmonium survival by comparing the calculated correlator to reference 
correlators constructed from models of low or zero temperature spectra
and relating the results to potential studies \cite{c1,c2,c3} have so 
far not really provided unambiguous information. 
Hence further work on the theory of in-medium charmonium behavior, in 
lattice studies  as well as in analytic approaches 
\cite{Bram,Nora,Laine,OP,Blai}, is clearly called for. 

\section{Charmonium Production in Hadronic Collisions}

\subsection{Elementary Collisions}

\begin{figure}[tbp]
\vskip0.5cm
\centerline{\psfig{file=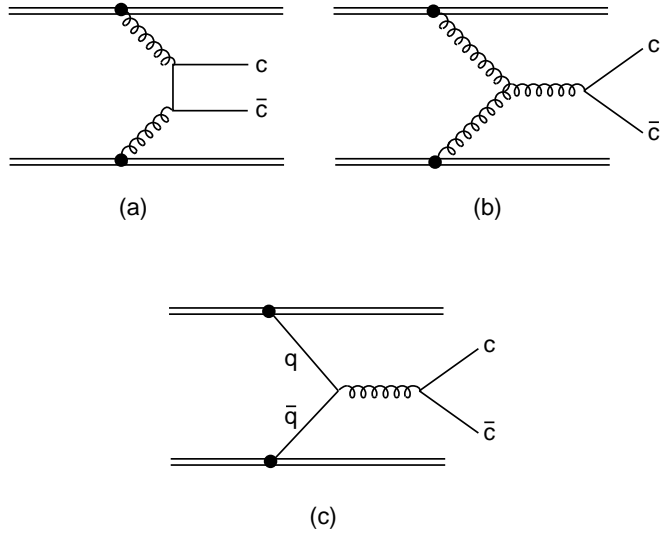,height=70mm}}
\caption{Lowest order diagrams for $\C$ production in hadronic
collisions, through gluon fusion (a,b) and quark-antiquark
annihilation (c).}
\label{5_1}
\end{figure}

The hadroproduction of charmonia occurs in two stages. The first stage
is the production of a $\C$ pair; because of the large quark mass, this
process is well described by perturbative QCD (Fig.\ \ref{5_1}).
A parton from the projectile interacts with one from the target; the
(non-perturbative) parton distributions within the hadrons are
determined empirically in other reactions, e.g.\ by deep inelastic
lepton-hadron scattering. The produced $\C$ pair is in general in
a color octet state. In the second stage, it neutralises its color
and then eventually forms physical resonances, such as \J, \X~ or \P. 
Color neutralisation occurs by interaction with the surrounding color 
field; this and the subsequent resonance binding are presumably of
non-perturbative nature. 

\medskip

On a fundamental theoretical level, color neutralization is not yet
fully understood. However, the color evaporation model \cite{CE}
provides a simple and experimentally well-supported phenomenological 
approach. In the evaporation process, the $c$ and the $\bar c$ can
either combine with light quarks to form open charm mesons ($D$ and
$\bar D$) or bind with each other to form a charmonium state.
The basic quantity in this description is the total sub-threshold
charm cross section $\R$, obtained by integrating the perturbative $\C$
production cross section $\sigma$ over the mass interval from 
$2m_c$ to $2m_D$. 
At high energy, the dominant part of $\R$ comes from gluon fusion (Fig.\
\ref{5_1}a), so that we have
\be
\R(s) \simeq \int_{2m_c}^{2m_D} d\hat s \int dx_1 dx_2~g_P(x_1)~g_T(x_2)~
\sigma(\hat s)~\delta(\hat s-x_1x_2s), 
\label{5.1}
\ee
with $g_P(x)$ and $g_T(x)$ denoting the gluon densities \cite{MRS}, 
$x_1$ and $x_2$
the fractional momenta of gluons from projectile and target,
respectively; $\sigma$ is the $gg \to \C$ cross section. In
pion-nucleon collisions, there are also significant quark-antiquark
contributions (Fig.\ \ref{5_1}c), which become dominant at low
energies. The basic statement of the color evaporation model is
that the production cross section of any charmonium state $i$
is a fixed fraction of the subthreshold charm cross section, 
\be
\sigma_i(s)~=~f_i~\R(s), \label{5.2}
\ee
where $f_i$ is an energy-independent constant to be determined empirically. 
It follows that the energy dependence of the production cross section 
for any charmonium state is predicted to be that of the perturbatively
calculated sub-threshold charm cross section. As a further consequence,
the production ratios of different charmonium states
\be
{\sigma_i(s)\over \sigma_j(s)} = {f_i\over f_j} = {\rm const.}
\label{5.3}
\ee
must be energy-independent. Both these predictions have been compared in 
detail to charmonium and bottomonium hadroproduction data over a wide range
of energies \cite{quarko}; they are found to be well supported, both in the 
energy dependence of the cross sections and in the constancy of the
relative species abundances.  Let us consider in more detail what 
this tells us about the hadronization of charm quarks.

\medskip

We recall that the relative abundances of light hadrons produced in 
hadron-hadron and $e^+e^-$ interactions follow the statistical pattern 
governed by phase space weights \cite{Hagedorn,Becattini}: the relative 
production rates are those predicted by an ideal resonance gas at the 
confinement/deconfinement transition temperature $T_c\simeq 175$ MeV. 
For two hadron species $i$ and $j$ that implies at all (high) collision 
energies
\be
R_{i/j} \simeq {d_i \over d_j} \left({m_i\over m_j}\right)^{3/2}
\exp-\{(m_i-m_j)/T_c\}
\label{rates}  
\ee
for the ratio of the production rates, with $d_i$ for the degeneracy
(spin, isospin) and $m_i$ for the mass of species $i$. For strange
particles, the rates (\ref{rates}) overpredict the experimental data;
this can, however, be accommodated by a common strangeness suppression 
factor $\gamma_s \simeq 0.5-0.7$, applied as $\gamma_s^n$ if the produced 
hadron contains $n$ strange quarks \cite{raf}. A recent alternative 
explanation for strangeness abundances is based on the assumption of 
a color event horizon with corresponding Unruh radiation \cite{HU}.

\medskip

For the hadroproduction of charm, such a statistical description does
not work, as seen in three typical instances \cite{quarko,Quack}: 
\begin{itemize}
\vspace*{-0.1cm}
\item{The total $\C$ cross section increases with energy by about a
factor ten between $\sqrt s = 20$ and 40 GeV, while the light hadron
multiplicity only grows by about 20\%. Hence the ratios of hadrons 
with and without charm are not energy-independent.} 
\vspace*{-0.2cm}
\item{From $p-p$ data one finds for \J~production a weight factor 
$f_{\j} \simeq 2.5 \times 10^{-2}$. Since the subthreshold 
$\C$ cross section is about half of the single $D$ production cross 
section \cite{Quack}, this implies $R_{(\j)/D} \simeq 10^{-2}$;
the ideal resonance gas gives with $R_{(\j)/D} \simeq 10^{-3}$
a prediction an order of magnitude smaller. Of the total charm production,
more goes into the hidden charm sector than statistically allowed.} 
\vspace*{-0.2cm}
\item{ For the production ratio of \P~to \J, which have the same charm 
quark infrastructure, one finds experimentally over a wide range of 
collision energies $R_{\p/(\j)} \simeq 0.23$. This energy-independent 
\P~to \J~ratio can be accounted for in terms of the charmonium masses and 
wave functions; it disagrees strongly with the statistical prediction, which 
gives with $R_{\p/(\j)} \simeq 0.045$ a very much smaller value. The same
holds true for the other measured charmonium states \cite{HSjpg}.}
\vspace*{-0.1cm}
\end{itemize} 

Charm production in elementary collisions thus does not seem to be of 
statistical nature. It appears to be determined by parton dynamics at 
an early stage \cite{Baier-R}, rather than by the phase space size at the 
confinement temperature.

\medskip

Although the color evaporation model provides a viable phenomenological
description of the hadroproduction of quarkonia, leading to correct 
quantitative predictions up to the highest energies under consideration,
it cannot predict the fractions $f_i$ of the hidden charm cross sections, 
and it can even less describe the space-time evolution of color 
neutralization. For charmonium production in p-A and A-B collisions, 
the latter is crucial, however, and hence a more detailed description 
of color neutralization is needed. 

\medskip

\begin{figure}[htb]
\centerline{\epsfig{file=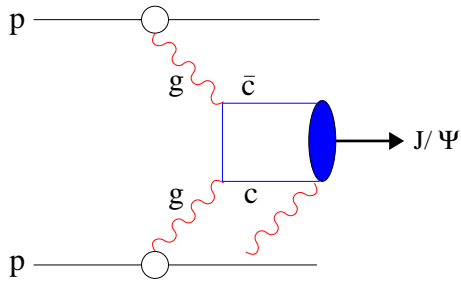,width=6cm}}
\caption{The evolution of \J~production}
\label{Jpsi-evo}
\end{figure}

\begin{figure}[htb]
\centerline{\epsfig{file=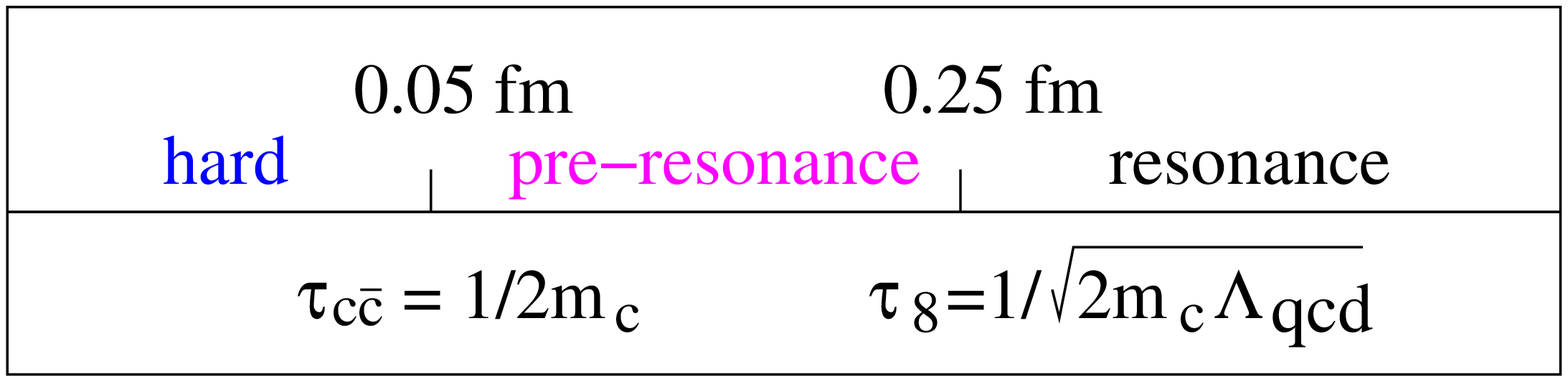,width=8cm}}
\caption{Scales of \J~production}
\label{scales}
\end{figure}

In the first step of the collision process, a $\C$ pair is formed
through a hard process, with very short formation time $\tau_{\C}$;
to reach subsequently \J~quantum numbers, the $\C$ must be in a color 
octet state.
The color octet model \cite{CO} proposes that this $\C$ then 
combines with a soft collinear gluon to form a color singlet $(\C\!-\!g)$ 
state. After a relaxation time $\tau_8$, this pre-resonant 
$(\C\!-\!g)$ state turns into the physical $\C$ mode by absorbing 
the accompanying gluon, with similar formation processes for $\chi_c$ and 
\P~production. The color octet model encounters difficulties if the 
collinear gluons are treated perturbatively, illustrating once more that 
color neutralization seems to require non-perturbative elements 
\cite{polarize,lansberg}. However, it does provide a conceptual basis for the 
evolution of the formation process (see Fig.\ \ref{Jpsi-evo}). The color 
neutralization time $\tau_8$ of the pre-resonant state can be estimated 
\cite{KS6}; it is essentially determined by the lowest momentum possible 
for confined gluons, $\t_8 \simeq (2m_c\l)^{1/2} \simeq 0.25$ fm.
The resulting scales in \J~formation are illustrated in Fig.\ \ref{scales}.
The formation time for the actual physical ground state \J~is presumably
somewhat larger than $\t_8$; although $r_{\j}/2 \simeq \t_8$, the
heavy $c$ quarks do not move with the velocity of light. For the larger 
higher excited states, the formation times will then be correspondingly 
larger.

\medskip

There is one further important feature to be noted for \J~hadroproduction. 
The \J's actually measured in hadron-hadron collisions have three distinct
origins: about 60 \% are directly produced 1S charmonium states, while
about 30 \% come from the decay $\x_c(1P) \to \j +$ anything, and the remaining
10 \% from $\p(2S) \to \j +$ anything \cite{feeddown}. 
Such feed-down also occurs in \U~production \cite{Upsi-feed}. In all cases,
the decay widths of the involved higher excited states are 
extremely small (less than one MeV), so that their life-times are very long. 
The presence of any medium in nuclear collisions would therefore affect 
these excited states themselves, and not their decay products. 

\subsection{\B$p-\!A$ Collisions}

In $p-\!A$ collisions, the presence of normal nuclear matter can affect 
charmonium production, so that such collisions provide a tool to probe 
the effect of confined matter.   
Nuclear effects can arise during the entire evolution of \J~production,
and several different phenomena have been studied in considerable detail.
We note in particular:
\begin{itemize}
\item{The presence of other nucleons in the nucleus can modify the 
initial state parton distribution functions, which enter in the perturbative
$\C$ production process illustrated in Fig.\ \ref{5_1}. 
\item{Once it is produced, the $\C$ pair can be dissociated in
the pre-resonance as well as in the resonance stage, due interactions 
with nucleons during its passage through the target nucleus.}}
\end{itemize} 
In both cases, the crucial quantity is the momentum of the charmonium state
as measured in the nuclear target rest frame. 

\medskip

Since we eventually want to probe the effect which the `secondary' medium 
{\sl produced} by nucleus-nucleus collisions has on charmonium production, 
it is of course essential to account correctly for any effects of the
nuclear medium initially present. Let us therefore first summarize 
the main features observed for charmonium production in $p-\!A$ collision
experiments \cite{Leitch}.

\begin{itemize}
\item{At fixed collision energy, quarkonium production rates per target
nucleon decrease with increasing $A$.}
\item{The production rates decrease for increasing \J~momentum as measured
in the nuclear target rest frame.}
\item{The nuclear reduction at $p-N$ mid-rapidity appears to become weaker
with increasing collision energy.}
\item{For fixed collision energy, mass number $A$ and \J~rapidity, the 
reduction appears to increase with the centrality of the collision.} 
\item{At sufficiently high momentum in the target rest frame, the different 
charmonium states appear to suffer the same amount of reduction, while at 
lower energy, the \P~is affected more than the \J.}
\end{itemize}

At present, there does not seem to exist a theoretical scenario able to
account quantitatively for all these observations. In fact, so far not even 
a common scaling variable for the different effects has been found.
Shadowing would suggest scaling in the fractional target parton 
momentum $x_2$, while absorption of the pre-resonance state would point 
to the fractional beam momentum $x_F$. Neither is in good accord with the 
data. The problem has recently been addressed in the context of 
parton saturation and color glass condensate formation \cite{K-T-cgc}.
Here we shall concentrate on indicating some operational methods 
to specify nuclear effects in $p-\!A$ collisions in a way that can be 
extended to $A-B$ collisions; for further discussions, we refer to reviews 
\cite{Leitch,Vogt}. The problems encountered in formulating a theoretical
description of charmonium production on nuclear targets underline again
the crucial importance of having $p\!-\!A$ data in order to arrive at 
a correct interpretation of $A\!-\!A$ results.

\medskip

We now return in some more detail to the two main aspects arising for
charmonium production in $pA$ collisions,
the (initial state) nuclear modification of the parton distribution 
function and the (final state) absorption of the nascent charmonium 
state during its traversal of the nucleus. 

\medskip

\begin{figure}[htb]
\centerline{\epsfig{file=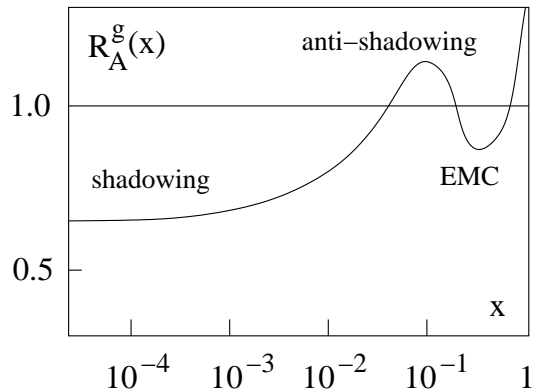,width=7cm}}
\caption{Nuclear modifications of the gluon distribution function}
\label{PDF}
\end{figure}

Assuming again gluon fusion as the dominant high energy process of
$\C$ formation, we have for \J~production in $pA$ 
collisions (see eqs.\ (\ref{5.1}) and (\ref{5.2}))   
\be
R_{\j}(s) \simeq f_{\j}
\int_{2m_c}^{2m_D} d\hat s \int dx_1 dx_2~g_p(x_1,\mu^2)~g_A(x_2,\mu^2)~
\sigma(\hat s)~\delta(\hat s-x_1x_2s), 
\label{shadow1}
\ee
where $\mu$ denotes the scale of the probe in the evaluation of the gluon
distribution functions in either the proton ($g_p$) or the nucleus ($g_A$).
The nuclear modification of the distribution functions have been studied
in detail in different approaches \cite{N-pdf}. In Fig.\ \ref{PDF}
we show a typical result for the ratio $R_A^g(x)$ of the gluon distribution 
function of a nucleon in a heavy nucleus, relative to that in a single
nucleon. It is shown as function of the Bjorken variable $x$, specifying 
the fraction of the nucleon momentum carried by the gluon.
We note that there are essentially four different
regimes. Near $x=1$, Fermi motion inside the nucleus leads to an enhancement;
following this is a suppression (the EMC effect) due to nucleon-nucleon
interactions. Around $x \simeq 0.1$, we then again have an enhancement 
(``anti-shadowing''), followed by a suppression at very small $x$ 
(``shadowing''). High energy collisions generally access regions 
$x \leq 0.1$, so that anti-shadowing and shadowing
are the phenomena of particular interest. Depending on the collision
energy and the momentum of the observed \J, the production rate can thus
be either enhanced or reduced by the initial state nuclear modification 
of the gluon distribution function. In order to determine the effect of
the final state (cold nuclear matter or, in $AA$ collisions, a possible
QGP) on charmonia, such initial state modifications must evidently be
brought under control.     

\medskip

\medskip

The second aspect is the dissociation of the nascent \J~in cold nuclear 
matter; the $\C$ pair has thus been formed, with whatever nuclear 
modifications of the parton distribution functions applied for its 
formation rate, and now traverses the remaining part of the nucleus. 
To specify the dissociation, we note \cite{K-L-N-S} that
according to the Glauber formalism of nuclear scattering theory,
a $\C$ pair formed at point $z_0$ in a target nucleus $A$ has a survival 
probability
\be
S_i^A = \int d^2 b~\! dz\! ~\rho_A(b,z)
\exp \left\{ -(A-1)\int_{z_0}^{\infty} dz'~ \rho_A(b,z')~
\sigma^i_{\rm diss} \right\},
\label{5.11}
\ee
where the integration covers the path, at impact parameter $b$,
remaining from $z_0$ out of the nucleus, and where $\sigma^i_{\rm diss}$ 
describes the overall ``absorption'' effect on the observed
charmonium state $i$ along the path. The result is then averaged over
impact parameter and path lengths. The traversed medium of nucleus $A$ 
is parametrized through a Woods-Saxon density distribution $\rho_A(z)$, 
and by comparing $S_i$ to data for different targets $A$, the effective 
dissociation cross section can be obtained for the \J~and \P~absorption 
in nuclear matter. The effect of the charmonium passage through the nucleus 
will arise from a superposition of the different stages; but if
part of the passage is carried out as physical resonance, higher 
excited states should lead to larger absorption cross-sections than the 
much smaller ground state \J. 

\medskip

The way charmonium production in $pA$ (or $dA$) collisions enters the
analysis is thus clear: with the initial state modifications of the parton
distribution functions taken into account, the dissociation cross section 
in cold nuclear matter has to be determined. With this given as a baseline,
one can then turn to $AA$ collisions and look for the effects of a newly 
produced medium.

\medskip

It should be noted here that in some studies \cite{K-L-N-S} the initial 
and final state effects have been parametrized jointly in the form 
of a common ``dissociation cross-section'', which includes both
sources of modification. With the help of this cross section, normal
nuclear matter effects were then calculated for nucleus-nucleus collisions.
Evidently such a procedure makes sense only if neither of the nuclear
modifications depend sensitively on the momentum of produced charmonium
state, or if the accessible momentum range is the same and very small 
both in $p-A$ and in nuclear collisions. For the SPS data, the latter
seems to be the case. 

\subsection{Nuclear Collisions}

\begin{figure}[htb]

\centerline{\epsfig{file=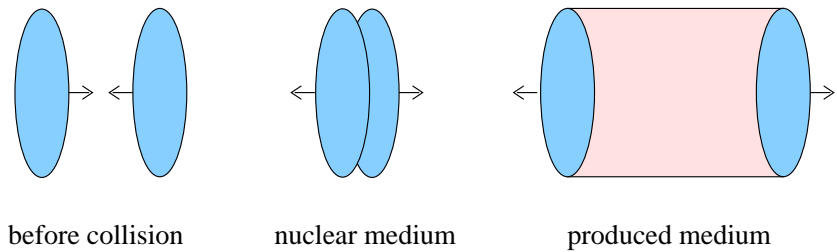,width=11cm}}
\caption{Collision stages}
\label{col-evo}
\end{figure}

The basic assumption in the attempt to create deconfined matter through
nuclear collisions is that the excited vacuum left after the passage of
the colliding nuclei forms a thermal medium. This picture is schematically
illustrated in Fig.\ \ref{col-evo}. A charmonium state produced in such
a collision will in its early stages first be subject to the possible
effects of the nuclear medium, just as it is in $p-\!A$ collisions,
and then, after the nuclei have separated, encounter the newly produced
medium.

\medskip

\medskip

The Glauber formalism used above to calculate the survival probability of 
an evolving charmonium state in a $p-A$ collision now has to be extended
to $A-B$ interactions \cite{K-L-N-S}. The survival probability at impact 
parameter $b$ now becomes
\vskip-0.2cm
$$
S^{AB}_i(b) = \int d^2s~\!dz~\!dz' \rho_A(s,z) \rho_B(b-s,z') ~\times
$$
\vskip-0.2cm
\be
\exp \left\{ -(A-1)\int_{z_0^A}^{\infty} dz_A~ \rho_A(s,z_A)~
\sigma^i_{diss} ~-(B-1)\int_{z_0^B}^{\infty} dz_B~ \rho_B(b-s,z_B)~ 
\sigma^i_{diss}\right\},
\label{5.13}
\ee
\vskip0.2cm
as extension of Eq.\ (\ref{5.11}). Here $z_0^A$ specifies the formation
point of the $\C g$ within nucleus $A$, $z_0^B$ its position in $B$.
With the dissociation cross sections $\sigma^i_{diss}$ determined in
$p-A$ collisions, Eq.\ (\ref{5.13}) specifies the `normal'
survival probability, i.e., that due to only the nuclear medium.

\medskip

To use quarkonia as probes for the produced medium, we now have to study 
how the behavior observed in $A-\!A$ collisions differs from this 
predicted pattern. Several possible and quite different effects have 
been considered as consequences of the produced medium on charmonium 
production. 
\begin{itemize}
\vskip0.2cm 
\item{Suppression by comover collisions: A charmonium state produced in 
a primary nucleon-nucleon collision can be dissociated through interactions
with the constituents of any medium subsequently formed in the collision.
Such dissociation could in principle occur in a confined \cite{hadro-co}
as well as in a deconfined medium \cite{K-S-OP}.}
\item{Suppression by color screening: If the produced medium is a hot
QGP, it will dissociate by color screening the charmonium states produced 
in primary nucleon-nucleon collisions \cite{M-S}. Since the different 
states have different dissociation temperatures, a sequential suppression 
pattern will specify the relevant thresholds \cite{K-M-S,K-S,KKS}.}
\item{Enhancement by regeneration: In the hadronization stage of the QGP, 
charmonium formation could occur through the binding of a $c$ with a $\bar c$ 
from different nucleon-nucleon collisions. If the total number of available 
$\C$ pairs considerably exceeds their thermal abundance, such statistical 
regeneration could enhance hidden relative to open charm production, as 
compared to hadron-hadron collisions \cite{PBM,Thews,Rapp,PBM-St}.}
\end{itemize}
In addition, the partonic initial state of the colliding nuclei, which 
leads to the formation of the produced medium and to that of 
charmonium states, will change its nature for large $A$ and sufficiently 
high $\sqrt s$; eventually, parton percolation (saturation or
color glass formation) can lead to a very different medium, 
with possible effects on production and binding of charmonia \cite{perco}.

\medskip

Is it possible for experiment to distinguish between these different
scenarios? Before turning to the experimental situation, we want
to discuss in some more detail the salient features of each approach. 

\subsubsection{Suppression by Comover Collisions}

If the charmonium state moves in a random scattering pattern through
the produced medium, its survival rate is approximately given by
\be
S_i = \exp\{-\sigma_i n \tau_0 \ln [n/n_f]\}
\label{comover}
\ee
with $\sigma_i$ denoting the dissociation cross section, $n$ the
initial density of the medium after a formation time $\tau_0$, and
$n_f$ the `freeze-out' density, at which the interactions stop.

\medskip

Since the cross section for \J~break-up through gluon collisions is 
large \cite{K-S-OP} and the gluon density high, there will be significant 
charmonium suppression in a deconfined medium, even if this is not
thermalized. In an equilibrium QGP, this dissociation is presumably
accounted for by color screening, provided the effect of the
medium on the width of the surviving states is also calculated.

\medskip

Charmonium dissociation by interaction with hadronic comovers has
received considerable attention in the past \cite{hadro-co}. However,
if one restricts the possible densities to values appropriate to
hadronic matter ($n_h \lsim 0.5$ fm$^{-3}$) and the cross sections
to those obtained in section 3, the effect of hadron dissociation
is negligible. Even a cross section increase to the high energy
limit still leads to less than 10\% effects. More recent analyses
\cite{Maiani} thus conclude that a hadronic medium will not result
in significant suppression.

\medskip

\begin{figure}[h]
\begin{minipage}[t]{7cm}
\epsfig{file=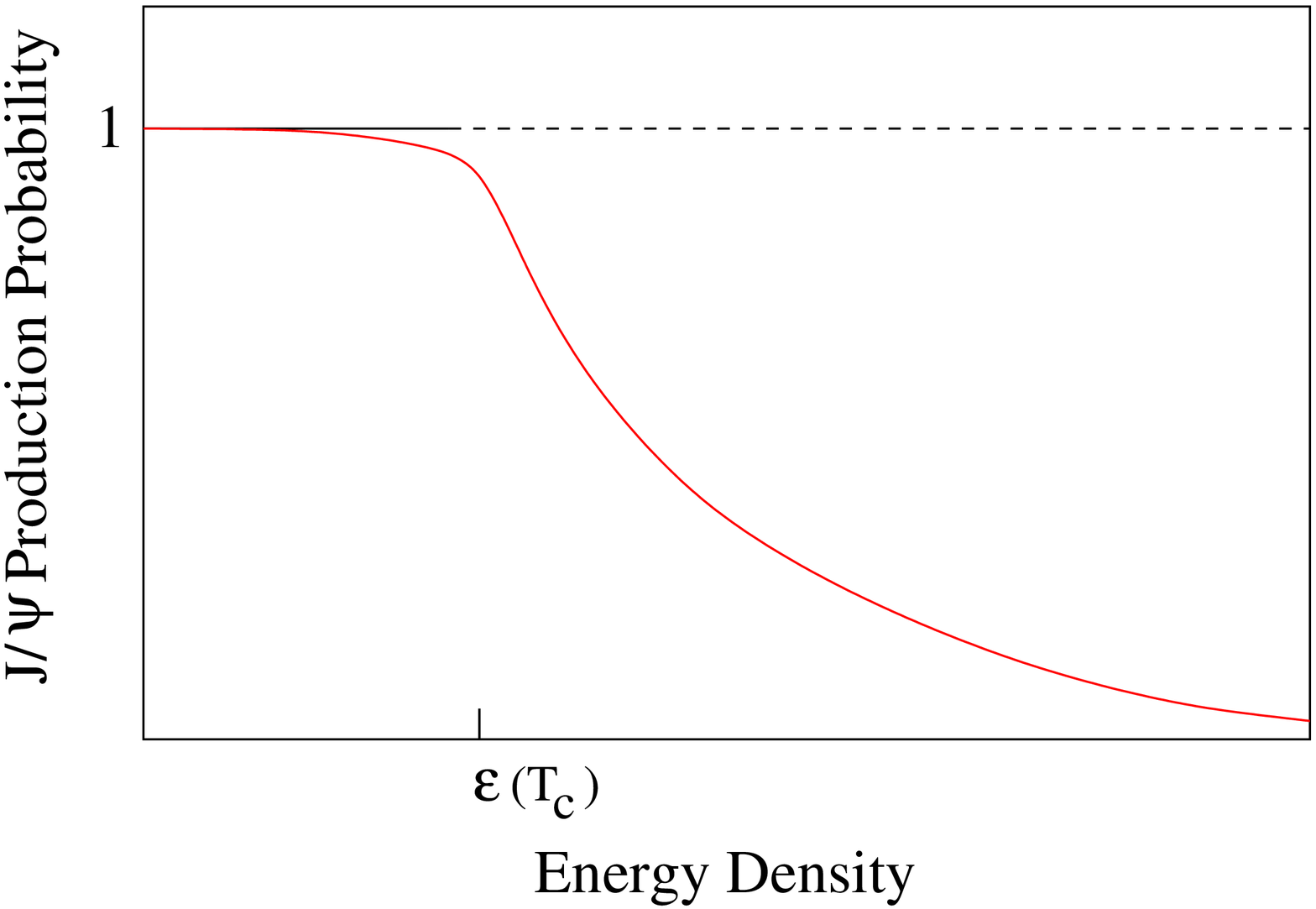,width=6.5cm}
\caption{\J~suppression by comover collisions}
\label{abs-cartoon}
\end{minipage}
\hspace{1.3cm}
\begin{minipage}[t]{7cm}
\vspace*{-4.6cm}
\hskip0.3cm
\epsfig{file=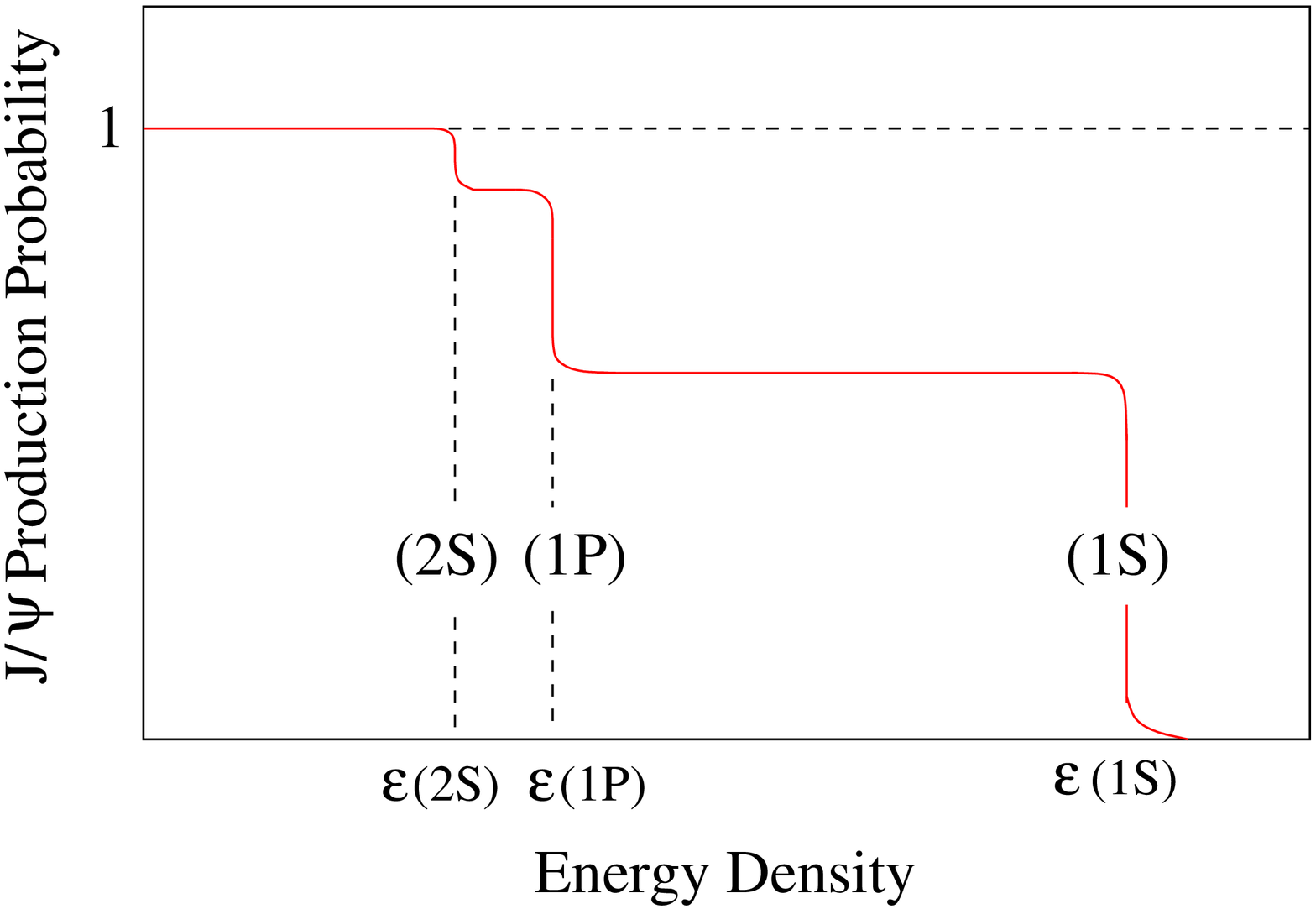,width=6.7cm}
\caption{Sequential \J~suppression by color screening}
\label{seq-cartoon}
\end{minipage}
\end{figure}

In Fig.\ \ref{abs-cartoon} we illustrate schematically the overall
behavior expected for \J~dissociation through comover collisions,
assuming that beyond a deconfinement threshold, the comover density 
increases with energy density in a monotonic fashion, with little or
no prior suppression in the hadronic regime. 

\medskip

\subsubsection{Suppression by Color Screening}

The theoretical basis of this effect has been considered in detail in
chapters 3 and 4; the color field between the heavy quarks becomes modified 
due to the presence of a medium of unbound 
color charges. The results obtained for this effect in statistical QCD 
are as such model-independent, once all calculational constraints are 
removed. What is speculative and model-dependent is its application to 
nuclear collisions: these do not necessarily produce the medium studied 
in thermal QCD, and the different evolution stages in nuclear collisions 
can introduce factors not present in the study of equilibrium thermodynamics, 
such as the oversaturation of $\C$ pairs just mentioned.

\medskip

If the medium produced in high energy nuclear collisions is indeed the
quark-gluon plasma of statistical QCD, and if charmonium production can
be treated as a distinct process within such an environment, then the
effect of color screening seems clear. The partitioning of the $\C$ pairs
produced in nucleon-nucleon collisions into hidden and open charm is
non-statistical, favouring the hidden charm sector because of dynamical
binding effects. Color screening destroys these and hence strongly 
suppresses charmonium production rates relative to those observed in 
elementary interactions. 

\medskip

A crucial feature of \J~suppression by deconfinement is its sequential
nature \cite{K-S,Gupta,D-P-S2,KKS}. In the feed-down production of \J, 
the produced medium affects the intermediate excited states, so that 
with increasing temperature or energy density, first the \J's originating
from \P~decay and then those from \X~decay should be dissociated. Only
considerably higher temperatures would be able to remove the directly
produced \J's. Such a stepwise onset of suppression, with specified
threshold temperatures, is perhaps the most characteristic feature 
predicted for charmonium as well as bottomonium production in nuclear
collisions. It is illustrated schematically in Fig.\ \ref{seq-cartoon},
where we have defined the \J~production probability to be unity if
the production rate suffers only the calculated nuclear suppression. 

\subsubsection{Enhancement through Regeneration}

In charmonium hadroproduction, \J's are formed because some of the 
$\C$ pairs produced in a given collision form a corresponding bound state.
In a collective medium formed through superposition of many nucleon-nucleon
(NN) collisions, such as a quark-gluon plasma, a $c$ from one NN collision
can in principle also bind with a $\bar c$ from another NN collision. 
This `exogamous' charmonium formation at a later evolution stage
could lead to enhanced \J~production,
provided the overall charm density of the medium at hadronization is 
sufficiently high and provided the binding force between charm 
quarks from different sources is large enough \cite{PBM,Thews,Rapp}.

\medskip

The production of $\C$ pairs in primary collisions is a hard process and
thus grows in $A-\!A$ interactions with the number of nucleon-nucleon
collisions; in contrast, the multiplicity of light hadrons grows roughly 
as the number of participant nucleons. Hence the relative abundance
of charm to non-charm quarks will be higher in $A-\!A$ than in $p-\!p$
collisions. Moreover, the $\C$ production cross section increases faster
with energy than that for light hadron production. The two effects
together imply that in the medium produced in energetic $A-\!A$ collisions, 
the ratio of charm to non-charm quarks is initially much higher than 
in a equilibrated QGP. Whether or not this results in a charmonium 
regeneration depends on two factors. On one hand, the initial charm 
oversaturation must be preserved, so that the total charm abundance 
in non-thermal. On the other hand, it is necessary that the binding
potential of random pairs into charmonia is sufficiently strong. 

\medskip

\begin{figure}[htb]
\centerline{\epsfig{file=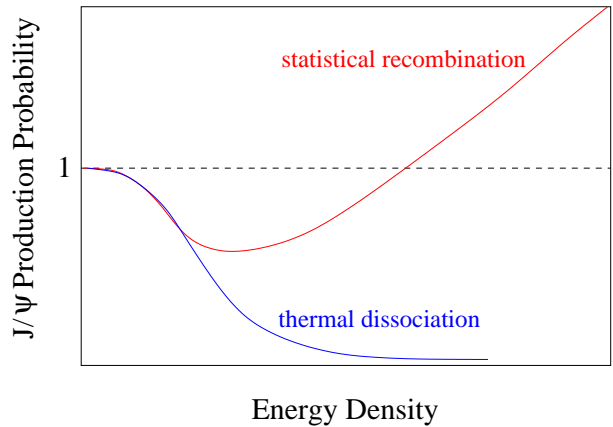,width=8cm}}
\caption{\J~enhancement by statistical regeneration} 
\label{recom}
\end{figure}

Charmonium regeneration in nuclear collisions will be addressed in a 
separate section of this Handbook \cite{PBM-St}. We only note here that
several studies have led to considerable enhancement 
factors \cite{PBM-St,Bob}, in the strongest form even predicting 
a large overall 
enhancement of \J~production in $A-\!A$ collisions relative to $p-\!p$ 
results scaled by binary collisions. One crucial prediction of the
approach is the {\sl increase} of the enhancement with centrality, as
shown in Fig.\ \ref{recom}, because of the corresponding increase in 
the number of collisions and hence of the number of $\C$ pairs. 
Another is that the distributions of the observed charmonia in
transverse momentum as well as in rapidity must be given as convolutions 
of the open charm distributions \cite{bob-recom}. What this means
for transverse momentum spectra will be addressed in the next subsection.

\subsection{Transverse Momentum Behavior}

The production pattern of charmonia as function of their transverse
momentum can provide information about the production process, the
evolution of the resonance formation and of that of the produced medium.
We begin with the production process.

\medskip 

The transverse momentum distribution of charmonia measured in $pA$
as well as in $AA$ collisions is generally broadened in comparison
to that in $pp$ interactions. The main effect causing this is the 
collision broadening of the incident gluons which fuse to make $\C$ pairs. 
A standard random walk analysis gives for the squared transverse momentum 
of the produced \J~\cite{K-N-S,early}
\be
\langle p_T^2 \rangle_{pA} = \langle p_T^2 \rangle_{pp} + N_c^A \delta_0
\label{pTpA}
\ee
in $p\!-\!A$ and to
\be
\langle p_T^2 \rangle_{AA} = \langle p_T^2 \rangle_{pp} + N_c^{AA} \delta_0
\label{pTAA}
\ee
in $A\!-\!A$ collisions. 
Here $N_c^A$ denotes the average number of collisions of a projectile parton 
in the target nucleus $A$, and $N_c^{AA}$ the sum of the average number of 
collisions of a projectile parton in the target and vice versa, at the given 
centrality. The parameter $\delta_0$ specifies the average ``kick'' which 
the incident parton receives in each subsequent collision. 
The crucial parameters are thus the elementary $\langle p_T^2 \rangle_{pp}$ 
from $p\!-\!p$ interactions and the value of $\delta_0$, determined by 
corresponding $p\!-\!A$ data; both depend on the collision energy. 
The $A$-dependence of $N_c^A$ as well as the behavior of $N_c^{AA}$ as 
function of centrality can be obtained through a Glauber analysis including
absorption in cold nuclear matter, thus specifying the ``normal'' centrality 
dependence of $\langle p_T^2 \rangle_{AA}$. In the absence of any 
anomalous supppression, this would be the expected behavior of the
average \J~transverse momentum. 

\medskip

\begin{figure}[htb]
\centerline{\epsfig{file=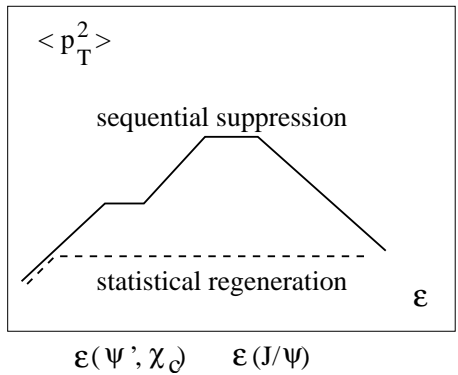,width=6cm}}
\caption{Transverse momentum broadening in sequential suppression
vs. statistical regeneration scenarios}
\label{pt-patterns}
\end{figure}

Given the sequential suppression by color screening, the \J's observed
for energy densities in a range above the onset of anomalous suppression,
$\e(2S),\e(1P) \leq \e \leq \e(1S)$ are the directly produced $1S$
states unaffected by the presence of the QGP. They should therefore still
show the normal broadening pattern (\ref{pTAA}), increasing linearly with
the number of collisions as long as $\e < \e(1S)$ \cite{KKS}. This broadening
is a memory of the initial state and hence essentially absent if the
\J~is formed only at the hadronization point, where such memory has
been destroyed. Thus \J~production through regeneration should show
a flat distribution as function of the number of collisions
\cite{bob-recom}, in contrast to the rise expected in the sequential 
suppression scenario. The behavior of  $\langle p_T^2 \rangle_{AA}$
thus should provide a clear indication of how the observed \J's were
produced. In Fig.\ \ref{pt-patterns} we illustrate the different
patterns expected.

\medskip

\medskip

A further interesting signal is how an anomalous suppression behaves
as function of $p_T$. If the nascent \J~is formed as a small color
singlet gradually expanding to its physical size, then it can be 
out of the deconfining medium either spatially \cite{K-P} or temporally
\cite{B-O} before suffering serious suppression, provided it has a high 
enough transverse momentum. The
``end'' of anomalous suppression as function of $p_T$ could thus
give indication on the size or life-time of the QGP.

\medskip

More recently, calculations have appeared \cite{AdS}
which are based on the AdS/CFT 
correspondence applied to QCD; they indicate the opposite effect, suggesting 
that \J's of high transverse momentum should suffer more suppression.  
It must be kept in mind, however, that binding and the dissociation of 
specific bound states are clearly non-conformal phenomena, so that such 
modelling is not necessarily valid.

\section{Summary of the Theoretical Status}

The theoretical work of the past years has done much to clarify the
questions that must be answered in order to reach final conclusions.

\medskip

In statistical QCD, improved MEM techniques, better statistics (from
faster computers) as well as larger lattices (on larger computers),
promise that in the foreseeable future we will have reliable first
principle calculations of the in-medium behavior of quarkonia.

\medskip

These will hopefully then be related to the results from heavy quark 
studies in finite temperature lattice QCD. In this work 
(see e.g. \cite{KFPZ,P-P,K-Z}),
there are at present still several unsolved questions, concerning
in particular gauge invariance and the specification of specific color 
states, and the possible in-medium transitions between color states.
Recent analytical work may provide help in dealing with them
\cite{Nora,Laine,OP,Blai}.  

\medskip

In parallel, potential theory studies are expected to provide further
insight into the melting of quarkonia in a hot QGP. One of the basic
issues to be addressed here is if and how the multi-component problem
near $T_c$ can be addressed. In particular, the role of the gluonic
dressing in heavy quark binding has to be taken into account \cite{HS-sqm}. 

\medskip

Once the in-medium behavior of quarkonia is clarified in statistical 
QCD, one can hope to address the behavior observed in high energy nuclear
collisions. If these indeed produce a QGP in the sense of statistical QCD,
sequential melting threshold could provide a direct quantitative
connection between theory and experiment. If, on the other hand,
the measured quarkonia are largely due to regeneration at hadronization,
i.e., to the combination of secondary $Q$ and $\bar Q$ from different 
nucleon-nucleon interactions, it is not evident how such a connection
can be obtained. It would, however, provide clear evidence for the
production of a deconfined thermal medium. 

\vskip3cm

%% file: NNkl.tex
\oddsidemargin 15pt
\topmargin 0pt
\headheight 00pt
\headsep 00pt
\textheight 235mm
\textwidth 160mm
\voffset=-0.5cm
\hoffset=-0.5cm
\def\jpsi{$J/\psi~$}
\def\J{$J/\psi~$}
\def\j{J/\psi~}
\def\X{$\chi_c$}
\def\x{\chi}
\def\P{$\psi'~$}
\def\p{\psi'}
\def\U{$\Upsilon$}
\def\u{\Upsilon}
\def\C{c{\bar c}}
\def\B{b{\bar b}}
\def\cg{c{\bar c}\!-\!g}
\def\bg{b{\bar b}\!-\!g}
\def\b{b{\bar b}}
\def\q{q{\bar q}}
\def\Q{Q{\bar Q}}
\def\e{\epsilon}
\def\t{\tau}
\def\l{\Lambda_{\rm QCD}}
\def\A{$A_{\rm cl}$}
\def\a{\alpha}
\def\N{$n_{\rm cl}$}
\def\n{n_{\rm cl}}
\def\S{S_{\C}}
\def\s{s_{\rm cl}}
\def\bb{\bar \beta}
\def\chiral{\psi {\bar \psi}}
\def\CMP{{ Comm.\ Math.\ Phys.\ }}
\def\NP{{ Nucl.\ Phys.\ }}
\def\PL{{ Phys.\ Lett.\ }}
\def\PR{{ Phys.\ Rev.\ }}
\def\PRep{{ Phys.\ Rep.\ }}
\def\PRL{{ Phys.\ Rev.\ Lett.\ }}
\def\RMP{{ Rev.\ Mod.\ Phys.\ }}
\def\ZP{{ Z.\ Phys.\ }}
\def\EPJ{{Eur.\ Phys.\ J.\ }}

\def\Et{$E_{{\rm T}}$}
\def\Ez{$E_{\rm ZDC}$} 
\def\pt{$p_{\rm T}$\ }
\def\PT{p_{\rm T}\ }
\def\TR2{p_{\rm T}^2}

\def\gev{GeV}
\def\psip{$\psi'$\ }
\def\sgabs{$\sigma_{\rm abs~}$}
\def\Sgabs{\sigma_{\rm abs~}}
\def\sqs{\sqrt s }
\def\npt{N_{\rm part~}}
\def\ncl{N_{\rm coll~}}
\def\nch{N_{\rm ch~}}
\def\Rab{R_{\rm AB} }
\def\Raa{R_{\rm AA} }
\def\Rcp{R_{\rm CP} }

\def\be{\begin{equation}}
\def\ee{\end{equation}}
\def\lsim{\raise0.3ex\hbox{$<$\kern-0.75em\raise-1.1ex\hbox{$\sim$}}}
\def\gsim{\raise0.3ex\hbox{$>$\kern-0.75em\raise-1.1ex\hbox{$\sim$}}}



\bigskip

{\bf \LARGE II Experiment}

\vskip1cm

\setcounter{section}{0}
\section{Charmonium Experiments at the CERN-SPS} 
\setcounter{equation}{0}
\setcounter{figure}{0}
\bigskip

At the CERN-SPS, experiments NA38, NA50 and NA51 have systematically
measured charmonium production with incident protons, Oxygen, Sulphur
and Lead beams at various incident momenta. The latest version of the 
detector~\cite{NA50NIM?} was based on a muon pair spectrometer
able to stand high intensity incident beam fluxes. The 
spectrometer was further equipped with three devices leading to 
three independent estimates of the centrality of the collision.        
The Electromagnetic Calorimeter measured \Et, the integrated flux of 
neutral transverse energy, mainly due to neutral pions produced in 
the collision. The Zero Degree Calorimeter measured \Ez, the forward 
energy carried by the beam spectator nucleons. The Multiplicity 
Detector counted the charged particles produced in the reaction. 
All these three measurements are related to the impact parameter of 
the collision. 

\medskip

The properties of the detectors allowed the collection of large 
samples of data with incident protons and ions. This, in turn,  
made possible the study of both \J and \P production with minimal 
systematic uncertainties, thanks to the simultaneous detection of 
Drell-Yan events which could be used as appropriate "debiasing"  
tools.  

\medskip

From the very beginning, it was taken for granted that 
the study of \J production from different nuclear targets 
would provide, by some kind of extrapolation, the 
appropriate reference baseline, i.e., the "normal" charmonium 
behavior, relative to which the heavy ion collision specific 
features would be easily identifiable. Ideally, the reference 
data would have had to be collected with proton beams of the 
same energy of that of the corresponding ion beams, namely 
200\,GeV, the energy of the Oxygen and Sulphur beams, and 
later 158\,GeV, the energy of the Lead and Indium beams. 
For low cross-section measurements like charmonium production, 
it was highly desirable that proton beams be primary beams, 
directly extracted from the accelerator. Because of various 
reasons, low energy primary proton beams were not available 
at CERN. It took 18 years since the start of the heavy ion 
program  until they finally became available, for about 3 days, 
and could thus be used by experiment NA60. 
   
\bigskip

\subsection{The Nuclear Dependence of Charmonium Production}

\bigskip

The uncontroversial feature observed since long in charmonium 
production measurements in $p-\!A$ reactions is that, at fixed 
collision energy, quarkonium production rates per target nucleon 
decrease with increasing~\,$\!A$, the target atomic mass number. 
This effect has been traditionally, 
and somewhat empirically, quantified  with the parametrization 
$\sigma_{pA} = \sigma_{0} \times A^{\alpha}$ 
which leads to measured values of ${\alpha}$ lower than unity.
More recently, it has been assumed that some kind of $\C$ state   
is created in the reaction which is dissociated, or rather prevented 
to finally form a bound state, through 
interaction with the surrounding nuclear matter. 
In other words, the $\C$ pairs are created proportionally to the number 
of nucleon-nucleon collisions or, equivalently, to~$\!A$. Some interact 
with the surrounding nuclear medium while moving through the target and 
are no more able to reach a bound state. The final number of charmonium 
bound states reaching the detector exhibits therefore an apparent 
suppression with respect to the original~$\!A$ dependence of the $\C$ 
pair production. An appropriate variable to parametrize 
the measured \J cross-sections should then be the number of nucleons 
that the created state can potentially interact with \cite{GershelHuefner}.
This number can be calculated from the product $\rho\times L$,  
where $\rho$
is the nuclear density distribution and $L$ the length of nuclear 
matter the $\C$ state traverses while escaping from the interaction 
region. The charmonium survival probability can then be calculated
as a function of an "absorption" or dissociation cross-section   
\sgabs,  using the Glauber formalism of 
nuclear scattering theory or, alternatively, a simplified "${\rho L}$" 
exponential parametrization. Comparison with the experimentally 
measurable survival probability allows to determine 
\sgabs.  
As a matter of fact, systematic experiments can even provide some 
guidance for the validity of this elaborated view of nuclear effects 
in charmonium production. 

\medskip

Let us underline here that the ${\alpha}$ parametrization can be
used to fit a set of several measurements made under same 
conditions with different nuclear targets. The procedure then 
becomes an hypothesis test of such a parametrization. 
It can also be assumed that nature behaves according to this 
law and apply the parametrization to only two different targets.
The procedure then would just provide an estimate of the numerical 
value of ${\alpha}$, taking for granted that such a 
parametrization correctly accounts for the elementary process.
Identical procedures can be followed under the 
dissociation cross-section interpretation and resulting 
parametrization.     
Finally, let us point out that 
${\alpha}$ and the more elaborated \sgabs  
are obviously correlated.

\subsection{Normal Charmonium Production}

\begin{figure}[h!]
   \begin{minipage}[t]{8cm}
   \resizebox{0.89\textwidth}{0.935\textwidth}{%
   \includegraphics*{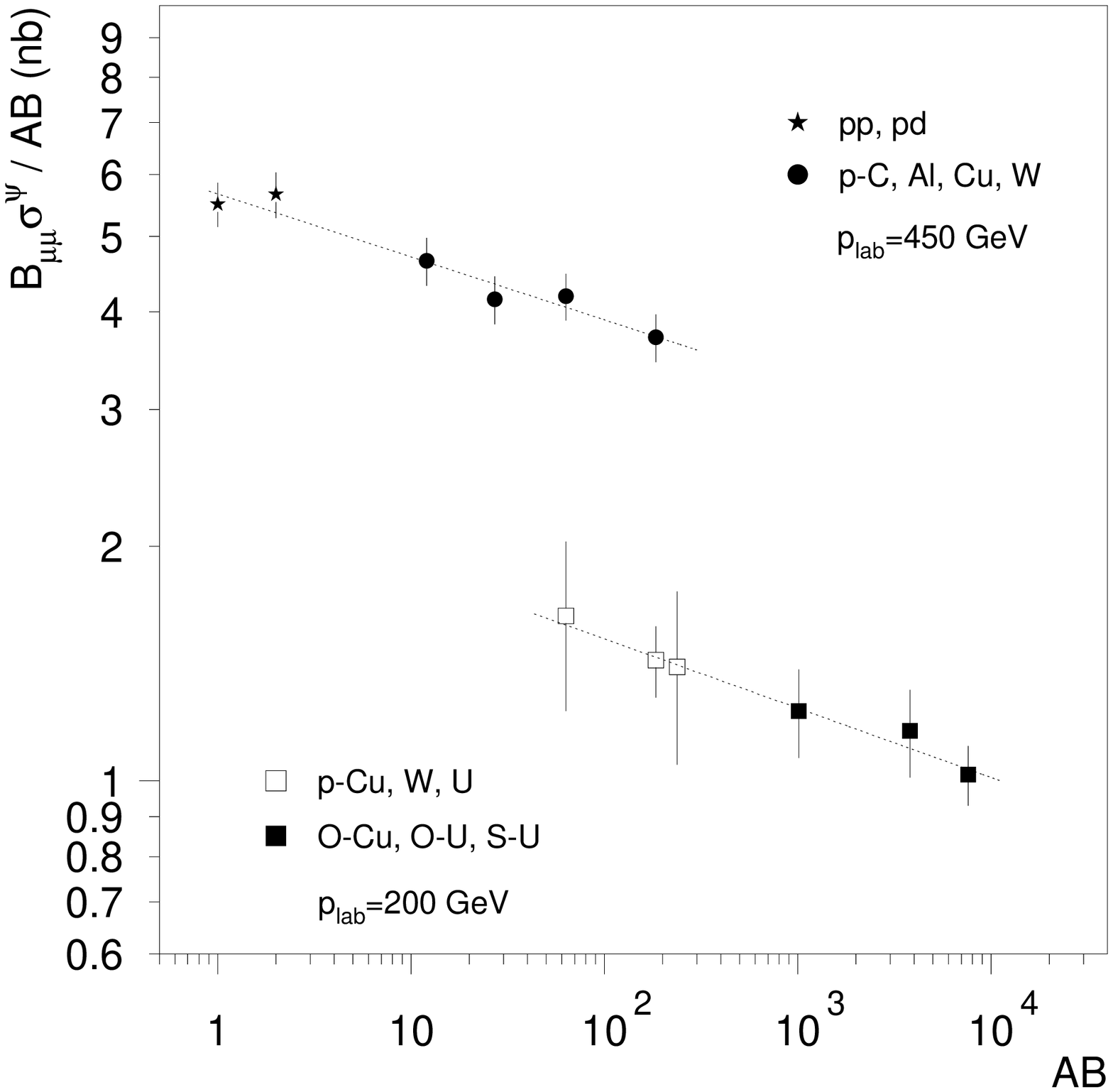}}
   \parbox{7.5cm}{\vspace*{-0.3cm} 
   \caption{The 450 and 200~GeV \J cross-sections per nucleon 
         (times branching ratio into dimuons, $B_{\mu\mu}$), in the rapidity 
         domain $ 3.0 < y_{\rm lab} < 4.0 $, as a function of ${A \times B}$. 
         The lines correspond to the best fit of the function 
\mbox{$ B_{\mu\mu}^{\j}/(\!A\times \!B) = \sigma_0 (\!A \times \!B)^{\alpha-1} $}.}
\label{ps450-200-NA38}}
   \end{minipage}  
   \hspace*{0.5cm}
   \begin{minipage}[t]{8cm}
   \vspace*{-7.45cm}
    \resizebox{0.89\textwidth}{0.93\textwidth}{%
   \includegraphics*{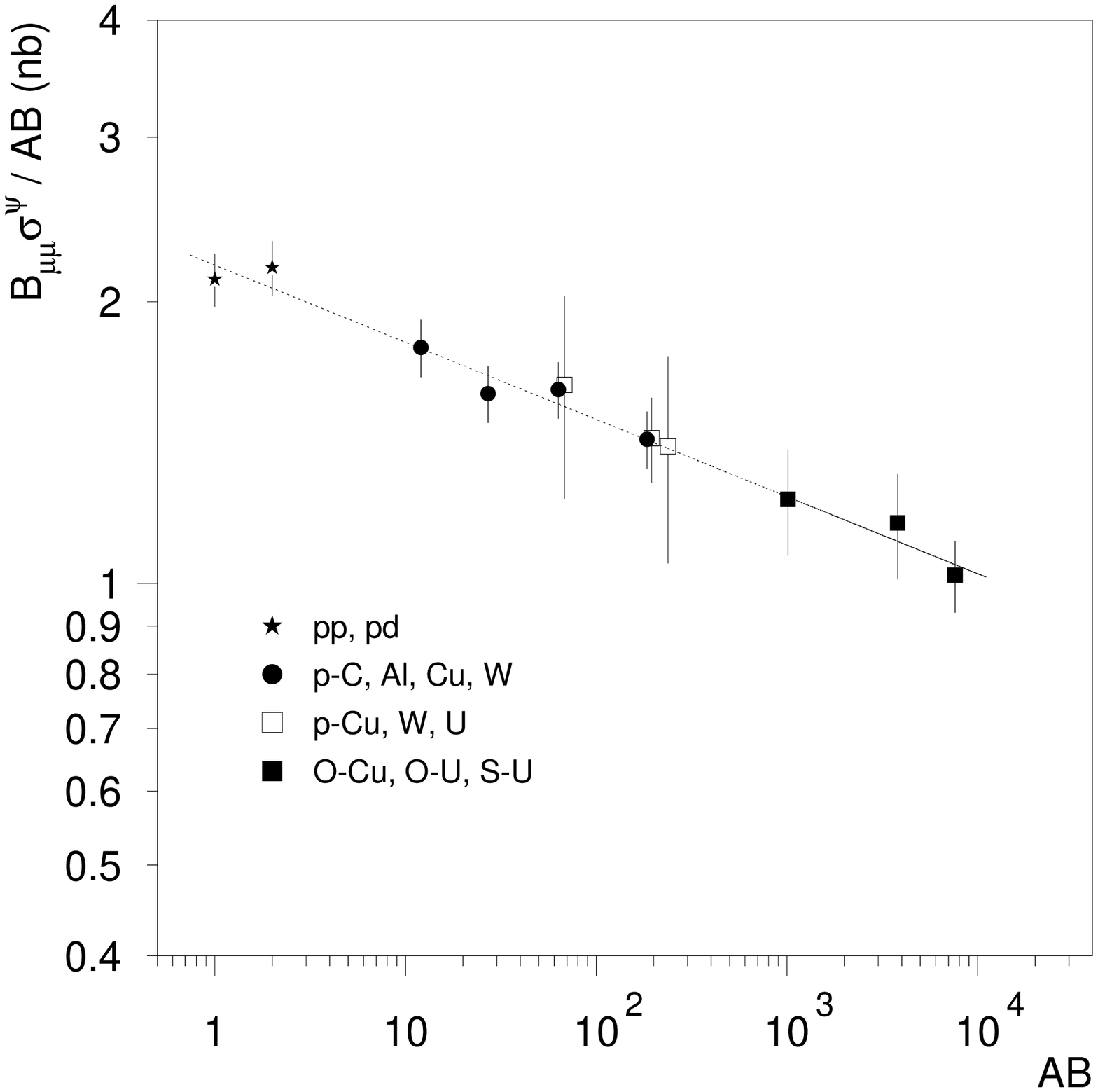}}
   \vspace*{-0.4cm}
   \parbox{7.5cm}{\vspace*{-0.35cm}
   \caption{\J cross-sections per nucleon, times b.r., plotted as a function 
         of the product of $ A \times B$ .
         The 450~GeV values are here rescaled to 200~GeV and recomputed 
         in the c.m.s. rapidity range 0.0-1.0. The line corresponds to 
         the best fit of the function 
         \mbox{$ B_{\mu\mu}^{\j}/(\!A\times \!B) = \sigma_0 (\!A \times \!B)^{\alpha-1} $}.}
\label{rescaled200-NA38}}
    \end{minipage} 
   \vspace{0.3cm}
    \end{figure}

The first significant samples of charmonium events produced 
in $p-\!A$ reactions were collected by experiment NA38 
in 1988, with 450\,GeV incident protons on various targets. 
Later on, they were complemented with measurements obtained, 
as a by-product, from experiment NA51 which provided the data 
allowing to extract the charmonium production cross-sections 
for p-p and p-d collisions at 450~GeV \cite{pppdNA51}. 
All these data, reanalyzed, with identical 
procedures as those used to analyze the heavy ion 
data, led to the production cross-sections
plotted in Fig.~\ref{ps450-200-NA38}
as a function of the atomic mass 
number of the target~\cite{pa-450-NA38}.

\medskip
 
\noindent Some time later, this set of measurements could be complemented 
with other results from the same experiment NA38, both with 
protons and with Oxygen and Sulphur beams incident on 
Copper and Uranium targets, all for collisions at 200~GeV per nucleon.    
When plotted as a function of the product of the atomic numbers 
$\!A \times \!B$, the comparison of the charmonium 
production cross-sections at 450 and at 200~GeV exhibits a 
remarkable feature: the fitted value of ${\alpha}$ was the same, 
namely \mbox{${\alpha}_{450}^{\j}$ = 0.919 $\pm$ 0.015}
and
\mbox{${\alpha}_{200}^{\j}$ = 0.911 $\pm$ 0.034}.
The remarkable compatibility of the two values justifies a 
simultaneous fit of the two sets of points, imposing a single 
${\alpha}$ exponent. This global fit leads to 
${\alpha}^{\j}$ = 0.918 $\pm$ 0.015.
The observed unexpected agreement between energies and 
reactions is a serious double hint that nuclear dependence is 
identical for 450~GeV and for 200~GeV and that, moreover, 
charmonium production in light ion-induced reactions, namely 
O-Cu, O-U and S-U, can be considered as normal, namely identical 
to the one observed in $p-\!A$ interactions. From the global fit 
can also be determined the ratio between the values of $\sigma_0$ 
for the 200 and 450 data sets. It amounts to 0.38 $\pm$ 0.04 
and results from the changes in both  $\sqrt s$ and in the rapidity 
domain covered by the two sets of data. The same factor can be used 
to rescale the 450~GeV \P cross-sections to the incident momentum 
and rapidity range  of the 200~GeV data. In fact the ratio between 
the \P and \J production cross-sections in $p-\!A$ collisions seems 
to be independent 
of $\sqrt s$, within the energy domain covered by the currently 
existing measurements \cite{CL-a-voir}.
The \J cross-sections per nucleon-nucleon collision, rescaled if necessary 
as described above, are shown in Fig.~\ref{rescaled200-NA38}.
The excellent agreement between the 200 and the 450~GeV results can be easily 
judged in the case of p-Cu and p-W collision systems for which both 
measurements exist.

\medskip

\begin{figure}[h!]
\begin{center}
\resizebox{0.49\textwidth}{0.49\textwidth}{
\includegraphics*{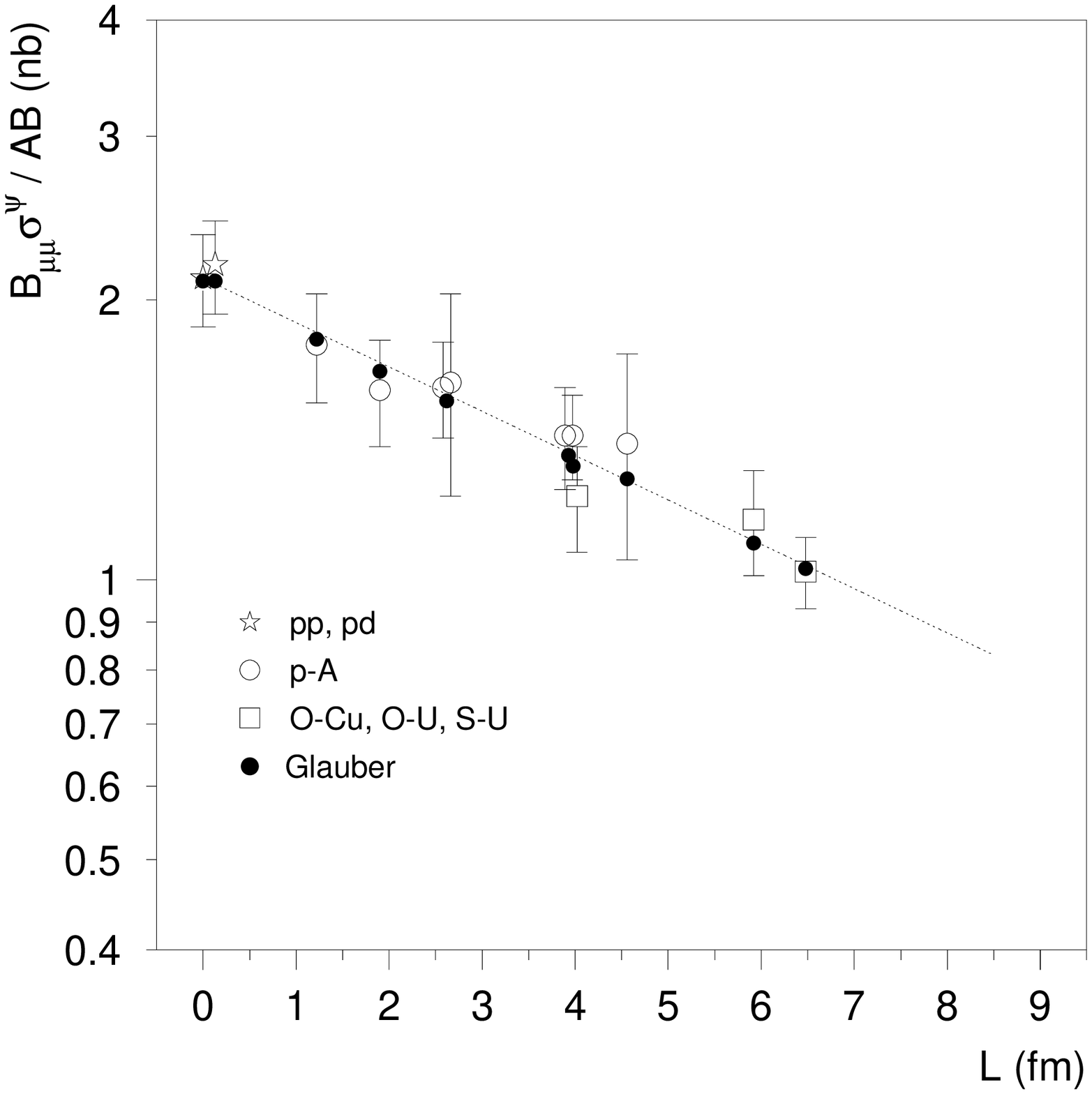}}
\end{center}
\vspace*{-0.8cm}
\caption{\J cross-sections per nucleon, times b.r., plotted as a function 
         of $L$. The 450~GeV values are rescaled to the 200~GeV kinematics. 
         The closed circles are the results of the "Glauber" calculation.}
\label{Ldependence200-NA38}
\end{figure}

\noindent Within the scenario of a dissociation cross-section, its numerical value 
can also be extracted from the data in the frame of a Glauber calculation. 
Fig.~\ref{Ldependence200-NA38} compares the results, rescaled to 200~GeV 
and to the 0.0-1.0 rapidity interval,  with  
an overall fit which leads to \mbox{$\Sgabs = 7.1 \pm 1.2$ mb}.     
The data exhibit a nice overall agreement with the "Glauber" calculated 
values and 
suggest good compatibility with the model, at their level of precision 
(systematic uncertainties have to be taken into account here). Assuming 
now the validity of the model, the most precise value of \sgabs can be 
determined using only the 3 most precise measurements, namely p-C, p-Cu 
and p-W as obtained at 450~GeV. The resulting value obtained from these 
cross-sections is \mbox{5.6 $\pm$ 0.4 mb}.

\medskip

The agreement of the Glauber calculation with the experimental data strongly 
suggests that the A-dependence of \J production can be properly accounted for 
by final state nuclear absorption up to and including S-U reactions. 
In particular, there is no indication for any new absorption or dissociation 
mechanism which would suppress the S-U value relative to the reference 
baseline established from the p-A systematics.  

\bigskip

The same experiments mentioned above did provide the corresponding 
first results for \P  production. Statistical errors, of 
the order of 10\%, were very much too large for precise conclusions.
In order to minimize systematic effects,  
the ratios  \P/\J measured at 450~GeV in p-C, p-Al, p-Cu and p-W can be 
considered, as taken under identical experimental conditions and subject 
to minimal systematic effects with respect to each other. 
Their values lead to  
\mbox{$\alpha^{\p}-\alpha^{\j} = -0.060 \pm 0.038$} 
already suggesting, although within large errors, a normal absorption 
for \P higher than for \J. 

\medskip

In the limited kinematical domain explored by the experiments,
the first set of data from experiments NA38 and NA51 led, within their 
uncertainties, to the following strong indications :
\begin{itemize}
\vspace*{-0.2cm}
\item \J and \P are absorbed in nuclear matter.
\vspace*{-0.2cm}  
\item For $J/\psi$, absorption is similar for 450~GeV 
p-A collisions and for 200~GeV p-A and light ion collisions, up to S-U.
\vspace*{-0.2cm}
\item The absorption is stronger for \P than for \J. 
\vspace*{-0.2cm}
\end{itemize} 
Much more precise results obtained later by experiment NA50 allowed to 
confirm these results both thanks to significantly increased statistics 
and to a thorough study of systematic uncertainties.

\bigskip

\begin{figure}[h!]
   
   \begin{minipage}[t]{8cm}
   \resizebox{0.89\textwidth}{0.89\textwidth}{%
   \includegraphics*{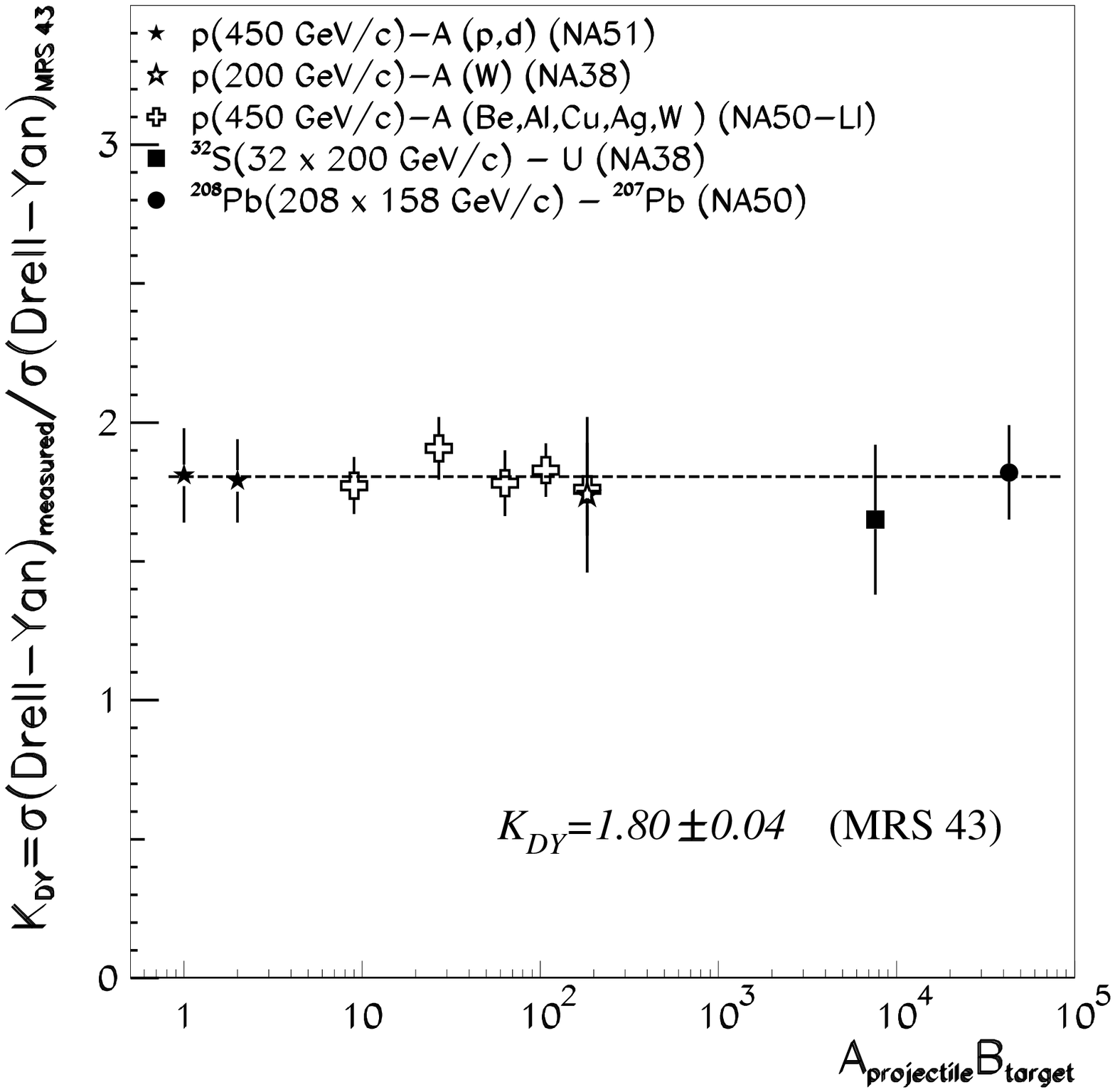}}
   \parbox{7.5cm}{\vspace*{0.1cm} 
   \caption{The Drell-Yan "K-factor", as determined experimentally, 
            from p-p up to Pb-Pb interactions 
           (from~\cite{qm2002}).}
   \label{kdy}}
   \end{minipage}  
   \hspace*{0.5cm}
   \begin{minipage}[t]{8cm}
   \vspace*{-6.9cm}
   \resizebox{0.89\textwidth}{0.93\textwidth}{%
   \includegraphics*{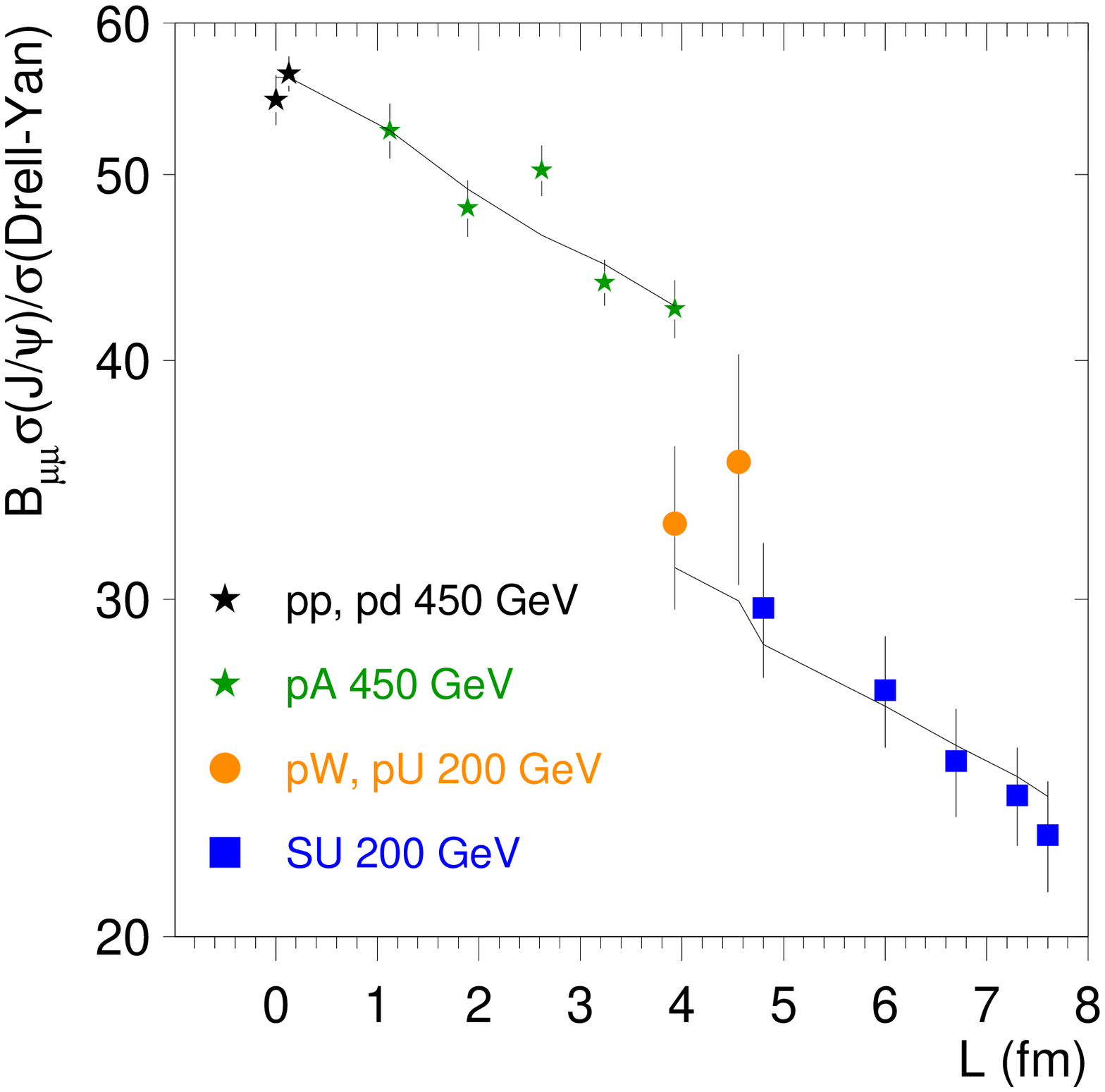}}
   \vspace*{-0.6cm}
   \parbox{7.5cm}{\vspace*{-0.5cm}
   \caption{The ratio $B_{\mu\mu}^{\j}\sigma_{corr}^{\j} / ~\sigma^{DY}$.
            The lines are the result of the simultaneous Glauber  fit to 
            all the data points.}
   \label{SSUnormality}} 
   \end{minipage} 
   \vspace{0.3cm}
   \end{figure}

In order to have full control on systematic effects, Drell-Yan muon pairs, 
always simultaneously measured with charmonium,  are the ideal tool despite 
their significantly lower production cross-section. 
They allow to use the ratio \J/~Drell-Yan, indeed with increased statistical 
erors but practically free from systematic uncertainties which are usually 
much more difficult to bring under control.
Moreover, they allow with minimal theoretical input the study of 
nucleus-nucleus collisions as a function of centrality. Indeed, NA50 has 
proven experimentally that Drell-Yan production is proportional to its 
theoretically computed value or, equivalently, to the number of 
nucleon-nucleon collisions, from p-p up to Pb-Pb interactions (Fig.~\ref{kdy}). 
It therefore exhibits no nuclear dependence.
%
%
The coherent study of the ratio \J / Drell-Yan, both for p-A collisions at 
450~GeV (the so-called low intensity (LI) although high statistics sample) 
and for some p-A and S-U interactions at 200~GeV, the latter as a function 
of centrality, leads to another estimate of the absorption 
cross-sections~\cite{Ale03A}. 
An independent "Glauber" fit on the S-U ratios gives 
\mbox{$6.3 \pm 2.9~\rm{mb}$} whereas a simultaneous fit to all the data 
gives \mbox{$4.3 \pm 0.6~\rm{mb}$}, as illustrated in Fig.~\ref{SSUnormality}.
This results further support previous hints that within the errors, 
no sizable additional suppression mechanism is present in S-U collisions 
with respect to p-A.

\subsection{The First Hints of an Anomaly in Pb-Pb Collisions}

\begin{figure}[h!]
\begin{minipage}[t]{8cm}
\resizebox{0.99\textwidth}{0.99\textwidth}{
\includegraphics*{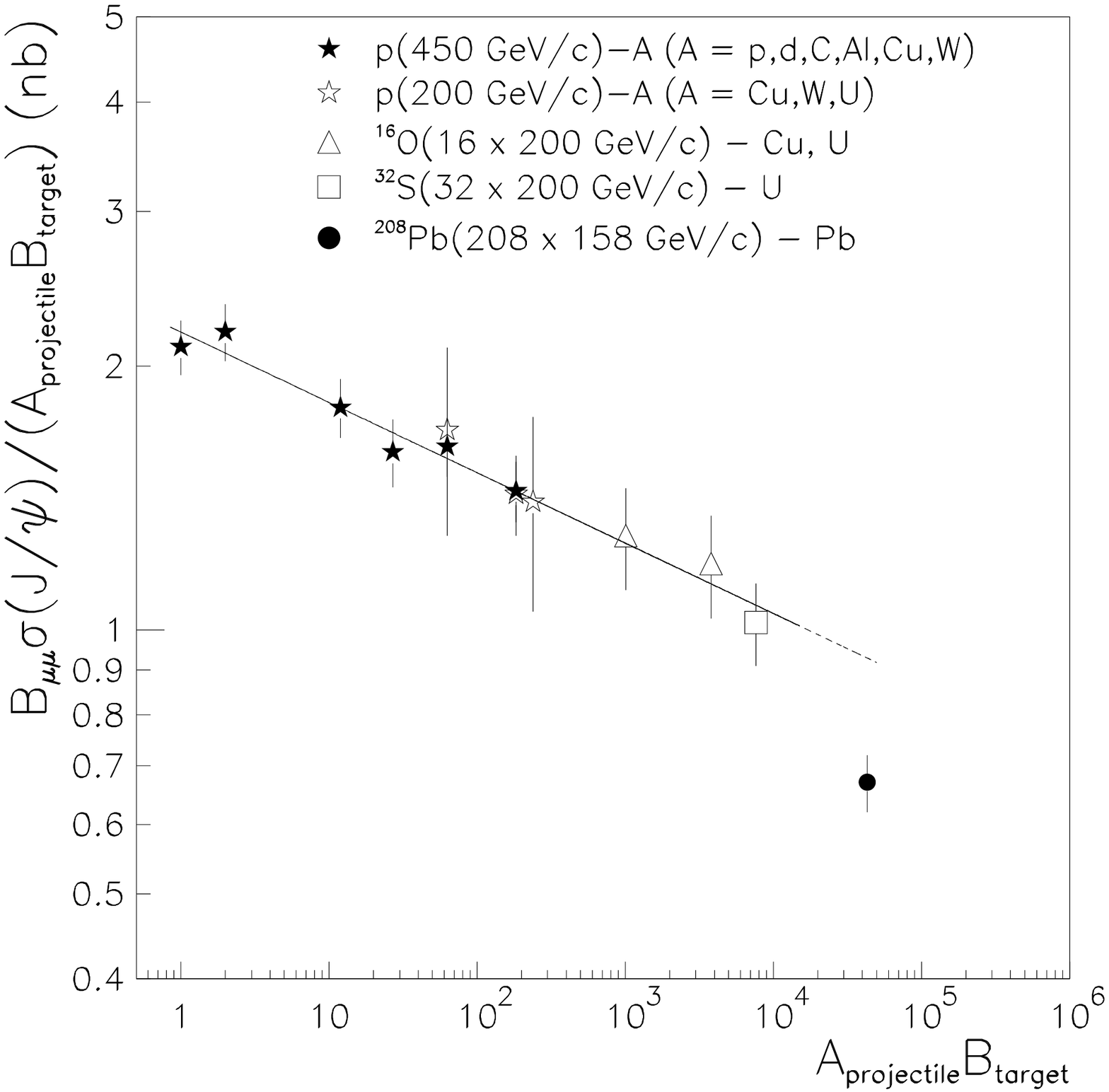}}
\end{minipage}
\begin{minipage}[t]{8cm}
\resizebox{0.99\textwidth}{0.99\textwidth}{
\includegraphics*{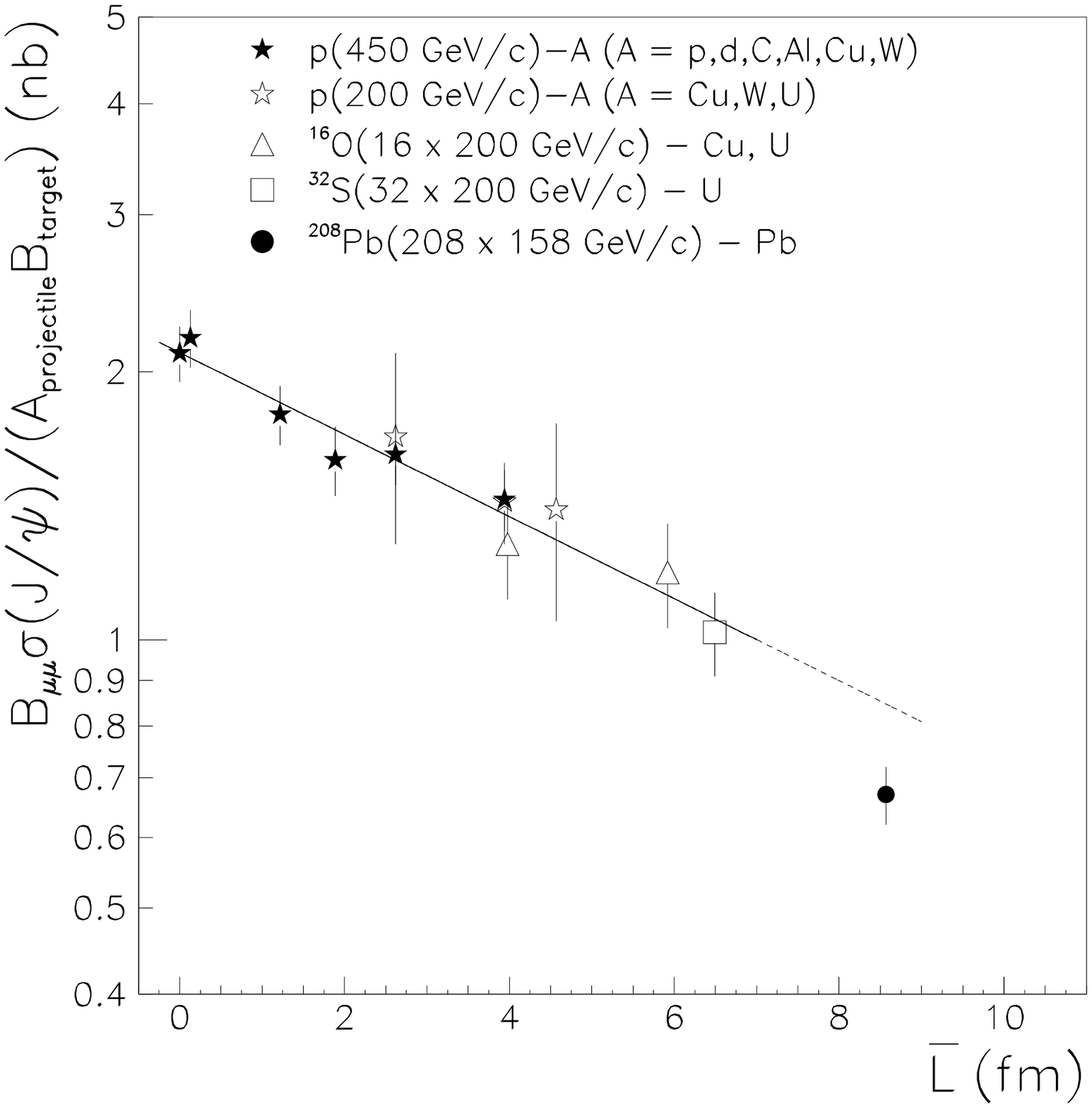}}
\end{minipage}
\caption{The \J cross-section per nucleon, times b.r., with the 450~GeV 
         results rescaled to the same kinematical domain as 200~GeV data, 
         using the "simultaneous"  fit method, (left) as a function of 
         A$\times$B and (right) as a function of L.
         The Pb-Pb result, obtained 
         at 158~Gev, is also rescaled to 200~GeV. }
\label{Integrated_Pb-Pb_anomaly}
\end{figure}

The experimental evidence of an unexpected \J suppression resulted 
from the first Pb-Pb data sample, collected in 1995~\cite{pap02}. 
It is illustrated on Fig.~\ref{Integrated_Pb-Pb_anomaly}. 
The anomalous character results from the comparison of the \J production 
cross-section per nucleon in Pb-Pb interactions with the assumed 
{\em normal} behaviour. The latter, as explained in detail above, 
was established by simple extrapolation of the behaviour obtained 
from the "simultaneous" fit method applied to proton and light 
ion-induced interactions.  

\medskip

The plot should be taken as a strong but only qualitative 
evidence because of the following {\em caveat}. The experimental reference 
is based here on the very first p-A and subsequent S-U results, which suffer 
from both non negligible statistical and also systematic errors. The latter 
have to be taken into account when using results from different setups 
(NA38, NA51). Moreover, it is also based on the implicit assumption contained 
in the "simultaneous" fit procedure of an energy independent normal absorption 
cross-section, as suggested by the results available at that time.       

\subsection{Anomalous $J/\psi$ Suppression in Pb-Pb Collisions}

From the early first indication, it took several years to learn how to make 
the measurement of charmonium suppression in Pb-Pb collisions. The goal was 
finally reached with three different sets of p-A measurements  performed from 
1996 until year 2000 and two sets of Pb-Pb data collected in 
1998 and 2000.

\medskip

\begin{figure}[th!]
\centering
\begin{tabular}{cc}
\resizebox{0.48\textwidth}{!}{%
\includegraphics*{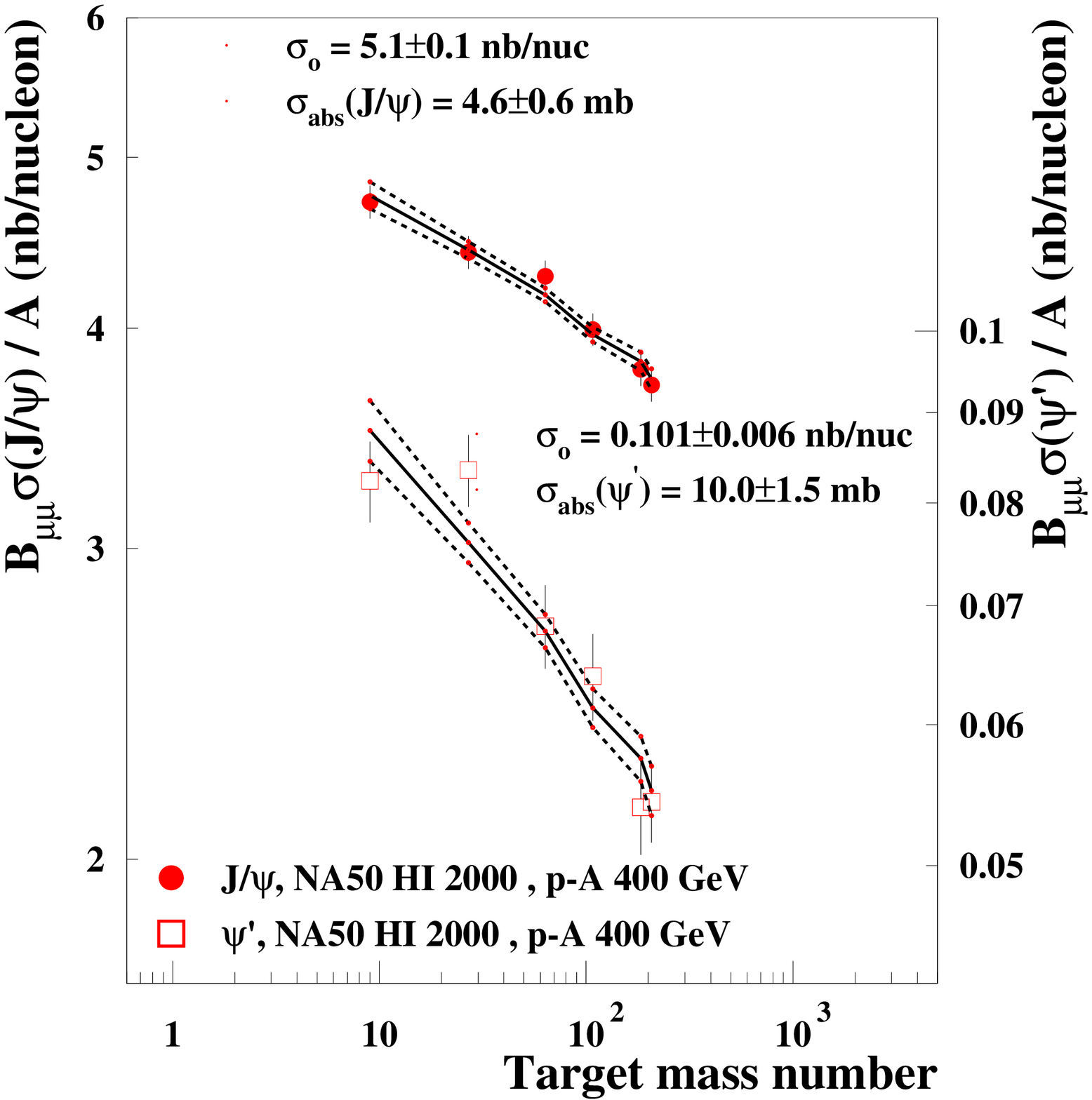}}
&
\resizebox{0.48\textwidth}{!}{%
\includegraphics*{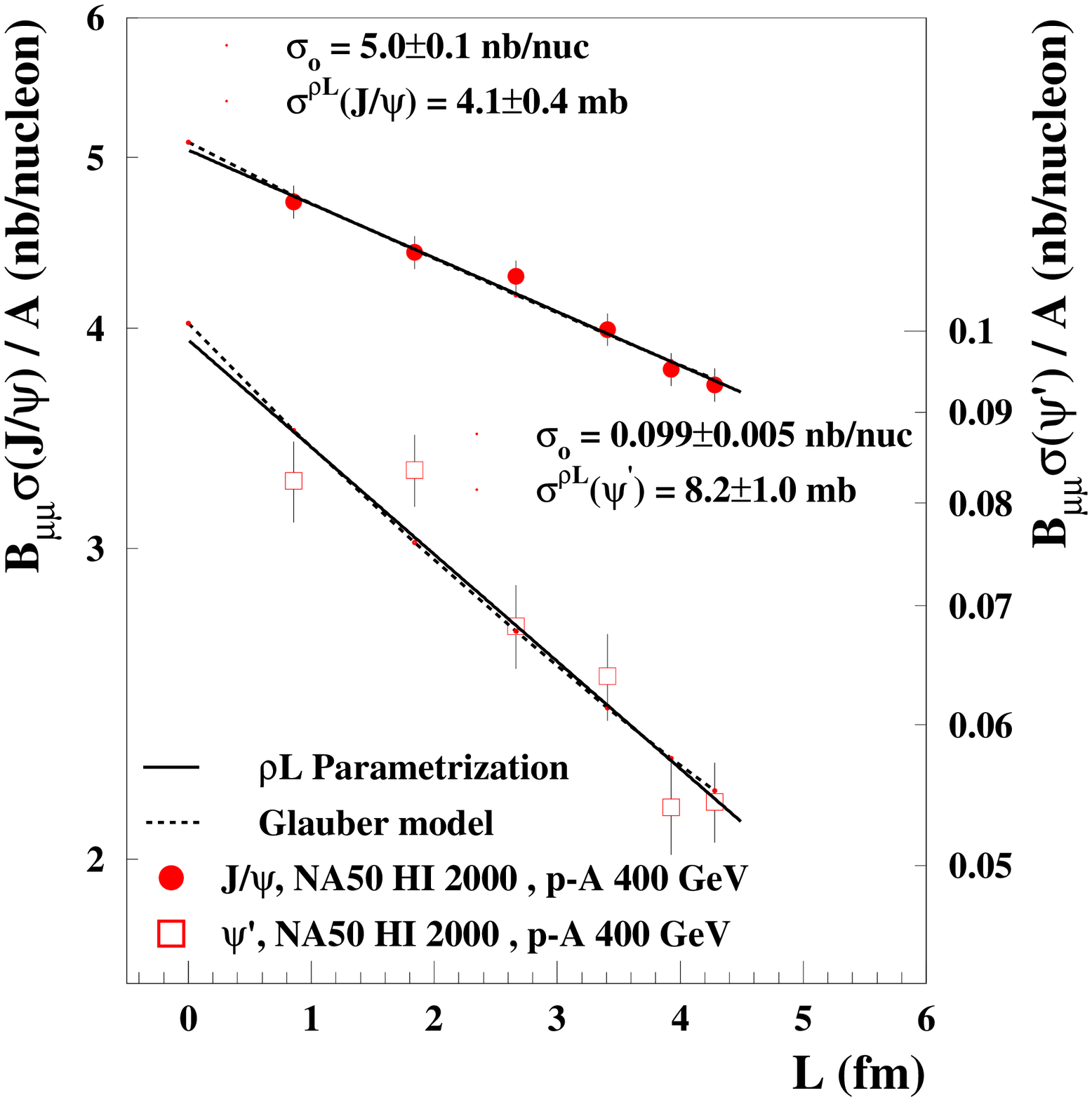}}
\end{tabular}
\vspace{-0.4cm}
\caption{The \J and \P cross-sections per nucleon adjusted by the
         "Glauber model" parametrization~(left) and compared to the $\rho L$ 
         parameterization (right) for the p-A data collected at 400\,\gev\
         in the rapidity range \mbox{$-0.425<y_{CM}<0.575$}.
         The band (left) represents the error associated to the normalization 
         and absorption cross-section uncertainties.
         Within the Glauber model, the fits extrapolation to A=1 or to
         $L=\langle \rho L \rangle / \rho_0$=0 leads to the value of $\sigma_0$.}
\label{fig:pa400glbrhoabs}
\end{figure}

\begin{figure}[h!]
\centering
\begin{tabular}{cc}
\resizebox{0.48\textwidth}{!}{%
\includegraphics*{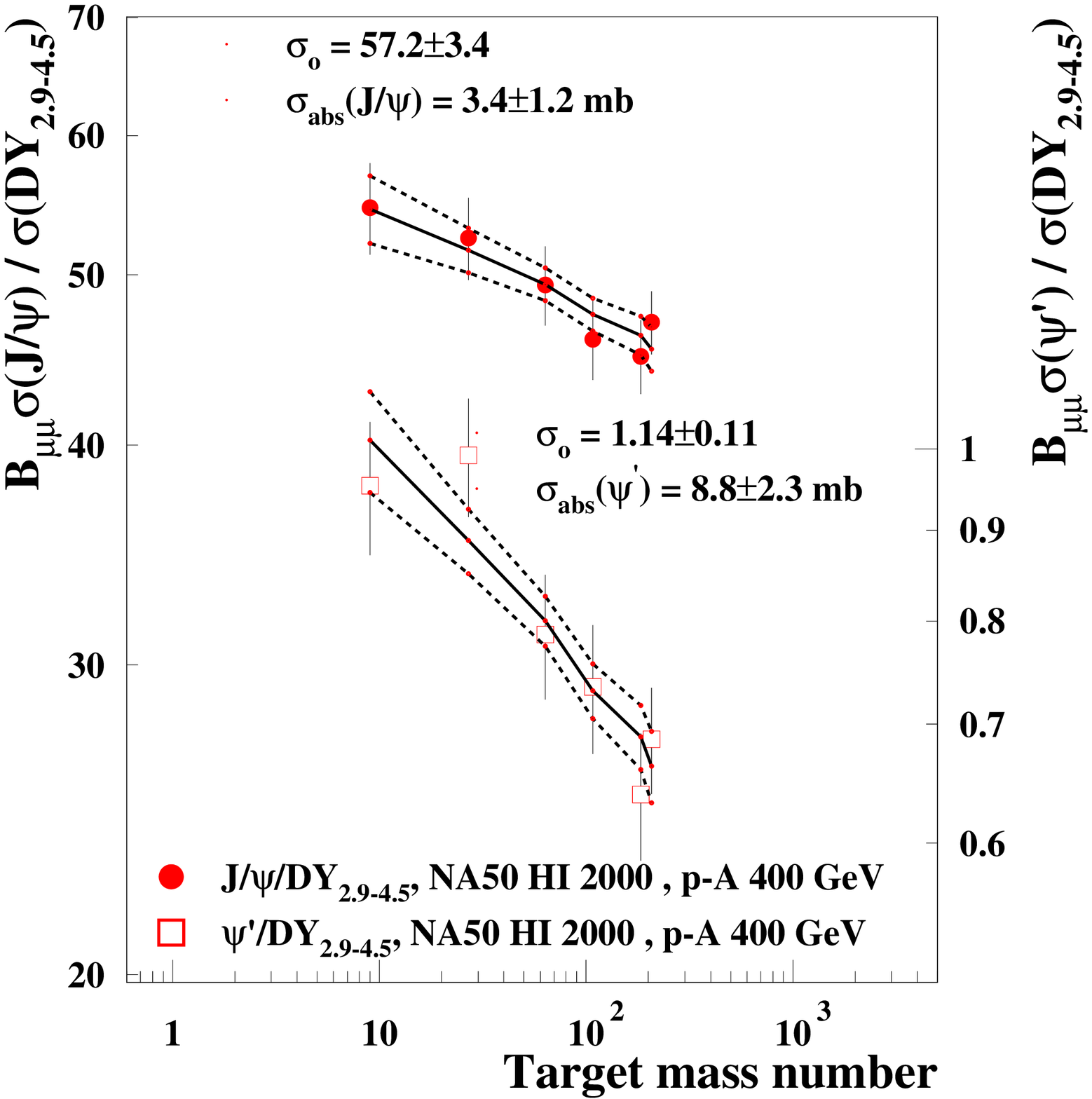}}
&
\resizebox{0.48\textwidth}{!}{%
\includegraphics*{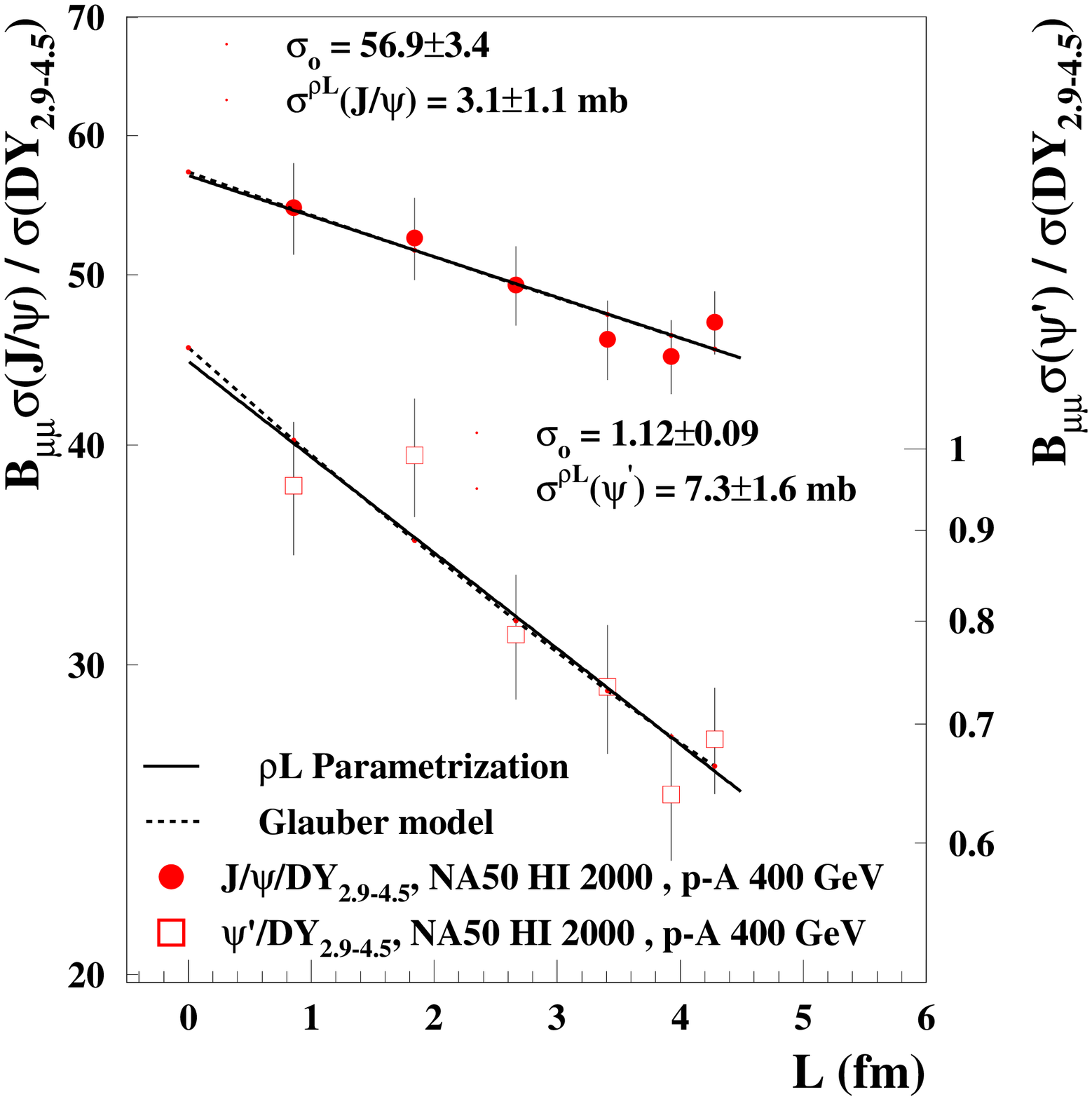}}
\end{tabular}
\vspace{-0.4cm}
\caption{Same as Fig.~\ref{fig:pa400glbrhoabs}
         for the ratios~ \jpsi/~DY and \psip/~DY. } 
\label{fig:pa400glbrhody}
\end{figure}

The results of the systematic set of measurements performed at CERN for 
450 and 400~GeV p-A reactions do prove that at these energies, both for 
\J and for \P :
\begin{itemize}
\vspace*{-0.2cm}
\item the agreement of the results with the "Glauber model parametrization" 
is extremely  good.
\vspace*{-0.2cm}
\item it becomes excellent when systematic uncertainties are perfectly under 
control. This is the case, for example, when the ratios \J/~Drell-Yan and 
\P/~Drell-Yan are used instead of the cross-sections themselves, or when 
absolute cross-sections measurements become independent from incident beam 
flux uncertainties
(which is the case for the NA50 measurements at 400~GeV).    
\vspace*{-0.2cm}
\end{itemize}    
\noindent
The two points are illustrated in 
Figs.~\ref{fig:pa400glbrhoabs} and 
\ref{fig:pa400glbrhody}~\cite{Goncalothesis,Goncalopap}.\\

\subsubsection{\J~suppression versus the centrality of the collision}

~The last sample of Pb-Pb collisions data collected by experiment NA50 
could benefit from the experience from previous data collections  
and consequent improvements of the detector. It also benefited 
from overall full coherence between data and simulations, which 
slightly affected normalizations. All the details and results can be 
found in~\cite{Ramello}.

\medskip
 
\begin{figure}[t]
\centering
\resizebox{0.43\textwidth}{0.4\textwidth}{%
\includegraphics{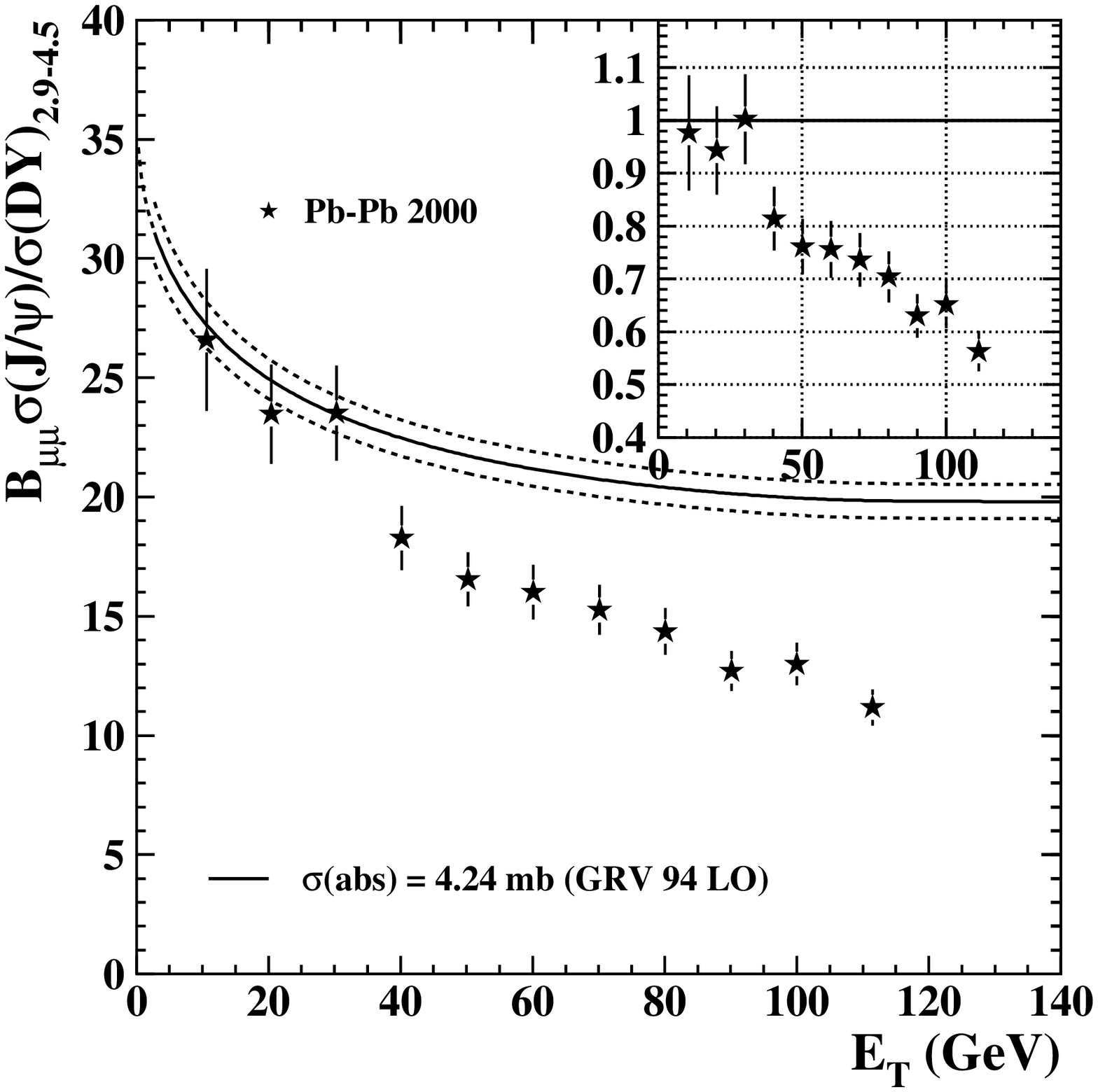}}
\vspace*{3mm}
\caption{The ratio \J/~Drell-Yan\, compared to the normal absorption
cross-section deduced from p-A and S-U measurements. The inset shows the ratio
\em{data\,/ \,(normal suppression)}. }
\label{etgrvoldref}
\end{figure}

As displayed in
Fig.~\ref{etgrvoldref}, as a function of the neutral transverse energy 
\Et~used here as the centrality estimator, the ratio of cross-sections~
{${B_{\mu\mu}\sigma_{J/\psi}/\sigma_{DY}}$} persistently decreases,
from peripheral to central collisions by a factor of $\sim 2.5$,
showing no saturation in the decrease even for the most central
collisions. The absorption pattern is here 
compared to the normal nuclear absorption curve as determined from the
most recent and accurate p-A results obtained in the same experiment
at 450 and 400~GeV together with results 
obtained from S-U data
collected at 200~GeV.  The technique of the ``simultaneous'' fit
described in section~3.2 leads to a normal absorption
cross-section $\Sgabs = 4.2\pm0.4~{\mathrm{mb}}$ and provides
part of the rescaling factor needed, from 450/400~GeV to 200~GeV. 
The advantage of using S-U results here is that it leads to minimal 
uncertainties in the energy rescaling factor which is "experimentally" 
determined, in part at least, as explained before. 
The curve is then further analytically rescaled to 158~GeV under
the assumption that \sgabs is energy independent. The
comparison with the normal absorption curve shows that the data behave
normally for the most peripheral collisions while increasingly departing from
this normal behaviour with increasing centrality.

\bigskip

\begin{figure}[t!]
\begin{minipage}[ht]{0.3\textwidth}
\resizebox{1.1\textwidth}{1.1\textwidth}{
\includegraphics{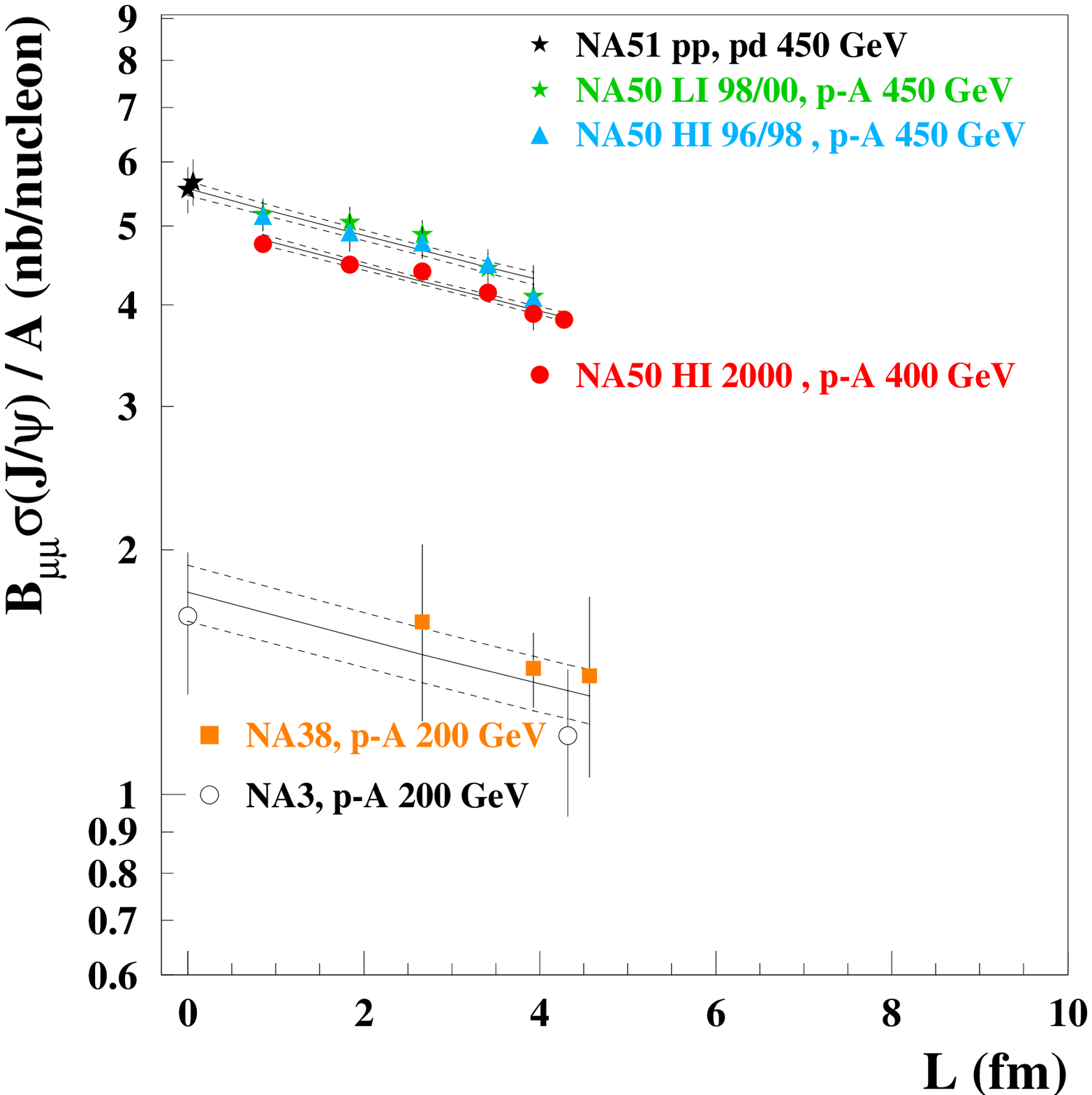}}
\end{minipage}
\hspace*{0.02\textwidth}
\begin{minipage}[t!]{0.3\textwidth}
\resizebox{1.2\textwidth}{1.2\textwidth}{
\includegraphics{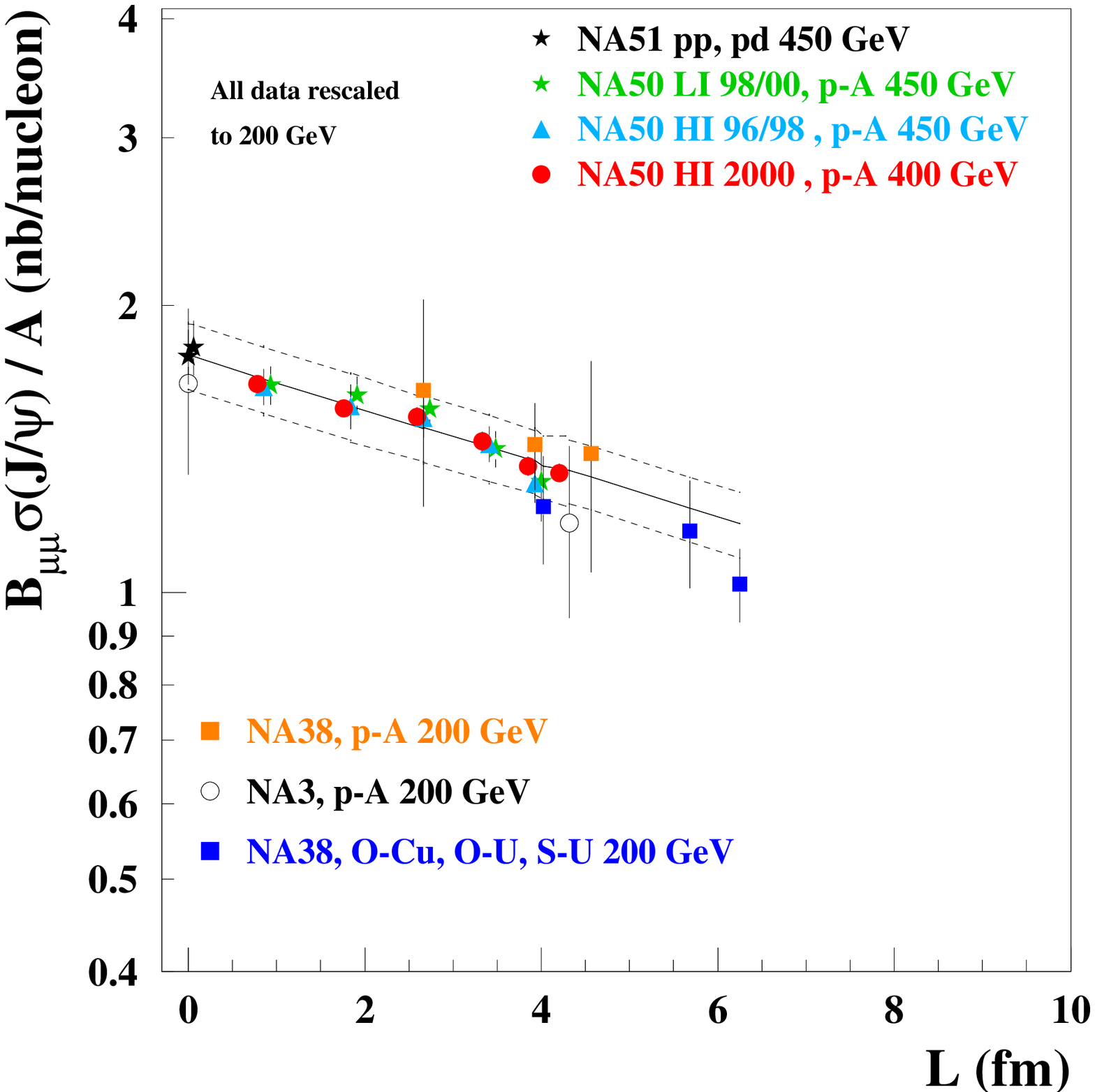}}
\end{minipage}
\hspace*{0.02\textwidth}
\begin{minipage}[t!]{0.3\textwidth}
\resizebox{1.2\textwidth}{1.2\textwidth}{
\includegraphics{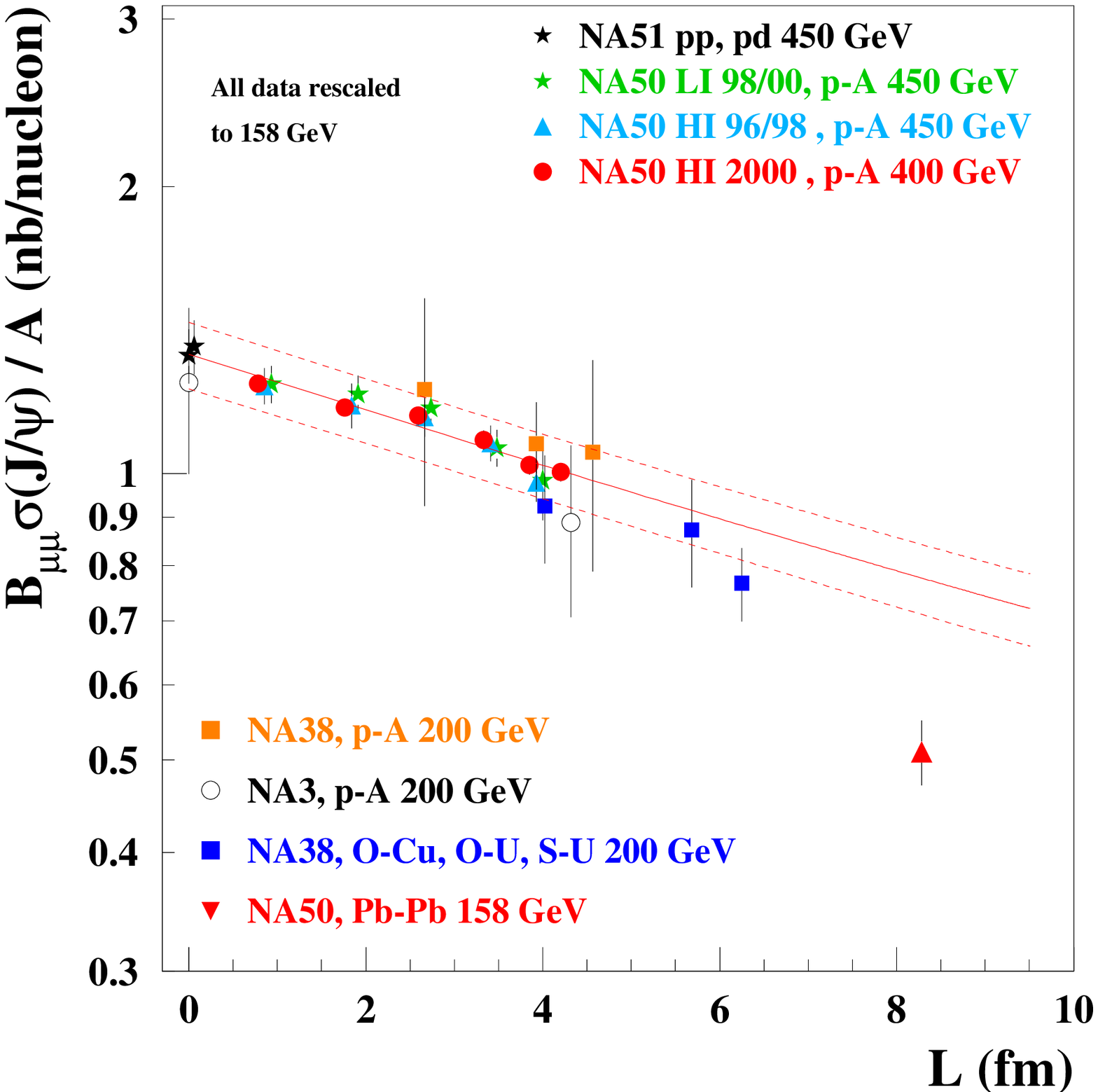}}
\end{minipage}
\vspace*{-1mm}
\caption{ \J cross-sections per nucleon in p-A collisions
          as a function of $L$. {\em Left:} Results obtained at 450, 
          400 and 200~GeV separately. {\em Center:} Same results 
          rescaled to 200~GeV and compared with results from light 
          nuclei collisions. {\em Right}: Same results rescaled to 
          158~GeV and further compared with \J production
          in Pb-Pb collisions.}
\label{gborg}
\vglue-0.5mm
\begin{minipage}[ht]{0.3\textwidth}
\centering
\resizebox{1.1\textwidth}{1.1\textwidth}{
\includegraphics{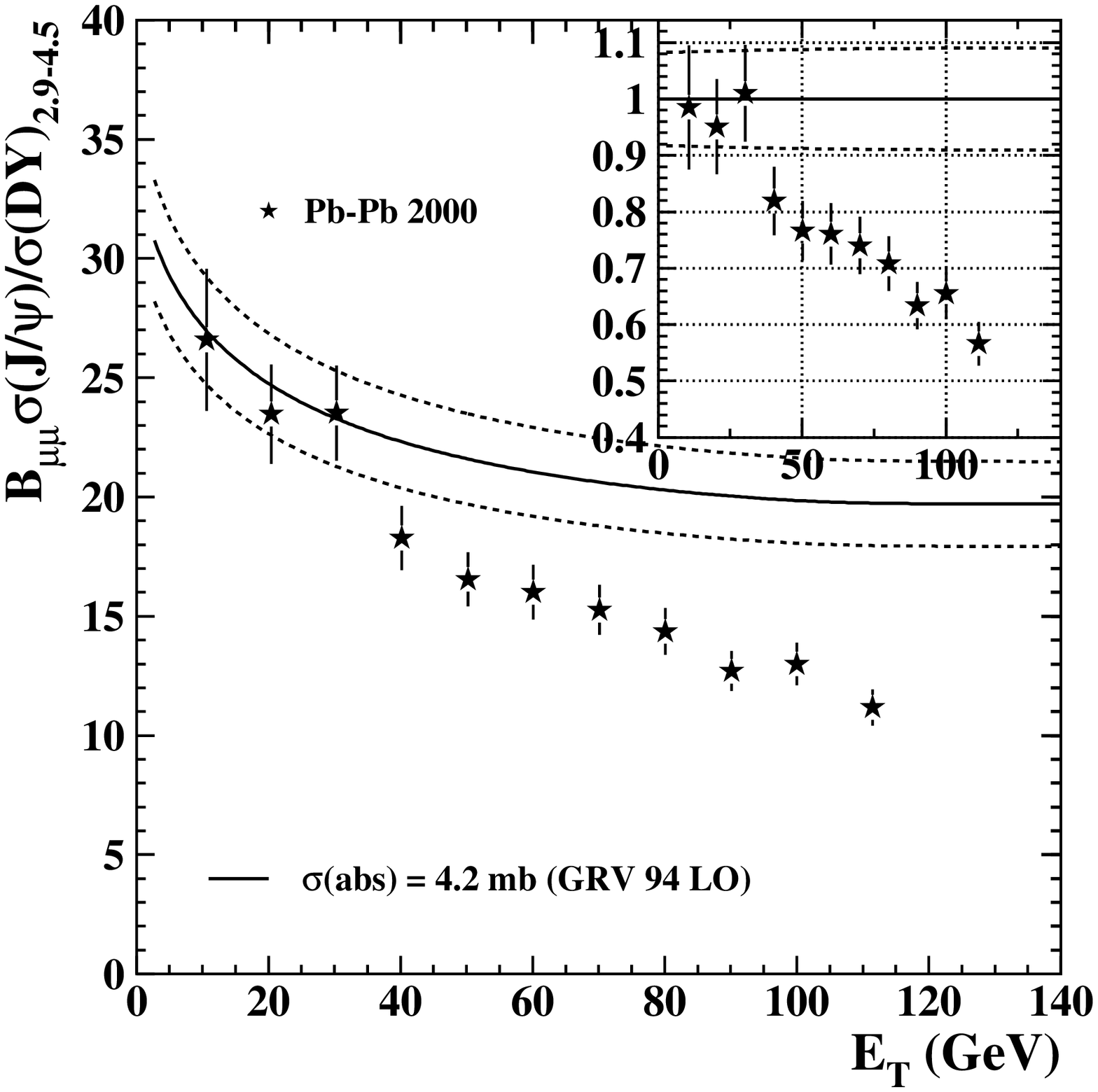}}
\end{minipage}
\hspace*{0.02\textwidth}
\begin{minipage}[t!]{0.3\textwidth}
\resizebox{1.1\textwidth}{1.1\textwidth}{
\includegraphics{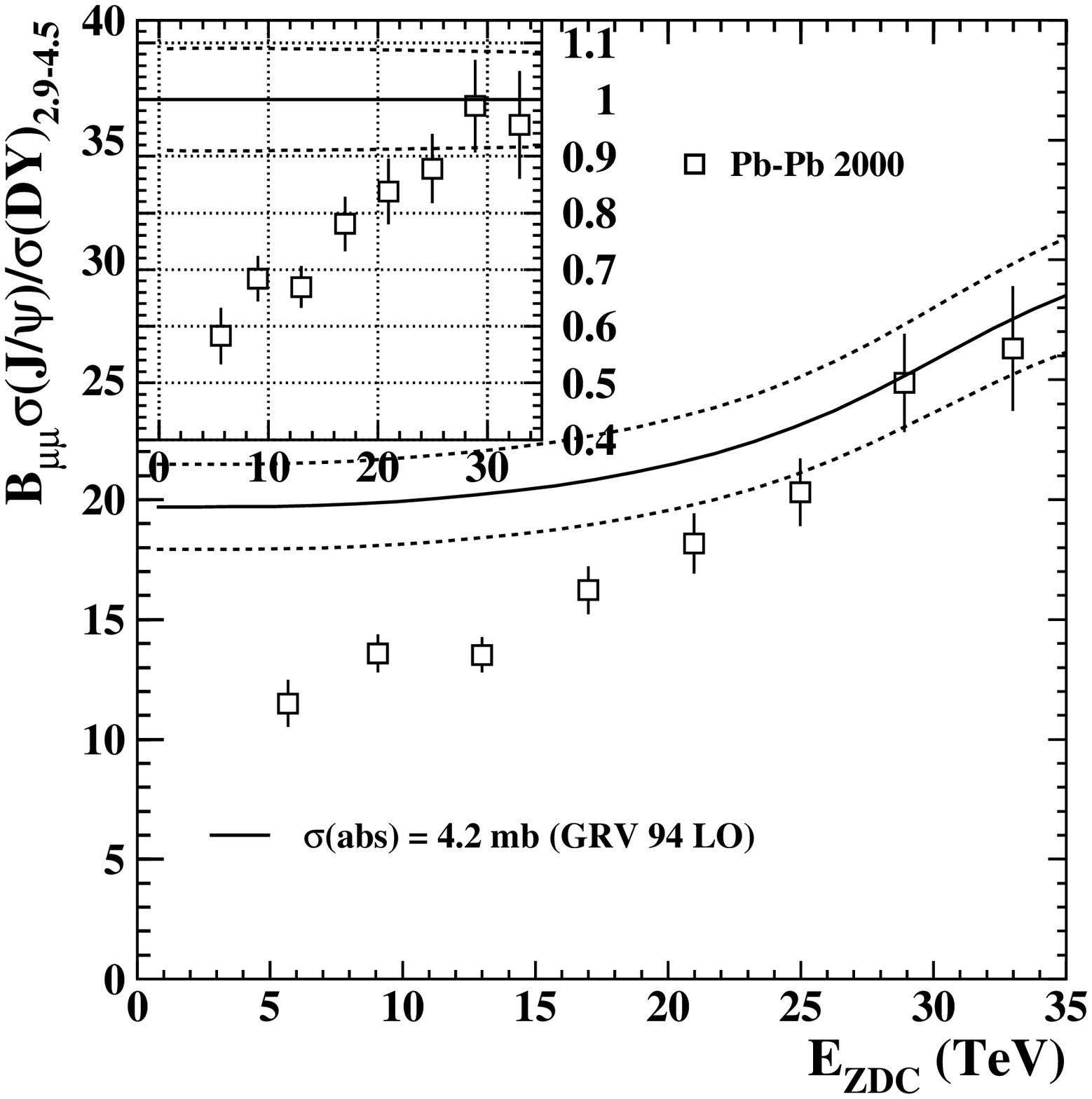}}
\end{minipage}
\hspace*{0.02\textwidth}
\begin{minipage}[t!]{0.3\textwidth}
\resizebox{1.1\textwidth}{1.1\textwidth}{
\includegraphics{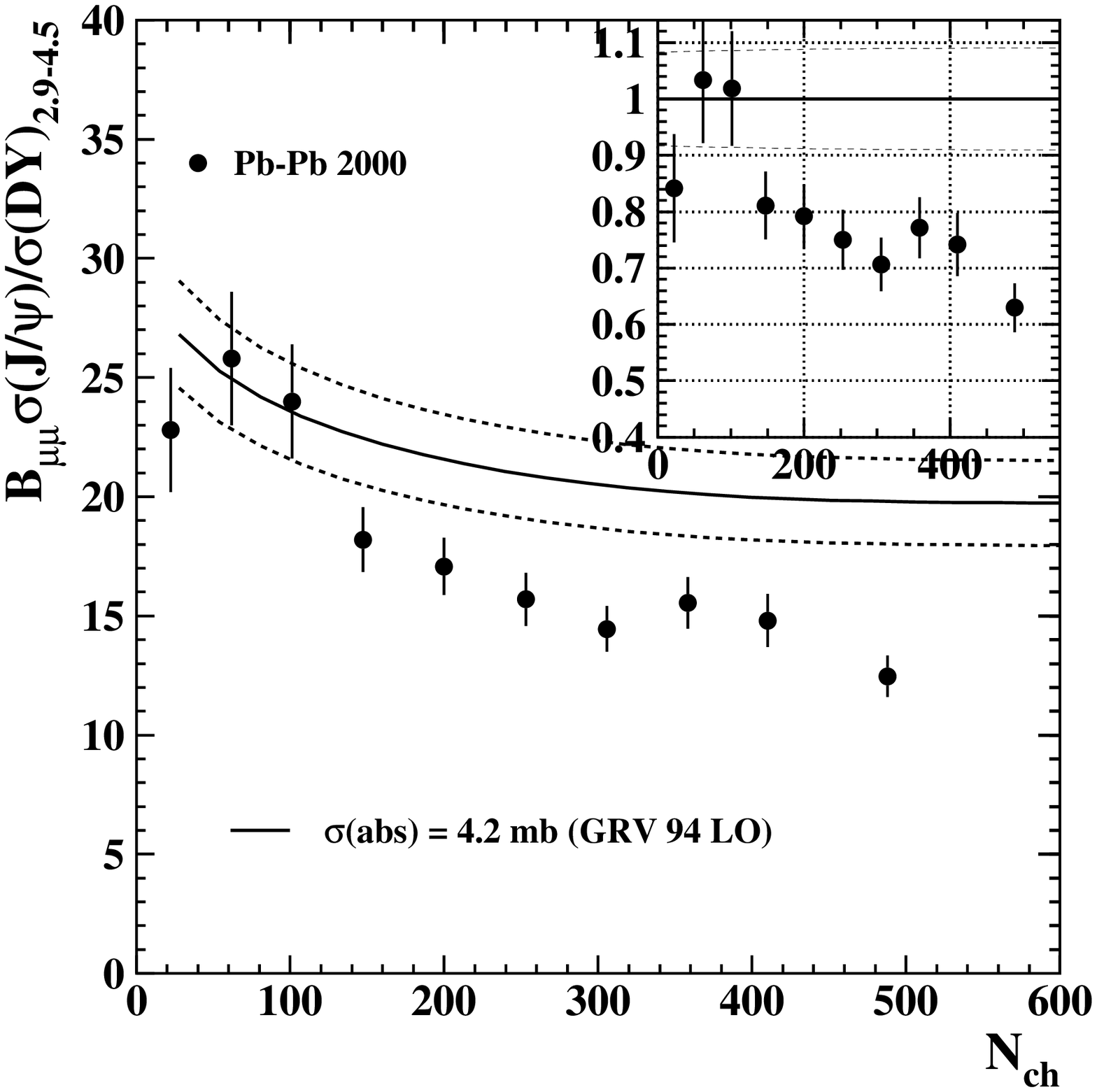}}
\end{minipage}
\vspace*{-1mm} 
\caption{The ratio of cross-sections \jpsi/~Drell-Yan in Pb-Pb
         collisions as a function of, from left to right, the 
         neutral transverse energy, the ``very forward'' hadronic 
         energy and the charged multiplicity. The curve displays 
         the normal suppression pattern as deduced from p-A 
         interactions only. The insets show the ratio 
        {\em data\,/\,(normal suppression)}.}
\label{3newref}
\vglue-0.5mm
\centering
\resizebox{0.7\textwidth}{0.35\textwidth}{%
\includegraphics{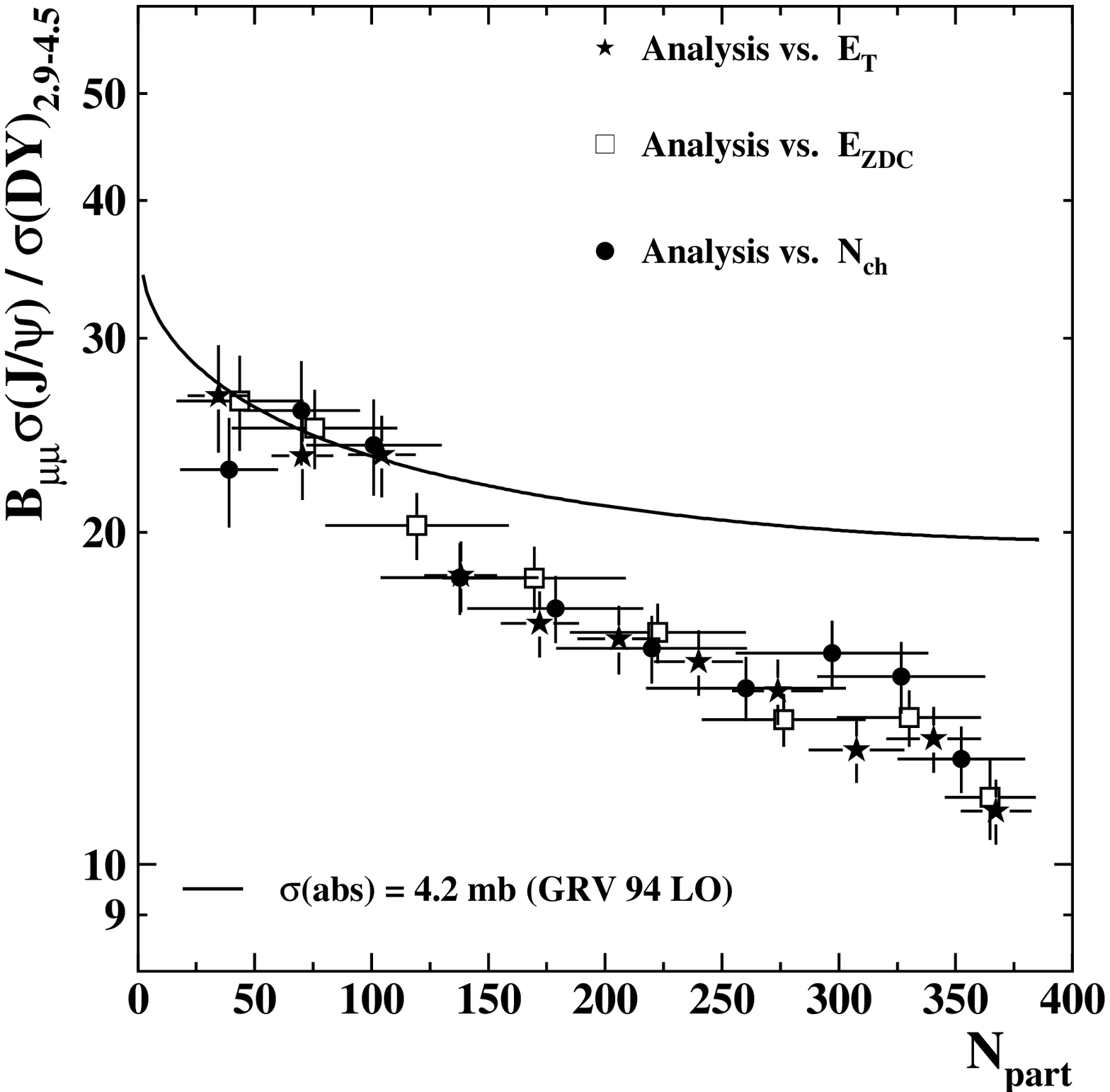}
\includegraphics{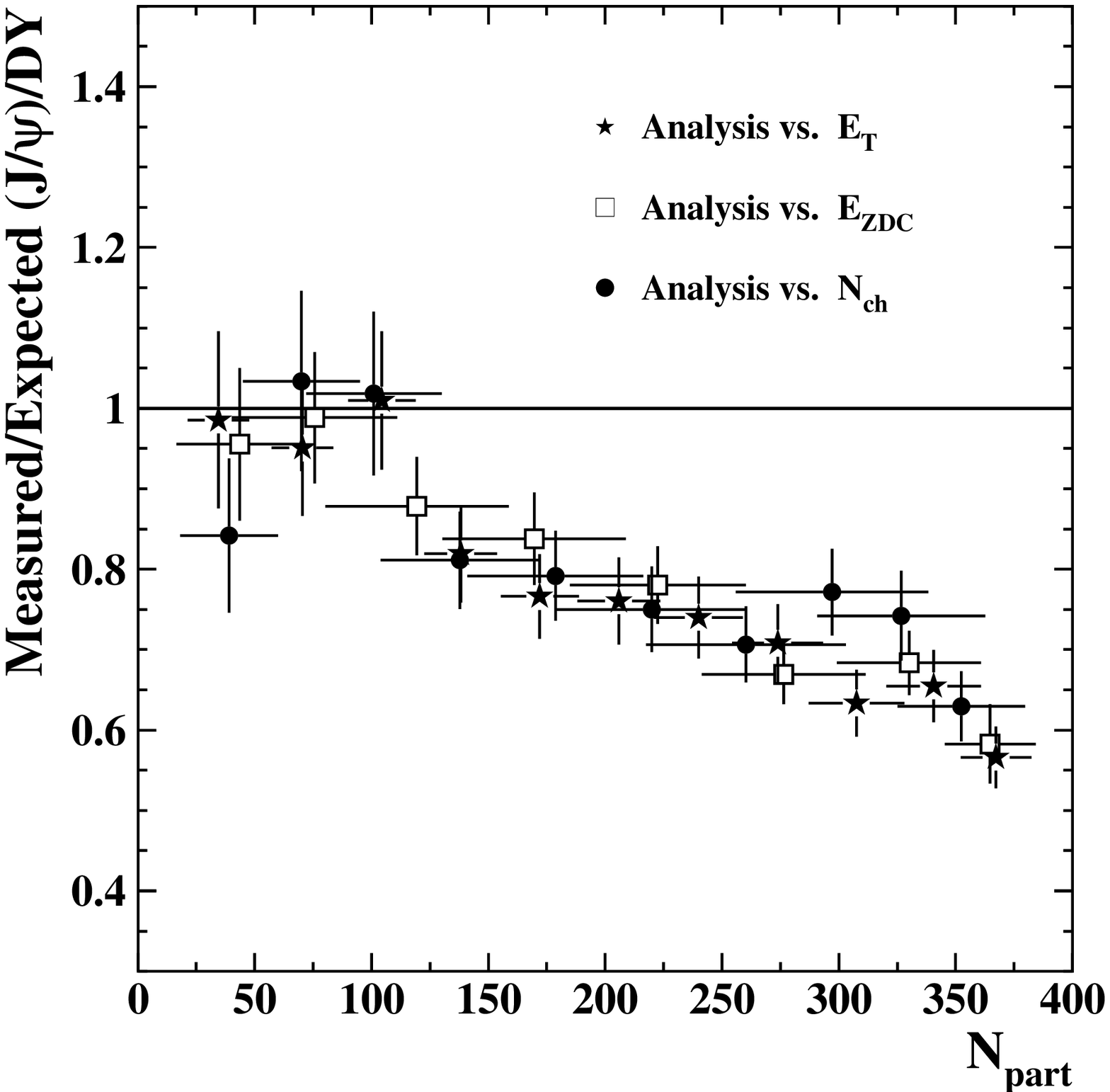}}
\vspace*{-3mm}
\caption{The \jpsi /~Drell-Yan cross-sections ratio as a function 
of $\npt$ from three analysis originally using the different 
centrality estimators, compared to (left) and divided by (right)
the normal nuclear absorption pattern. 
}  
\protect\label{npart}
\end{figure}

It has been pointed out that
this normal suppression reference could be biased by the use 
of S-U data in its determination since co\-mo\-ving produced hadrons, 
for example, could already affect \jpsi
production in S-U reactions. A new determination of the reference
curve has therefore been implemented as detailed hereafter.  Use has
been made of the most precise results on absolute \jpsi pro\-duc\-tion
cross-sections obtained by NA50 in p-A collisions at 450 and 400~GeV
together with the results obtained at 200~GeV, again in p-A collisions
exclusively, by experiments NA38~\cite{papNA3802} and
NA3~\cite{papNA3}. 
The Glauber "simul\-taneous fit'' method detailed in section~3.2
leads then to an absorption cross-section  
$\sigma_{abs} = 4.1\pm0.4~{\mathrm{mb}}$ which, it is worth noting, is
in excellent agreement with the value of $4.2\pm0.4~{\mathrm{mb}}$
obtained from the ratio of cross-sections \jpsi/~Drell-Yan.
Incidentally, from the precisely measured ratio of 
\J cross-sections in p-p and p-Pt by experiment NA3 at 200~GeV,
one gets, from a completely independent determination, using the same
Glauber approach, an absorption cross-section of 
$\sigma_{abs} = 4.1\pm1.0~{\mathrm{mb}}$ when restricted to the 
rapidity interval of NA50 Pb-Pb data, which, independently,  
strongly supports the same numerical value for $\sigma_{abs}$ 
at 450/400 and 200~GeV incident proton energies.  
As detailed in~\cite{HP04}, 
Figure~\protect{\ref{gborg}} (left and center) illustrates the method
and further shows that results from O-Cu, O-U and even S-U data
exhibit, within errors, a so-called normal p-A like behaviour.
\mbox{Figure}~\protect{\ref{gborg}} (right) confirms, that with respect to a
pure p-A reference curve, \jpsi is ``anomalously'' suppressed in Pb-Pb
collisions at 158~GeV.  It is worthwhile underlining here that the new
reference curve for the ratio \jpsi/~Drell-Yan at 158~GeV is partly deduced
by analytical rescaling: the \jpsi cross-section is rescaled from 200
to 158~GeV with a Schuler-type formula~\cite{Schuler, Goncalothesis} 
and, on the other side, the
appropriate factor is applied to the Drell-Yan cross-section to
rescale it from 450 to 158~GeV. This pure p-A ``normal'' reference curve
has identical shape as the one which was also making use of S-U data. It is
globally lower by a factor of 0.6\% whereas its experimental
uncertainty is increased by a factor 2.
The most recent results obtained from \makebox{Pb-Pb} can now be compared 
with this new exclusively p-A based reference curve.  As a function of 
three independently measured quantities tagging the centrality of the
collision (see section~2), 
the \jpsi survival pattern
exhibits very similar trends, as shown in Fig.~\ref{3newref}.

\medskip

The three independent analyses are displayed together in 
Fig.~\protect\ref{npart} after conversion of their originally 
used centrality estimator to the average 
number of participant nucleons in the reaction, $N_{\rm part}$.
The remarkable agreement between the results of the three analyses, 
underlines the coherence, equivalence and deep understanding  
of the data provided by the three independent detectors used to 
estimate the centrality of the collision. 

\medskip

It is worth recalling that \Et\ and the charged particle rapidity
density at mid-rapidity are directly related to the energy density
reached in the collision, through the Bjorken formula~\cite{Bjorken83}, 
while the forward energy measured in the ZDC is more strongly correlated 
to the geometry of the collision, and a simple robust estimator of
$N_{\rm part}$.
Note that at SPS energies, for a given interacting system, 
both \Et ~and $N_{\rm ch}$\ are linearly proportional to
$N_{\rm part}$, as expected in the framework of the wounded 
nucleon model, up to the most central collisions (see, for instance, 
Refs.~\cite{K-L-N-S} and~\cite{DNDETA2}).

\subsubsection{The transverse momentum dependence of \J~suppression}

\begin{figure}[h!t]
\begin{minipage}[t]{7.5cm}
\resizebox{0.99\textwidth}{1.19\textwidth}{%
\includegraphics{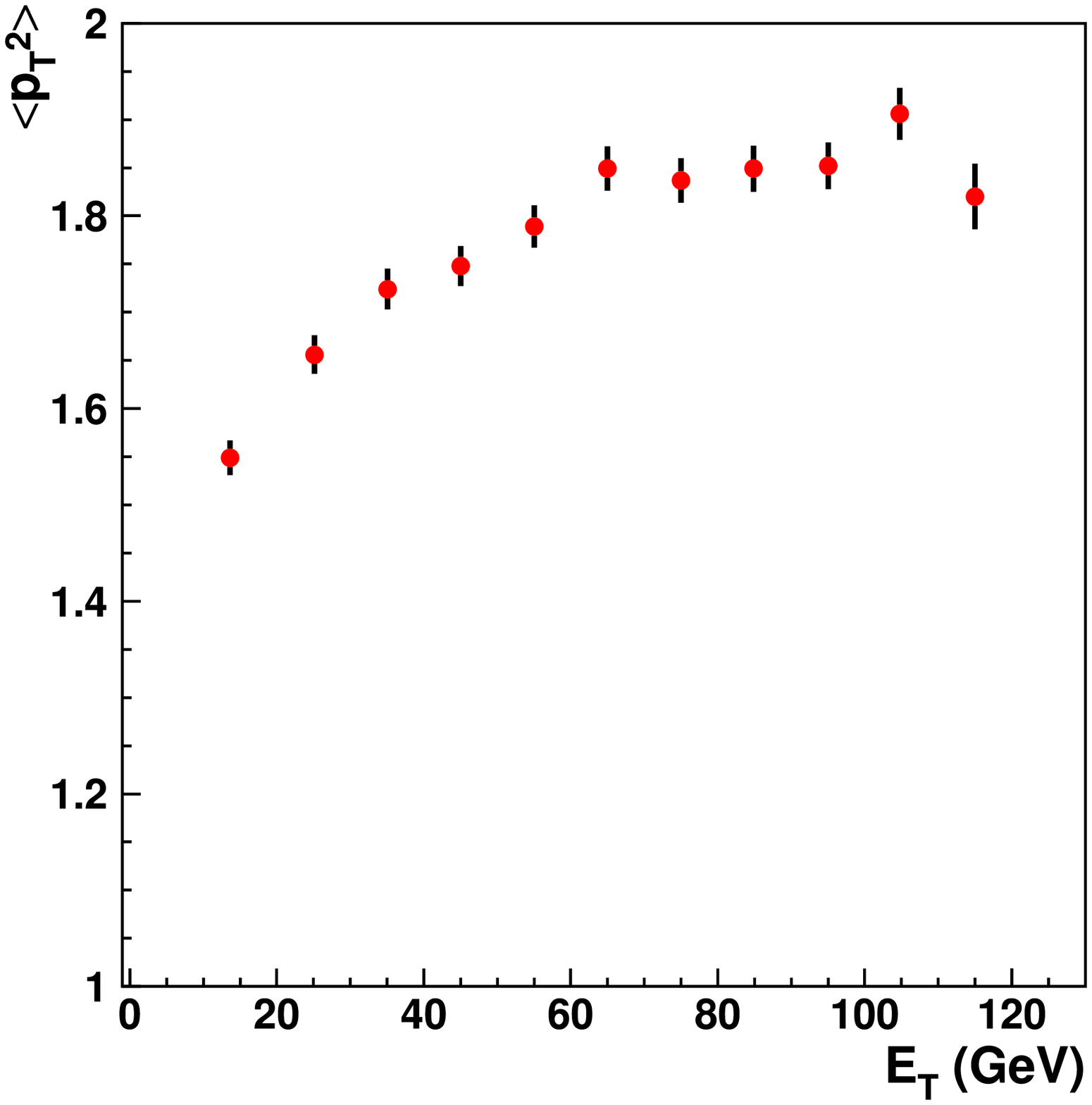}}
\vglue -60pt
\caption{The average $<\TR2>$ as a function of centrality, estimated from \Et,  
         for the surviving \J's in Pb-Pb collisions at 158~GeV.}
        \label{pt2-Et}
\end{minipage}
\hspace{0.3cm}
\begin{minipage}[t]{7.5cm}
\vspace*{-8.90cm}
\resizebox{1.09\textwidth}{1.19\textwidth}{%
\includegraphics{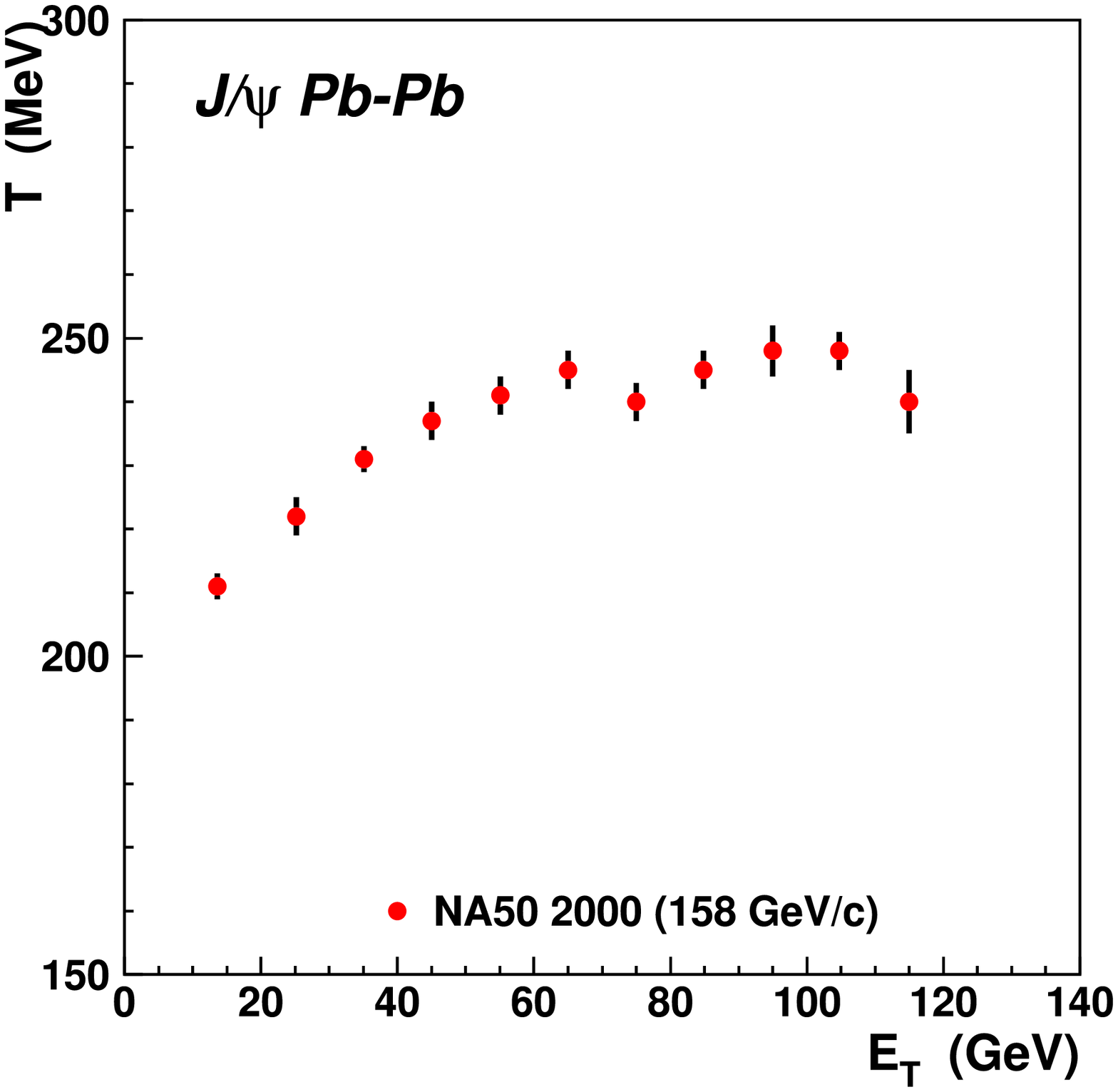}}
\vglue -58pt
\hspace*{0.6cm}
\parbox{7.5cm}{
\caption{The transverse mass inverse slope parameter T of the surviving \J's as a function of centrality,
         estimated from \Et, in in Pb-Pb collisions at 158~GeV.}
        \label{T-Et}}
\end{minipage}
\end{figure}

The \pt spectra of the surviving \J's have been studied by experiment 
NA50~\cite{pappt} and, in particular, from their most precise sample 
of Pb-Pb interactions at 158~GeV collected in
year 2000~\cite{NatachaBaldinKurprag}.
The value of $<\TR2>$ for the surviving \J's is plotted in 
Fig.~\ref{pt2-Et} as a function of \Et. The overall behaviour 
exhibits a steady increase with centrality, from the most peripheral 
collisions and up to \mbox{\Et $\simeq$ 60~GeV}. This value corresponds 
to $L \simeq 8.23~{\rm fm}$ and $\epsilon = 2.97~{\rm GeV}/{{\rm fm}^3}$. 
The data also show that, for high centralities, no increase is seen
up to the most central collisions. 

\begin{figure}[h!]
\begin{minipage}[t]{7.5cm}
\epsfig{file=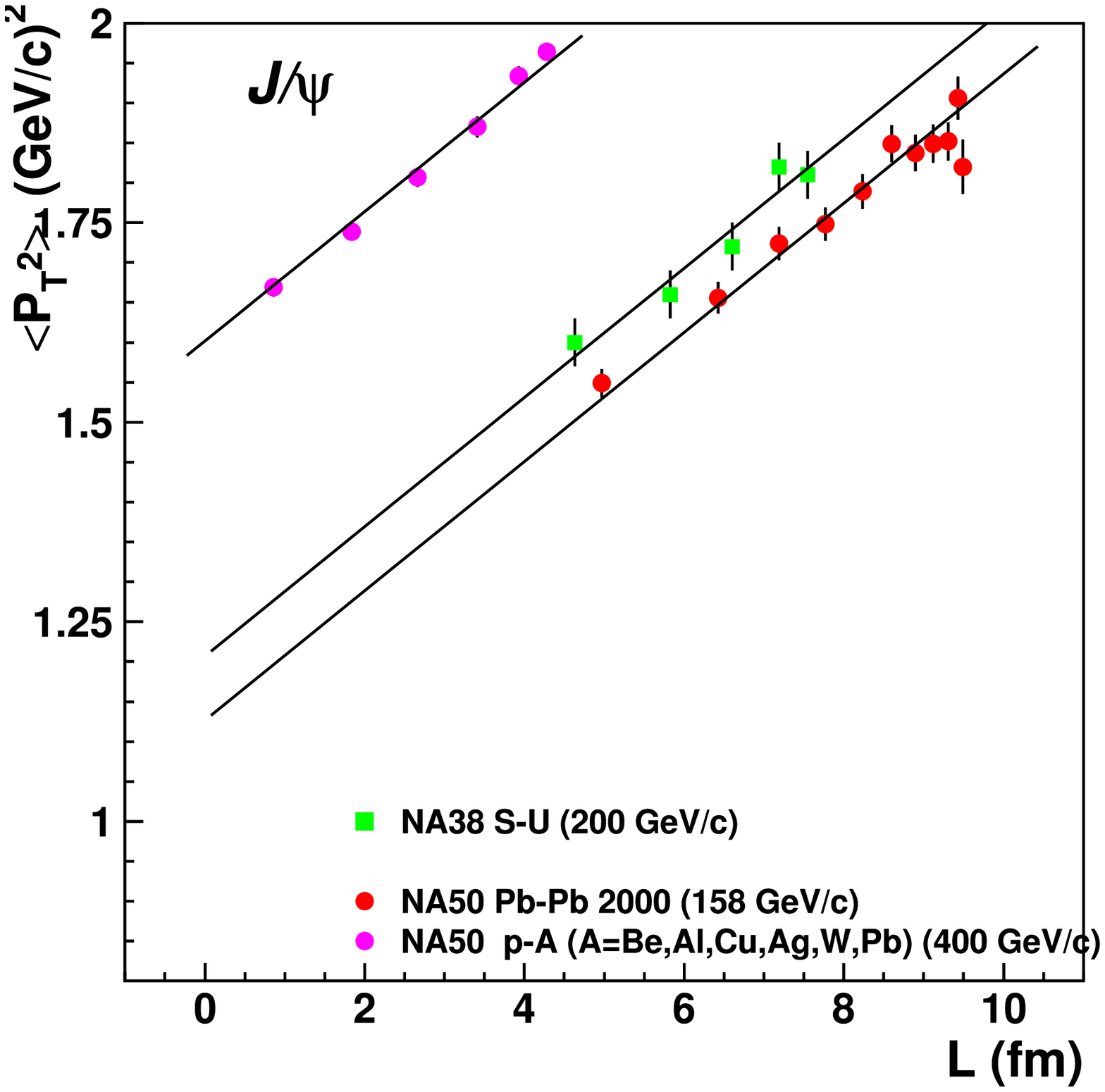,width=7.5cm,height=9.0cm}
\vspace*{-2.7cm}
\caption{The average $<p_{\rm t}^2>$ of the surviving \J's as a function of centrality, 
         here L, for p-A~(400~GeV), S-U~(200~GeV) and Pb-Pb~(158~GeV) with a 
         "simultaneous" common linear fit.}
\protect\label{transv-sep}
\end{minipage}
\hspace{1.0cm}
\begin{minipage}[t]{7.5cm}
\vspace*{-9.55cm}
\hskip-0.3cm
\epsfig{file=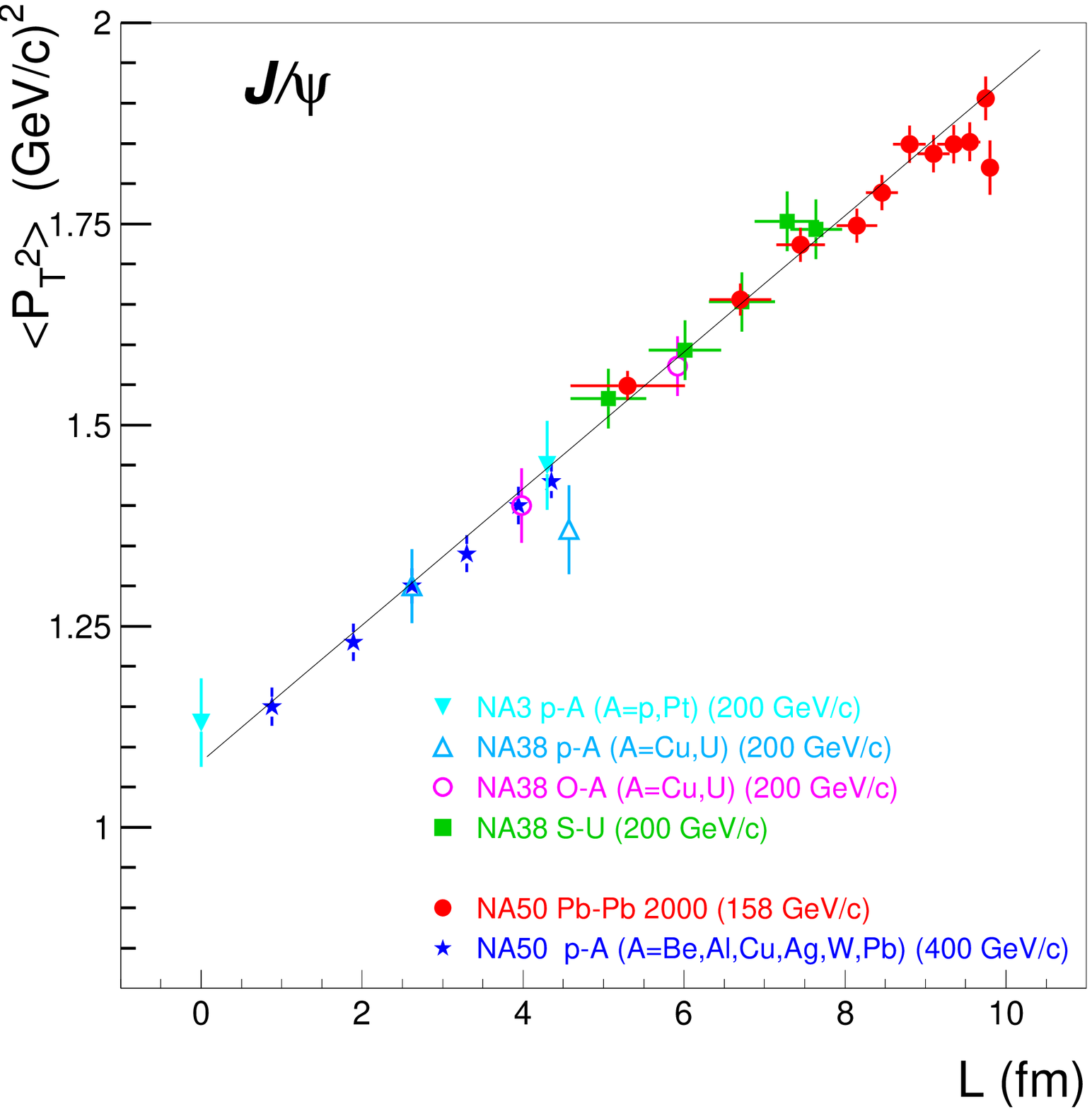,width=7.89cm,height=7.5cm}
\vspace*{-0.65cm}
\caption{Same as Fig.~\ref{transv-sep}, inclu\-ding results on p-A~(200~GeV) from NA3 and NA38,  
         O-Cu and O-U~(200GeV) from NA38. All results "experimentally" rescaled 
         to the kinematical window used for Pb-Pb~(158~GeV).}
\label{transv-allfitted}
\end{minipage}
\end{figure}

\noindent 
In order to investigate the underlying mechanism that leads to the observed 
increase of  $<\TR2>$ for the more peripheral collisions,  a comparison is 
made between different colliding systems as shown in Figs.~\ref{transv-sep}
and~\ref{transv-allfitted}. As a function of L,  the same linear dependence 
is observed whatever the colliding system, with a slope apparently independent 
of the energy of the colliding nuclei.
The first observation of this behaviour at 200~GeV based on results
from NA3~\cite{papNA3} and  NA38~\cite{NA38natcha, ptSU} have been successfully 
interpreted in terms of initial-state parton multiple 
scattering
and analytically described by \mbox{$<\TR2> = <\TR2>_{pp} + a_{gN} \times L$.} \\ 
\vspace*{10pt}

\vspace*{-0.3cm}

\vspace*{-5pt}
\noindent
All the samples of \J considered here are roughly in the same {$ y_{\rm lab} $} 
range, namely \mbox{$ 3 < y_{\rm lab} < 4 $}. The values of  $a_{gN}$ are nicely compatible for all colliding 
systems and energies. Using the technique of the "simultaneous fit"
for the slope and leaving free the normalization factor to account for the 
energy change of each set of data leads to the following results:
\begin{itemize}
\vspace*{-0.2cm}
\item$<\TR2>_{pp}$ increases linearly with $\sqs$, 
the total energy in the nucleon-nucleon center of mass system. 
\vspace*{-0.2cm}
\item As a function of L, the mean squared transverse momentum of the \J
exhibits the same increasing slope for p-A, O-Cu, O-U, S-U and peripheral
Pb-Pb interactions as shown in Fig.~\ref{transv-allfitted}, but saturates
for the most central Pb-Pb collisions.
\vspace*{-0.2cm}
\item The common fit of the increasing slope 
leads to $a_{gN} = 0.081 \pm 0.003$.
\vspace*{-0.2cm}
\end{itemize}

\noindent In an attempt to link this transverse momentum to the rate measurements, 
it can be empirically noticed that, as already discussed above, the measured
\J suppression rate has also the same, so called normal, L dependence from 
p-p up to S-U reactions. The specific anomalous suppression only appears 
for the most central Pb-Pb reactions, namely for \mbox{L $>$ 7~fm.}      
This peculiar transverse momentum 
behaviour is also clearly seen from Fig.~\ref{T-Et}
which shows the transverse mass inverse slope parameter $T$, for the 
surviving \J's, as a function of \Et. $T$ is deduced from a fit of the 
transverse mass distributions with the function
\mbox{${1/T}\cdot {M_{\rm T}^2} \cdot {K_1(M_{\rm T}/T)}$}, where $K_1$ is the 
modified Bessel function.
The inverse slope, $T$, is related to the effective
temperature of the system in thermal models of particle
production
~\cite{b28fromnatacha, ptSU}. 
The measurements  show that T increases with centrality for peripheral  
Pb-Pb reactions and then, slightly after anomalous suppression sets in,
exhibits a trend compatible with no further increase up to the most
central collisions. 

\begin{figure}[h!t]
\vspace*{30pt}
\centering
\resizebox{0.68\textwidth}{0.650\textheight}{%
\includegraphics{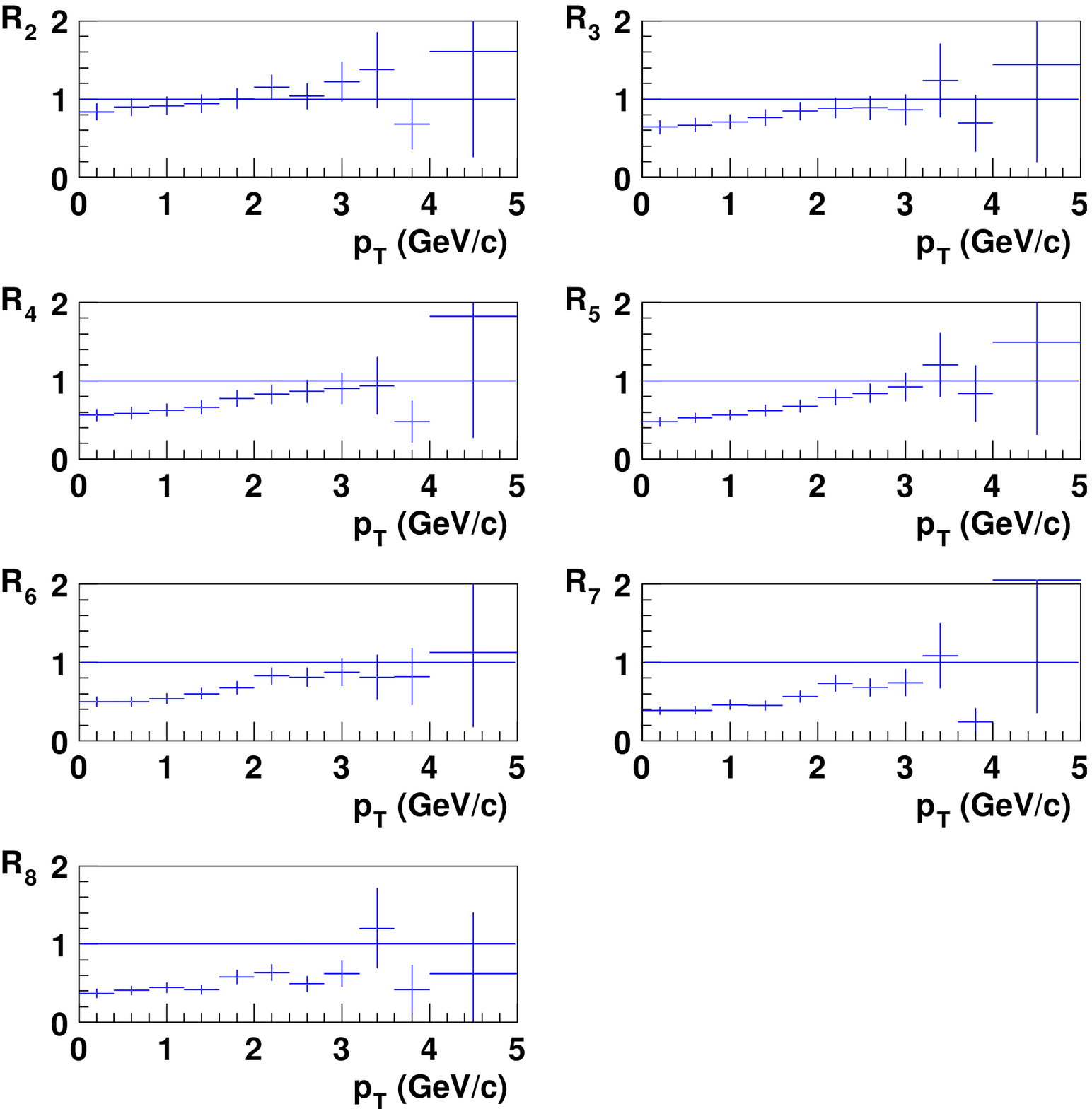}}
\caption{The ratio of the \jpsi \pt distribution in a given bin of
         centrality to the same distribution for the most peripheral 
         bin. With respect to the latter, the low \pt part of the 
         distributions depletes with increasing centrality.}
\label{ptfig}
\vspace*{3mm}
\end{figure}

\medskip

\noindent The \pt dependence of \jpsi suppression has been further 
studied in order to investigate what are the \pt features of the 
suppressed \J's, as a function of the centrality of the collision.   
In order to be independent from any external input 
related to the \pt distribution of the suppressed \J's, NA50
has limited its study to the ratio of the \pt spectra obtained in 
various centrality bins with respect to the spectrum obtained 
in the most peripheral one (a procedure which will be later 
known as $R_{CP}$). 
For statistical reasons of this two dimensional study,    
\jpsi and Drell-Yan events have been considered within
eight different centrality bins only.
Within each centrality bin, the \pt distribution
of the \jpsi has been studied, normalized to the total number of
Drell-Yan events with the same centrality, i.e. to the "average 
number of nucleon-nucleon collisions" in the considered bin. 
Thus, the integral of the \pt spectrum in each of the bins is the 
ratio \J/\,Drell-Yan, as would be plotted, for example, 
in Fig.~\ref{etgrvoldref} dividing the whole centrality range in 
only 8 bins (instead of 11). The ratios $F_i$ and $R_i$ are thus defined 
as:

\begin{displaymath}
F_i = {dN^i_{J/\psi}/dp_T \over N^i_{{DY}\left(M>4.2~\mathrm{GeV}/c^2\right)} } 
~~~~~~~~~~~~{\rm and}~~~~    R_i = { F_i \over F_1 }
\end{displaymath}

\noindent where {$i$} is the {\large{$i^{th}$}} centrality bin,
{\normalsize ${dN^i_{J/\psi}/dp_T}$} is the number of {\normalsize
{$J/\psi$}} events of a given {\large{$p_T$}} in centrality bin $i$ and
{\normalsize ${N^i_{DY}}_{(M>4.2~\mathrm{GeV}/c^2)}$} \,is the 
{\em total} \,number of Drell-Yan events with 
{\normalsize $M>4.2~\mathrm{GeV}/c^2 $} in centrality bin $i$.

\medskip

\noindent Figure~\ref{ptfig} displays the \pt dependence of the seven ratios
$R_i$. It shows that with increasing centrality, the lower the \pt
the higher the suppression.
Moreover and within statistical uncertainties, it suggests that, 
for transverse 
momenta larger than 2.5~GeV/$c$, the shape of the \pt spectrum of \J  
is independent of centrality and very close to the one observed in the most
peripheral collisions. 
It should be underlined here that, as all $R_{CP}$ studies, this approach 
does not allow to easily disentangle between transverse momentum spectrum 
modifications due to normal and to abnormal suppression separately.  
As an illustration, a similar study made for S-U reactions which are only 
subject to normal \J suppression exhibits nevertheless qualitatively 
similar features~\cite{ptSU}.    

\begin{figure}[b!]
\centering
\resizebox{0.925\textwidth}{0.462\textwidth}{%
\includegraphics{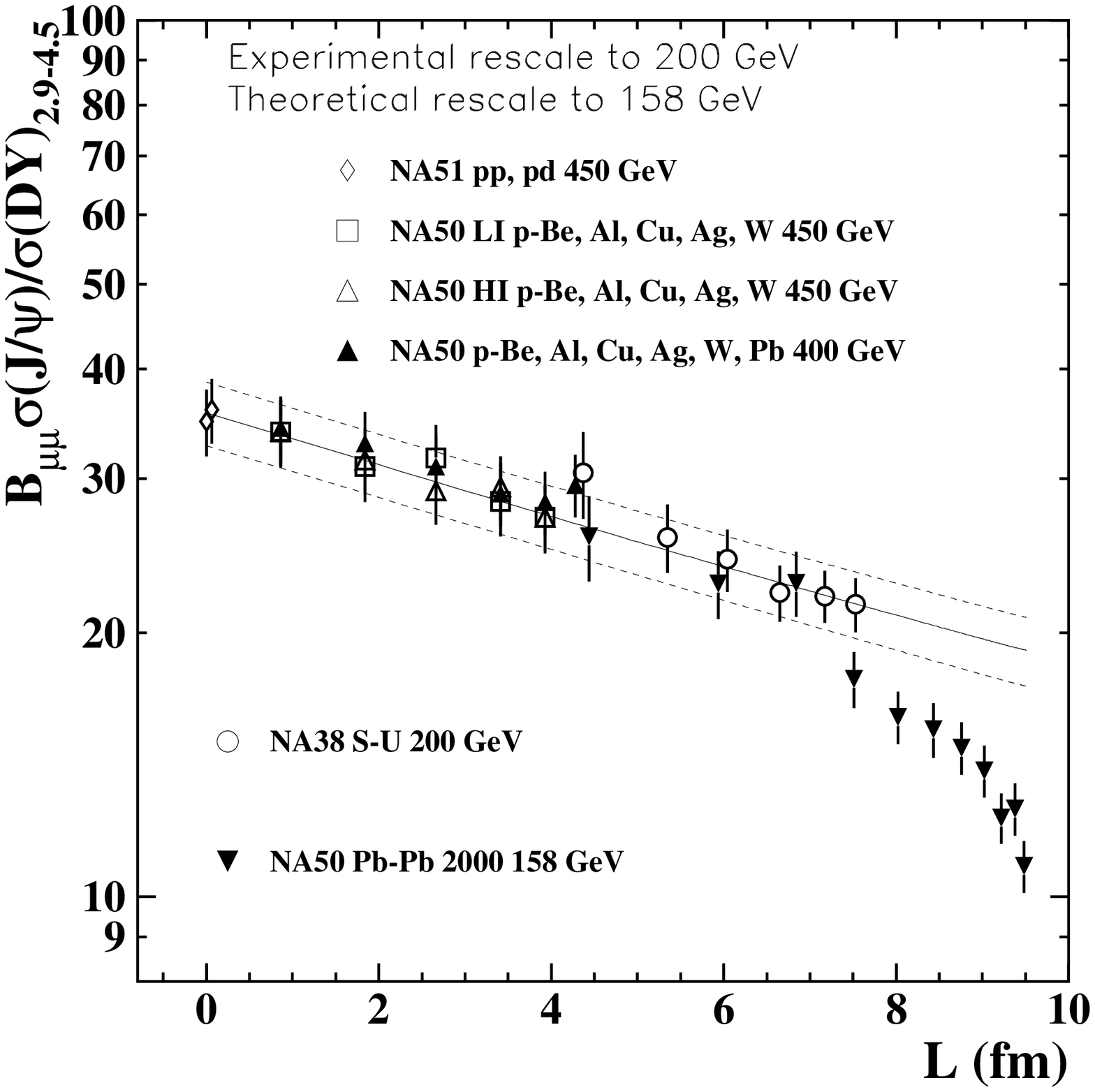}
\hspace*{50pt}
\includegraphics{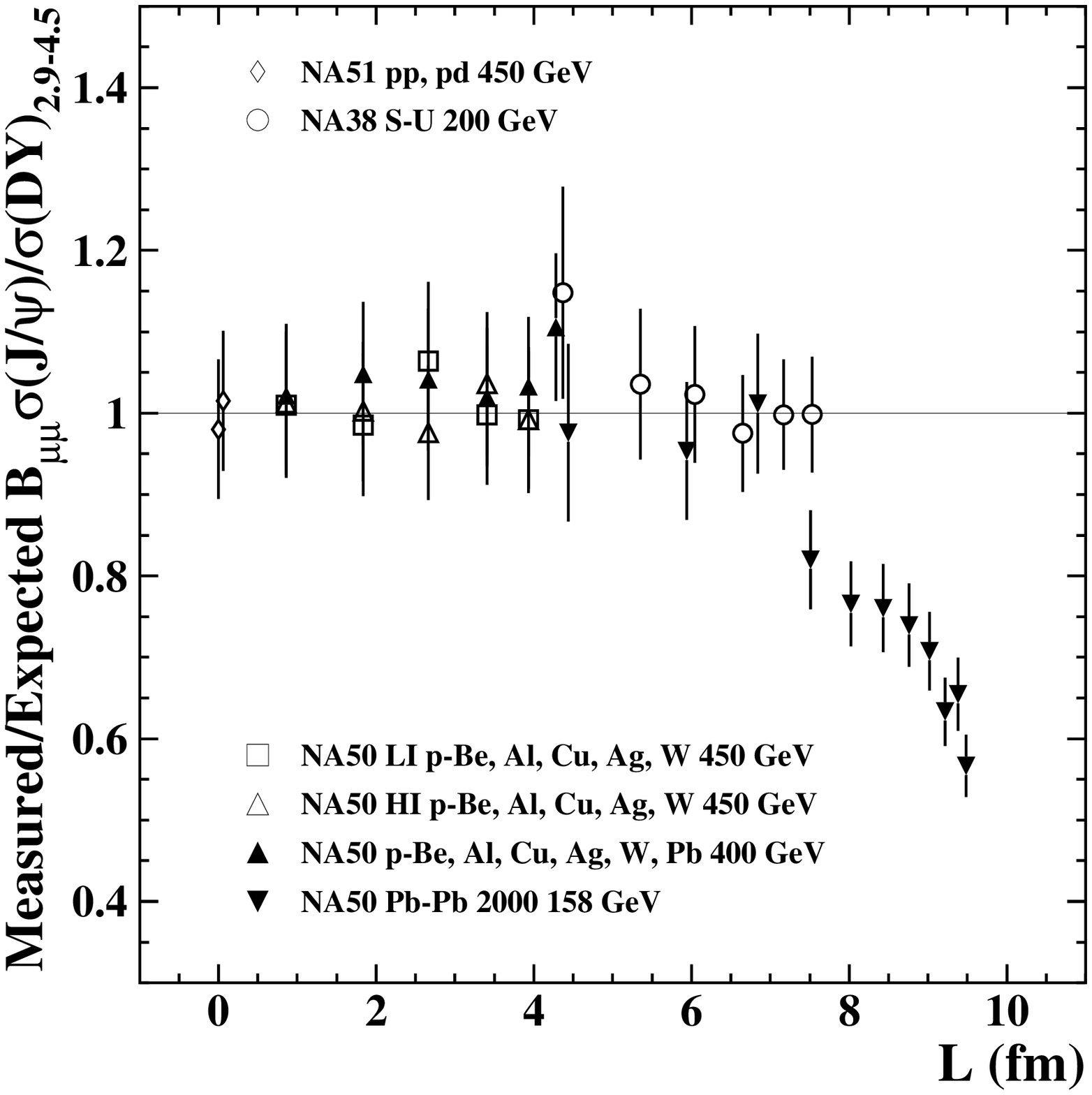}}
\vspace*{-3mm}
\caption{The \jpsi/~Drell-Yan ratio of cross-sections vs. $L$, for
         several collision systems, compared to (left) and divided by (right)
         the normal nuclear absorption pattern. All data are rescaled to
         158~GeV/nucleon.}  \protect\label{allsys}
\end{figure}

\subsubsection{\J~survival pattern: from p-p to Pb-Pb }
%

\begin{figure}[t!]
\centering
\resizebox{0.925\textwidth}{0.462\textwidth}{%
\includegraphics{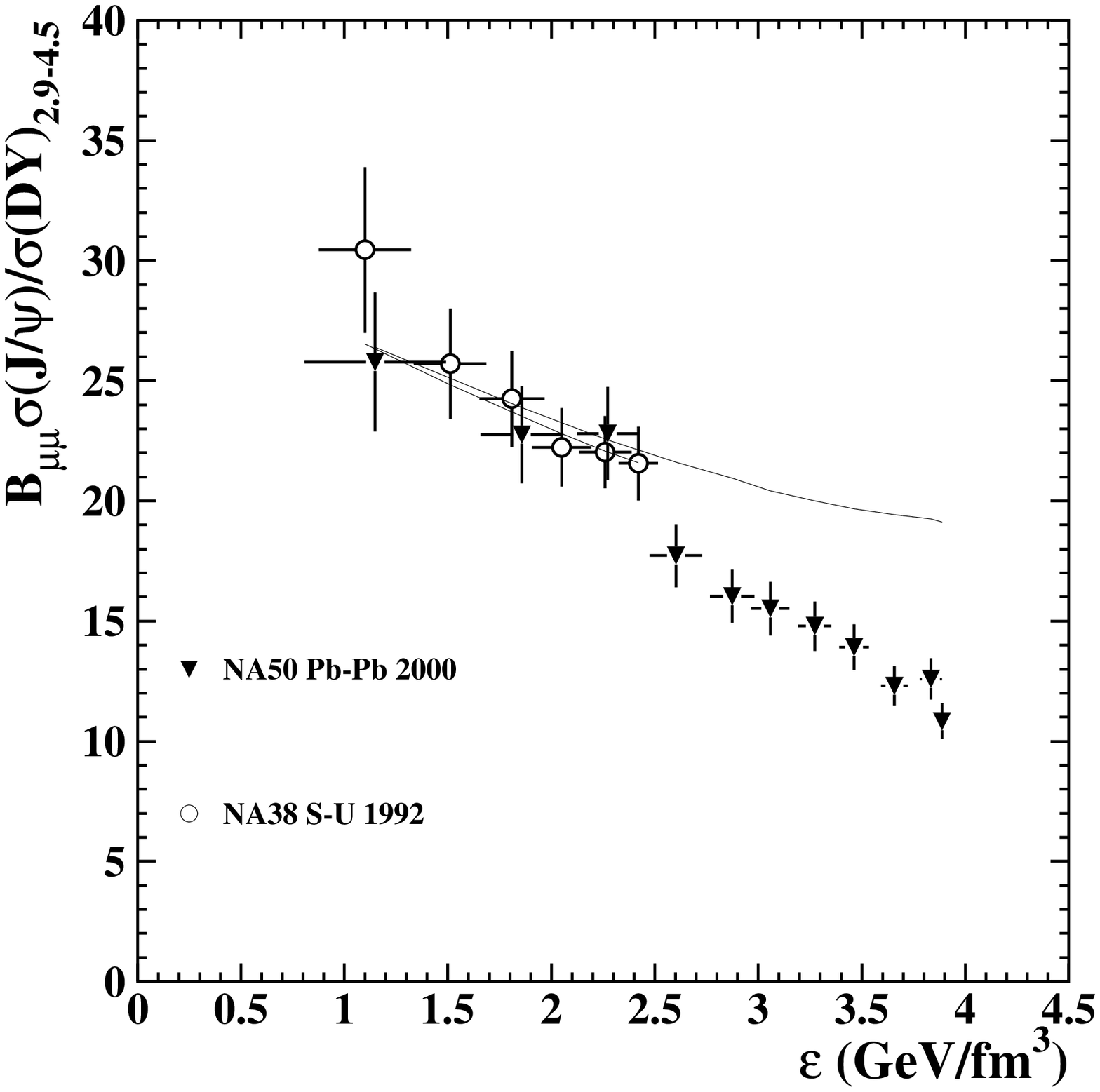}
\hspace*{50pt}
\includegraphics{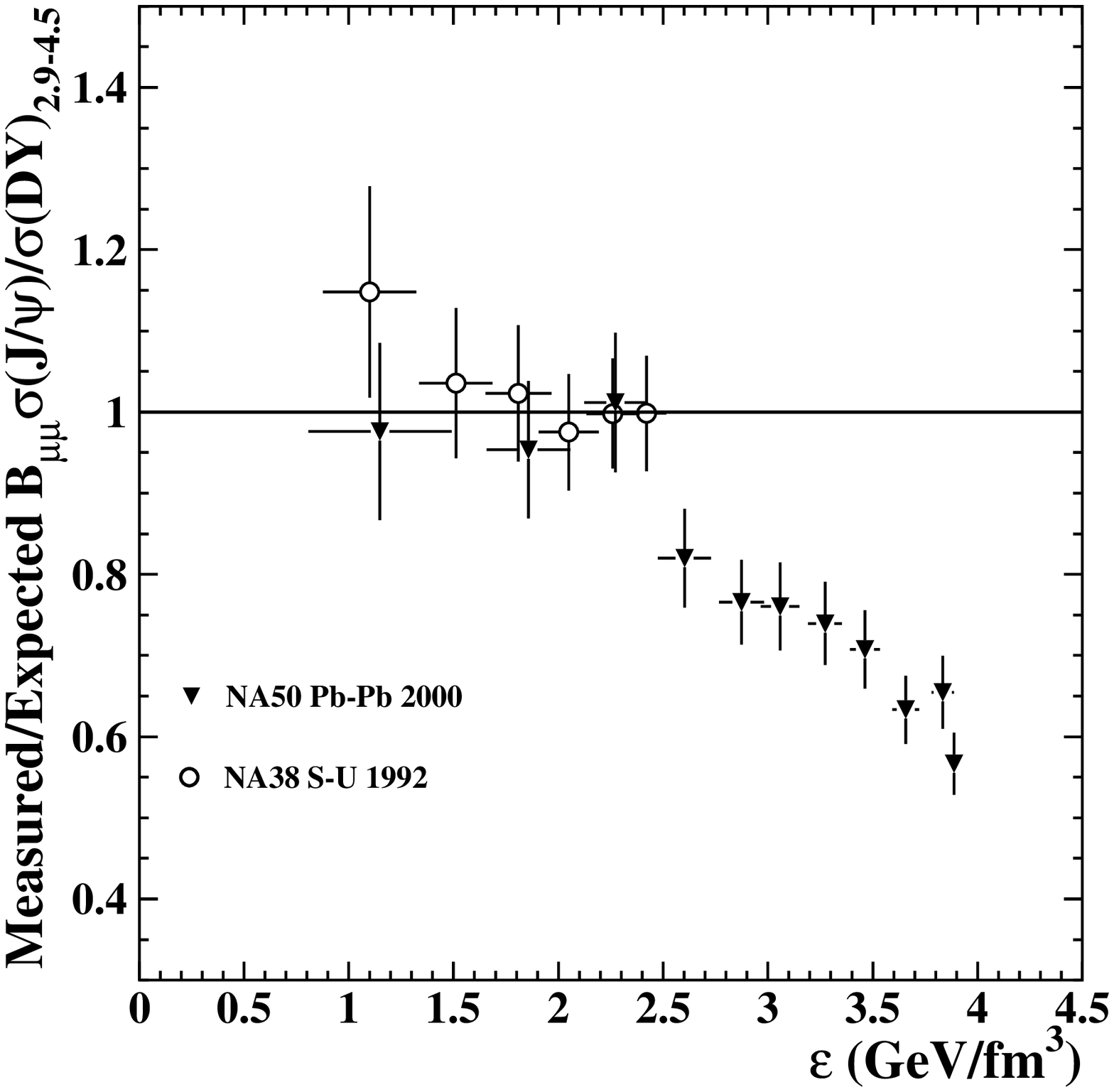}}
\vspace*{-3mm}
\caption{Same as Fig.\ref{allsys} for S-U and Pb-Pb collisions only,
         as a function of energy density. The absorption curves for S-U and Pb-Pb
         in the {\sl{left}} panel of the figure are slightly different because the 
         relation between energy density and $L$\, (obtained from a Glauber 
         calculation) depends on the colliding nuclei. }
\label{pbdens}
\end{figure}

Figure~\protect\ref{allsys} shows the ratio of cross-sections
\J/~Drell-Yan for the collision systems studied up to now, 
from the lightest p-p up to the heaviest Pb-Pb~\cite{Ramello}.
When necessary, 
they have been reanalyzed in order to ensure a fully coherent 
data selection and treatment. They have also been rescaled to an 
incident beam momentum of 158~GeV/$c$ when required. This rescaling,
both for energy and rapidity coverage, 
has used the ``simultaneous fit'' method to bring the results 
obtained at higher momenta down to 200~GeV/$c$ and an ana\-lytical
calculation to further bring the 200~GeV results down to 158~GeV.
Figure~\ref{pbdens} shows the same ratio of cross-sections
\jpsi/~Drell-Yan obtained in S-U and in Pb-Pb collisions as a
function of $\epsilon$, the energy density averaged over the whole
transverse area of the collision. 
The latter is obtained with the Bjorken~\cite{Bjorken83}
formula from the total transverse energy deduced from the 
measurement of its neutral component in the electromagnetic 
calorimeter~\cite{Helenathesis}.
It shows that the departure from the
normal nuclear absorption curve sets in for energy densities
around 2.5~GeV/fm$^3$, just above the values reached in the most 
central S-U collisions.

\medskip

\noindent The numerical values of the different variables used as centrality 
estimators are given in Table~\ref{Tab1pap2000}
for the $Et$~bins chosen above for Pb-Pb collisions at 158~GeV.
They have been calculated with the Glauber model and assuming an initial 
formation time $\tau_0=1~\rm fm/\rm c$.    
The values for S-U collisions at 200~GeV are given in 
Table~\ref{tab6paphelena}.  
They  include, in the errors, the resolution of the electromagnetic 
calorimeter. The authors~\cite{Helenathesis} consider that the relative 
scale between  S-U and Pb-Pb is robust, while the absolute values could 
be affected by a 20\%-25\% systematic uncertainty.

\vskip0.8cm

\begin{table}[ht]
\centering
\begin{tabular}{c cc cc cc cc cc}
\hline
Bin 
& \multicolumn{2}{c}{$E_{\rm T}$ (GeV)} 
& \multicolumn{2}{c}{$N_{\rm part}$}
& \multicolumn{2}{c}{$b$ (fm)}
& \multicolumn{2}{c}{$L$ (fm)} 
& \multicolumn{2}{c}{$\epsilon$ (GeV/fm$^3$)}\\
& range & average 
& average & rms 
& average & rms 
& average & rms 
& average & rms\\
\hline
 $1$ &   3--15  &  10.6 &  34 & 13 & 11.8 & 0.7 & 4.44 & 0.72 & 1.15 & 0.34\\ %
 $2$ &  15--25  &  20.4 &  70 & 13 & 10.3 & 0.5 & 5.94 & 0.42 & 1.86 & 0.20\\ %
 $3$ &  25--35  &  30.3 & 104 & 14 &  9.2 & 0.4 & 6.84 & 0.33 & 2.27 & 0.15\\ %
 $4$ &  35--45  &  40.2 & 138 & 16 &  8.2 & 0.4 & 7.51 & 0.27 & 2.60 & 0.13\\ %
 $5$ &  45--55  &  50.2 & 172 & 17 &  7.3 & 0.4 & 8.02 & 0.23 & 2.87 & 0.11\\ %
 $6$ &  55--65  &  60.1 & 206 & 18 &  6.5 & 0.4 & 8.43 & 0.20 & 3.06 & 0.09\\ %
 $7$ &  65--75  &  70.1 & 240 & 19 &  5.6 & 0.5 & 8.76 & 0.17 & 3.27 & 0.08\\ %
 $8$ &  75--85  &  80.1 & 274 & 20 &  4.8 & 0.5 & 9.02 & 0.15 & 3.46 & 0.07\\ %
 $9$ &  85--95  &  90.1 & 308 & 21 &  3.9 & 0.6 & 9.22 & 0.12 & 3.66 & 0.06\\ %
$10$ &  95--105 & 100.0 & 341 & 20 &  2.8 & 0.7 & 9.38 & 0.11 & 3.83 & 0.05\\ %
$11$ & 105--150 & 111.5 & 367 & 15 &  1.7 & 0.7 & 9.48 & 0.06 & 3.89 & 0.04\\ %
\hline
\end{tabular}
\caption{Centrality classes for Pb-Pb interactions at 158~GeV based on the 
         transverse energy measurement. For each class are listed the 
         $E_{\rm T}$ range and weighted average, together with the average 
         and rms values of $N_{\rm part}$, $b$, $L$ and $\epsilon$.}
\protect\label{Tab1pap2000}
\end{table}

\begin{table}[h!]
\centering
\begin{tabular}{c c c c c c c c c }
\hline Bin
& \multicolumn{2}{c}{$E_{\rm T}$ (GeV)} 
& \multicolumn{2}{c}{$N_{\rm part}$}
& \multicolumn{2}{c}{$L$ (fm)} 
& \multicolumn{2}{c}{$\epsilon$ (GeV/fm$^3$)}\\
& range & average 
& average & rms  
& average & rms 
& average & rms\\
\hline
 
                                           
1  &  13--28 & 22.1 & 31.3 & 9.1  & 4.37 & 0.54 &  1.04 & 0.22 \\ 
2  &  28--40 & 34.3 & 50.0 & 9.7  & 5.35 & 0.43 &  1.46 & 0.18 \\
3  &  40--52 & 46.3 & 66.7 & 10.8 & 6.04 & 0.41 &  1.76 & 0.16 \\
4  &  52--64 & 58.3 & 83.5 & 11.7 & 6.65 & 0.41 &  2.01 & 0.15 \\
5  &  64--76 & 70.2 & 98.8 & 11.0 & 7.17 & 0.38 &  2.22 & 0.13 \\
6  &  76--88 & 81.7 & 109.0 & 8.2 & 7.53 & 0.29 &  2.38 & 0.09 \\
\hline
\end{tabular}
\caption{\label{tab6paphelena} Same as previous table, for
  S-U collisions at 200~GeV.}
\end{table} 

\section{Features of $\psi'$ Suppression at SPS Energies}

As known since long, \P production is also affected by nuclear effects 
in p-A collisions. This is now quantitatively supported with precise 
results from NA50 based on two different samples of data collected 
at 450 and 400~GeV  with incident protons on Be, Al, Cu, Ag, W and Pb
targets~\cite{Rubenpap, Rubenthesis, Goncalopap, Goncalothesis}.  
The results, obtained with a Glauber parametrization separately at 450 
and 400~GeV, are quite compatible. Their combination leads to an 
absorption cross-section of
\mbox{8.3$\pm$0.9~mb} from the values of production cross-sections 
per nucleon and \mbox{7.7$\pm$0.9~mb} from the ratios \P/~Drell-Yan, 
with minimal systematic uncertainties in this case~\cite{Goncalopap}.    
These values can  be compared  to 
\mbox{4.18$\pm$0.35~mb} as obtained for \J.  

\medskip

\begin{figure}[h!]
\begin{minipage}[t]{8cm}
\resizebox{0.99\textwidth}{0.99\textwidth}{
\includegraphics*{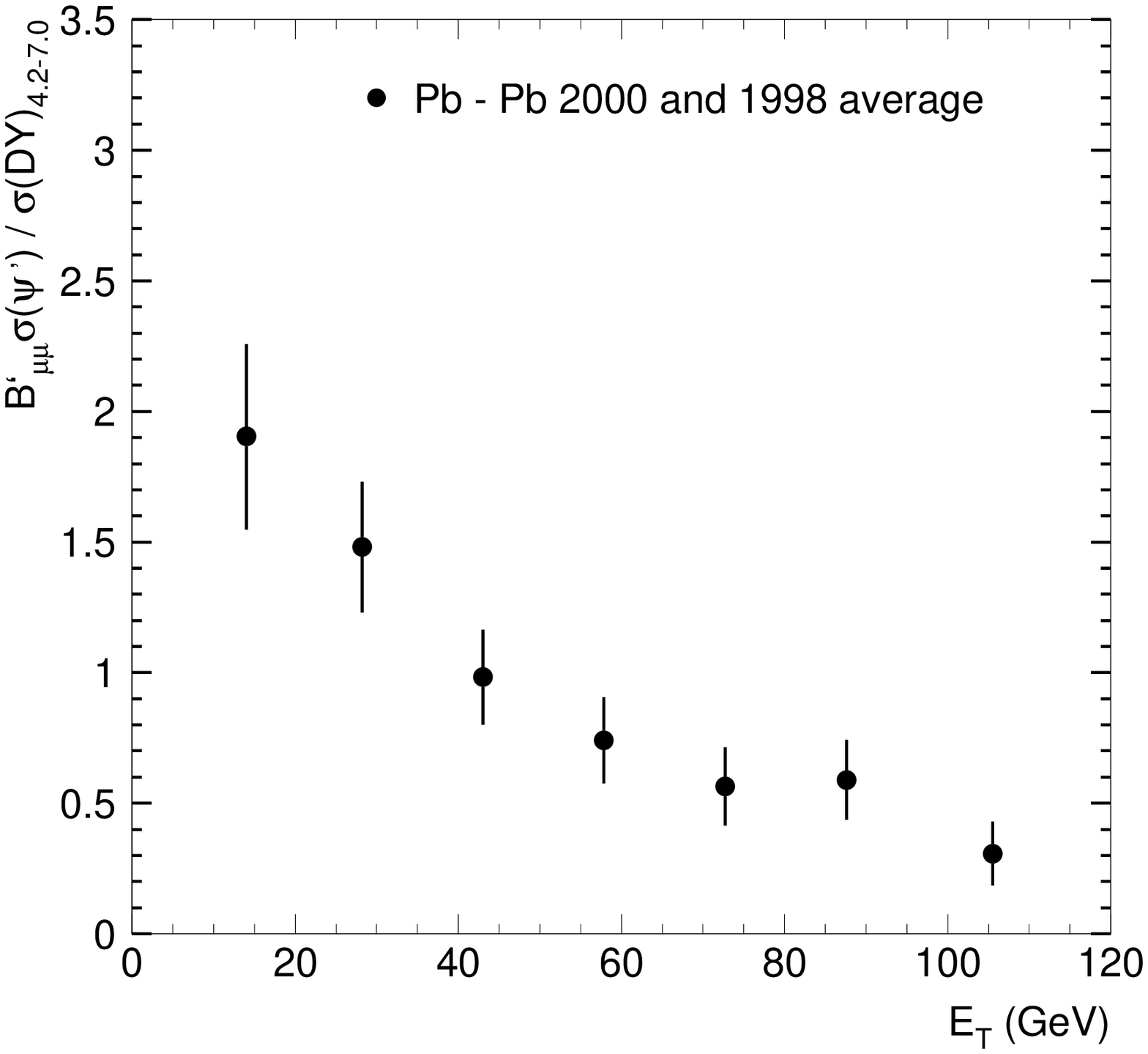}}
\vspace*{-1.4cm}
\parbox{215pt}{
\caption{The ratio of cross-sections 
         $B'_{\mu\mu}\sigma(\psi')/\sigma($DY$)$
         as a function of $E_T$, with DY in the mass range 
         \mbox{4.2-7.0~GeV/c$^2$}.}
\protect\label{psipDy-Et}}
\end{minipage}
\hspace*{10pt}
\begin{minipage}[t]{8cm}
\resizebox{0.99\textwidth}{0.99\textwidth}{
\includegraphics*{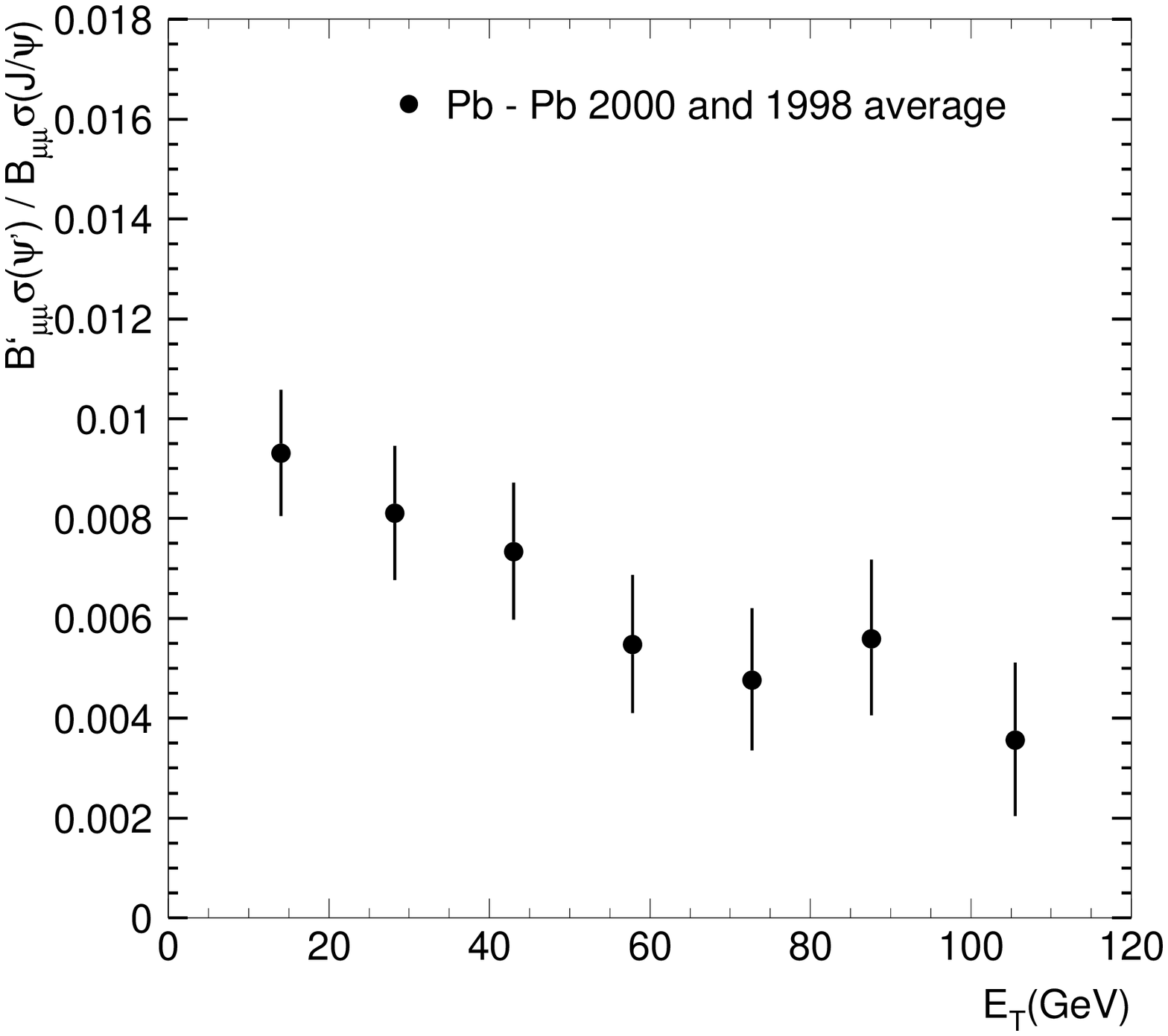}}
\vspace*{-1.4cm}
\parbox{215pt}{
\caption{The ratio $B'_{\mu\mu}\sigma(\psi')/B_{\mu\mu}\sigma($J$/\psi)$
         as a function of $E_T$.}
\protect\label{psippsi-Et}} 
\end{minipage} 
\end{figure}

\noindent The \P study~\cite{Helenapap} made with Pb-Pb interactions at 158~GeV 
is based on only 1285 \P events. It shows that both with respect to 
Drell-Yan, and also with respect to the \J, \P production 
decreases with increasing centrality, as displayed in 
Figs.~\protect\ref{psipDy-Et} and
\protect\ref{psippsi-Et}.   

\noindent
It is worthwhile noting here that 
the Drell-Yan mass range chosen here to normalize the $\psi'$
yield is \mbox{4.2--7.0 GeV/c$^2$}. Scaling up the results to the mass
range used in the J/$\psi$ suppression studies, namely 
\mbox{2.9--4.5 GeV/c$^2$}, requires to  apply the factor 7.96, 
which is the ratio between the Drell-Yan cross-sections in the two mass 
domains. 
In order to investigate whether $\p$ production exhibits any abnormal 
suppression pattern in nucleus-nucleus collisions, the production 
cross-sections per nucleon-nucleon collision (or, equivalently, normalized 
to Drell-Yan ) are compared, after appropriate rescaling to the same 
kinematical domain, between p-A, SU and Pb-Pb 
collisions~\cite{Goncalopap, Goncalothesis}.

\begin{figure}[h!]
\centering
\begin{tabular}{cc}
\resizebox{0.47\textwidth}{!}{%
\includegraphics*{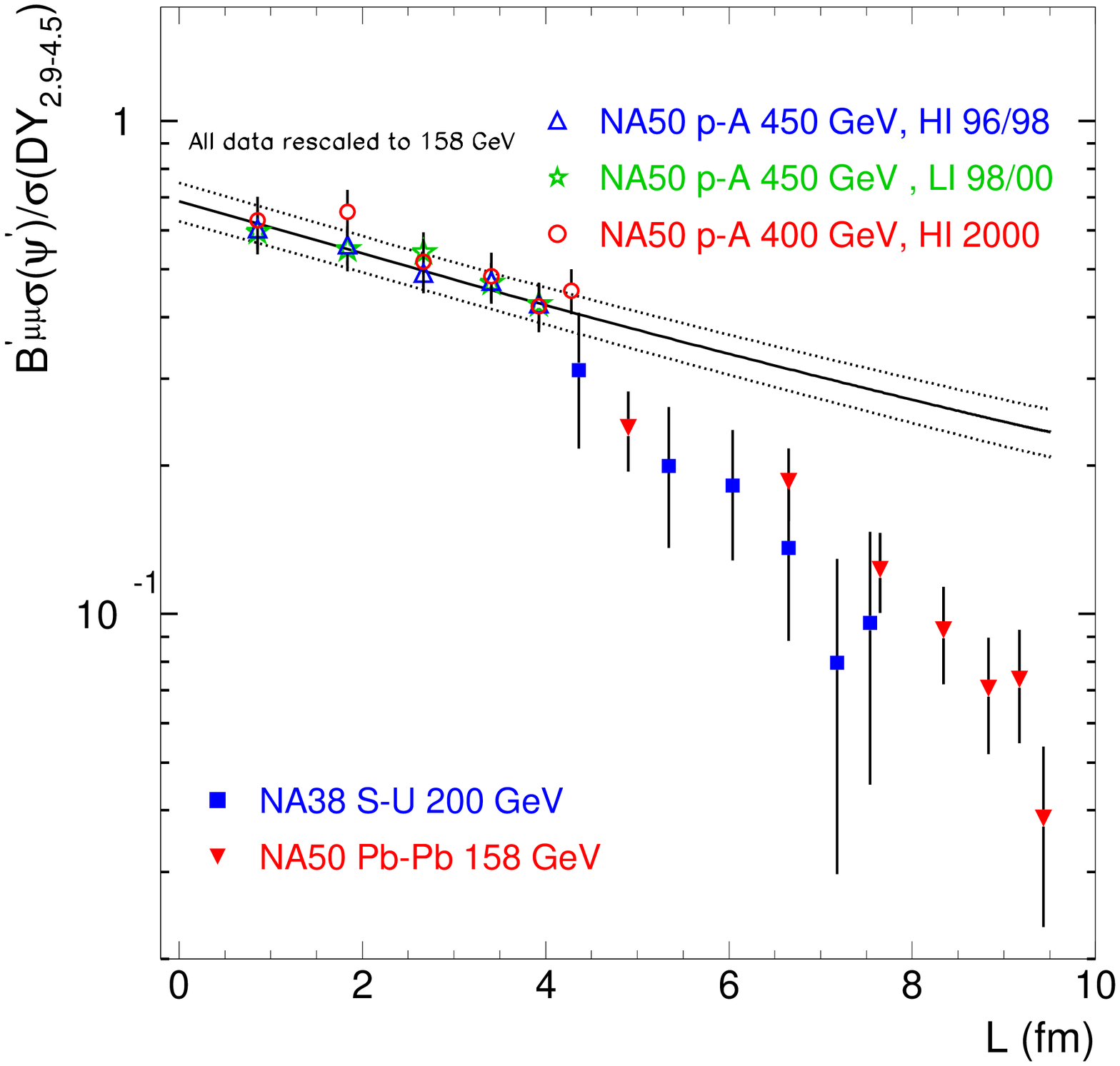}}
&
\resizebox{0.47\textwidth}{!}{%
\includegraphics*{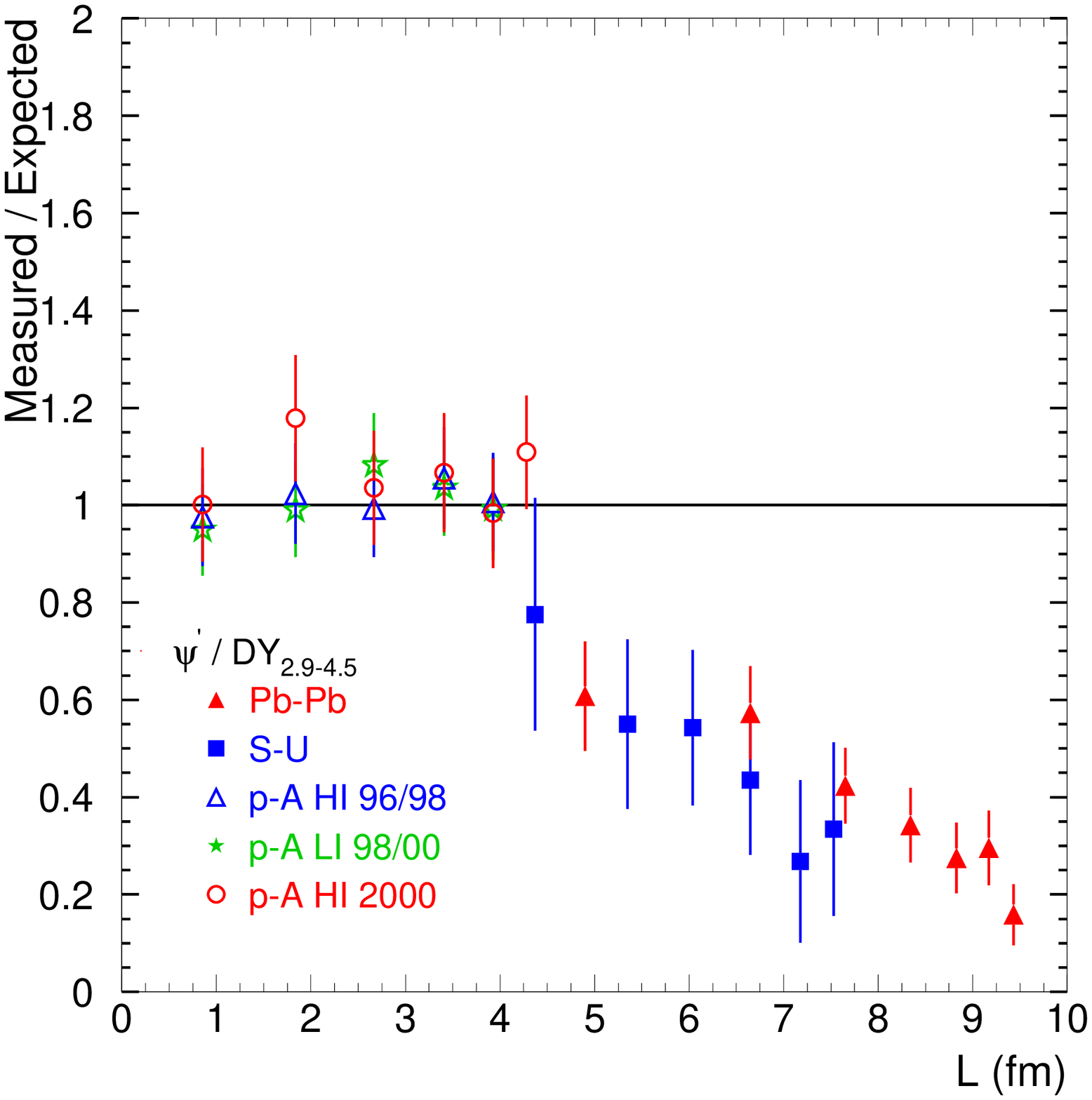}}
\end{tabular}
\vspace{-0.4cm}
\caption{The $\psi'$/~Drell-Yan ratio of cross-sections vs. L, for
         several collision systems, compared to (left) and divided by (right)
         the normal nuclear absorption pattern. All data are rescaled to
         158~GeV/nucleon.}  
\label{psipdyl}
\end{figure}
\begin{figure}[!h]
\begin{minipage}[t]{6.8cm}
\resizebox{1.0\textwidth}{1.0\textwidth}{%
\includegraphics*{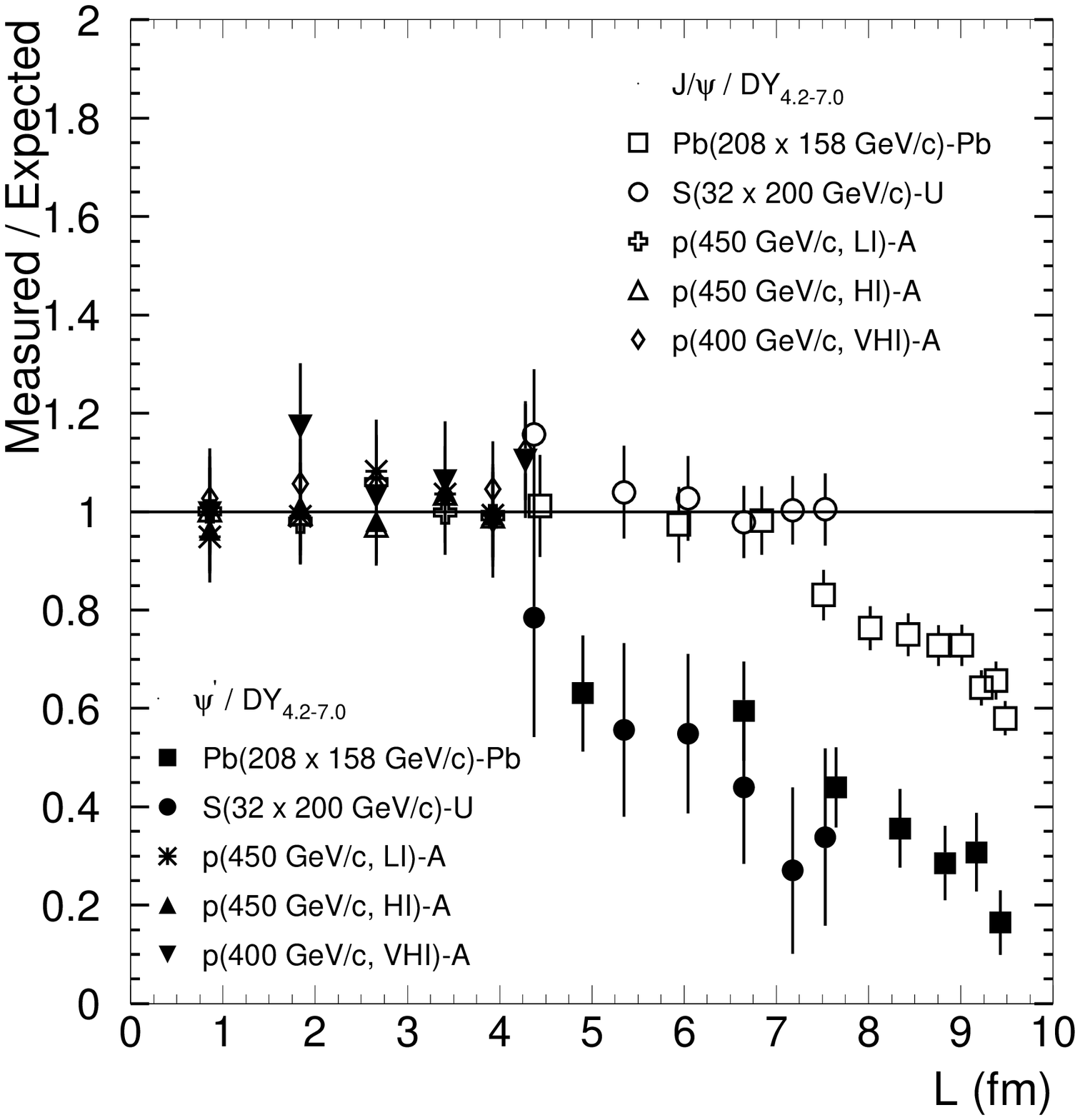}}
\vglue -15pt 
\caption{\label{prop}
         Same as Fig.~\ref{psipdyl} (right) for \J and \P on same L scale.}
\end{minipage}
\hspace*{30pt}
\begin{minipage}[t]{6.8cm}
\vspace*{-195pt}
\resizebox{1.15\textwidth}{1.15\textwidth}{%
  \includegraphics[width=\linewidth]{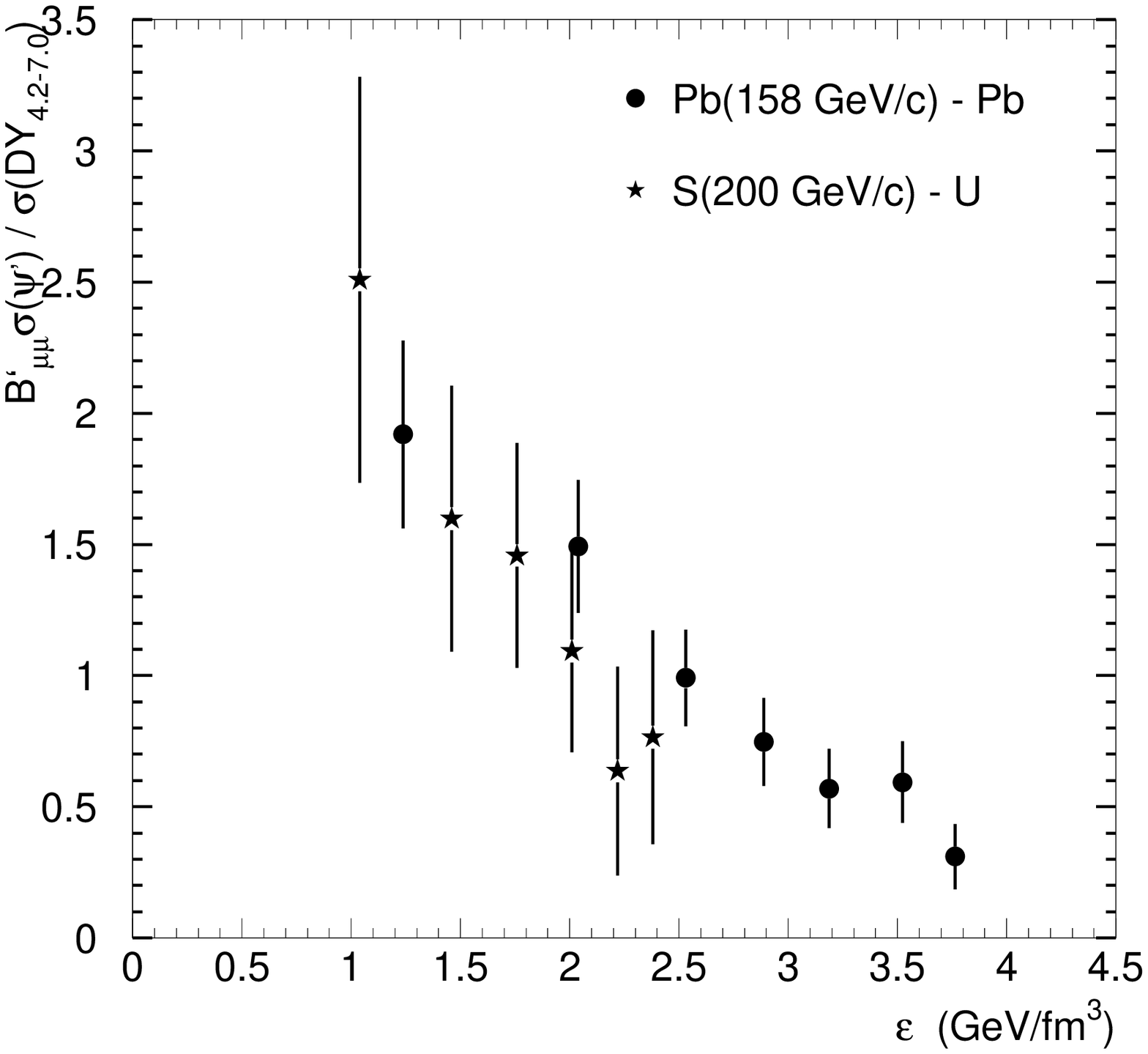}}
\vglue -40pt
\parbox{7.8cm}{
\caption{\protect\label{psipdyndens}The ratio 
         \mbox{$B'_{\mu\mu}\sigma(\psi')/\sigma($DY$)$} as
         function of the energy density. }}
\end{minipage}
\end{figure}

\medskip

\noindent The left panel of Fig.~\ref{psipdyl} shows the ratio 
\mbox{$B'_{\mu\mu}\sigma(\psi')/\sigma($DY$)$} as a function of L.  
The measured suppression patterns suggest the following features:
\begin{itemize}
\vspace*{-0.2cm}
\item A fair agreement with exponential behaviours, with two different 
   regimes, one for proton and a different one for ion-induced reactions.
\vspace*{-0.2cm}
\item A similar centrality dependence for S-U and Pb-Pb interactions.
\vspace*{-0.2cm}
\end{itemize}

\noindent Using the approximate exponential parametrization to describe 
$\psi'$ absorption as a function of L, the fit of the data leads 
to an absorption cross-section of ~7.3~$\pm$~1.6~mb 
in p-A collisions, while 
a much higher value, 19.2~$\pm$~2.4~mb, is obtained for ion-ion collisions
(S-U and Pb-Pb fitted simultaneously).
The right panel of Fig.~\ref{psipdyl} shows, as a function of L, the ratios
between the measured \P yields, normalized to Drell-Yan,
and the corresponding expected ``normal nuclear absorptions'', as extrapolated
from p-A measurements. 
The difference between the suppression patterns for \J and \P is shown in 
Fig.~\ref{prop}. 
The \P suppression for S-U and Pb-Pb collisions is shown in 
Fig.~\ref{psipdyndens} as a function of the energy density which 
is calculated as described in detail in~\cite{Helenathesis}. 
The numerical values of L, $\npt$ and the energy density $\epsilon$
used for the \P studies in Pb-Pb collisions are given in 
Table~\ref{tab5paphelena}.

\begin{table}[h!]
\centering
\begin{tabular}{c c c c c c c c c}
\hline Bin
& \multicolumn{2}{c}{$E_{\rm T}$ (GeV)} 
& \multicolumn{2}{c}{$N_{\rm part}$}
& \multicolumn{2}{c}{$L$ (fm)} 
& \multicolumn{2}{c}{$\epsilon$ (GeV/fm$^3$)}\\
& range & average 
& average & rms  
& average & rms 
& average & rms\\
\hline


1 &  3--20 & 13.9 &  44.6 & 17.7 & 4.90 & 0.84 &  1.24 & 0.37 \\ 
2 & 20--35 & 28.2 &  96.7 & 18.0 & 6.65 & 0.44 &  2.04 & 0.20 \\
3 & 35--50 & 43.0 & 147.1 & 19.3 & 7.65 & 0.31 &  2.53 & 0.14 \\
4 & 50--65 & 57.8 & 197.7 & 20.6 & 8.34 & 0.24 &  2.89 & 0.11 \\
5 & 65--80 & 72.7 & 248.6 & 22.0 & 8.83 & 0.18 &  3.19 & 0.09 \\
6 & 80--95 & 87.6 & 299.2 & 23.0 & 9.17 & 0.14 &  3.52 & 0.07 \\
7 & 95--150& 105.9& 353.3 & 22.4 & 9.43 & 0.10 &  3.76 & 0.06 \\
\hline
\end{tabular}
\caption{\label{tab5paphelena}Mean free path crossed by the
  $c\bar{c}$ pair inside the nucleus, number of parti\-cipating nucleons and energy
  densities for each Pb-Pb $E_{T}$ range.}
\end{table}

\section{More Results from SPS and RHIC}

After the systematic studies carried out by experiments NA38, 
NA50 and NA51, two other experiments were aimed at extending 
the knowledge on charmonium suppression in ultrarelativistic 
heavy ion interactions.

\medskip

\noindent At the CERN-SPS, in a fixed target experiment, the NA60 
collaboration uses basically the same muon spectrometer as 
NA50~\cite{NA50NIM?} except for the centrality estimate which 
is based exclusively on the very forward calorimeter 
measurements.
The NA50 detector is upgraded with a silicon pixel telescope, 
located in the target region, which leads to an extremely 
precise determination of the origin of the measured tracks
and, consequently, to a better dimuon mass resolution.  
Data have been collected to study  In-In collisions with an 
incident In beam of 158~GeV/c~\cite{NA60pap} and also  p-A 
reactions with incident proton beams of both 400 and 
158~GeV/c. 

\medskip

\noindent At the RHIC collider at BNL,
the PHENIX experiment studies
p-p~\cite{Phenixpp1, Phenixpp2}, \mbox{d-Au}~\cite{PhenixdAu1, PhenixdAu2}, 
Cu-Cu~\cite{PhenixCuCu} and Au-Au~\cite{PhenixAuAu1, PhenixAuAu2} collisions 
at much higher energies, namely 200~GeV in the c.m.s. and, consequently, at 
higher energy densities than those reached  at the CERN-SPS. 
It makes use of a detector specifically designed to study heavy ion 
collisions at RHIC~\cite{PhenixNIM}.\\
Some of the results obtained by these experiments in their studies of 
\J suppression are summarized hereafter.  

\subsection{\J~Suppression in In-In Collisions at 158~GeV}
 
\begin{figure}[h]
\begin{minipage}[t]{8cm}
\resizebox{0.99\textwidth}{0.99\textwidth}{
\hspace*{-26pt}
\includegraphics*{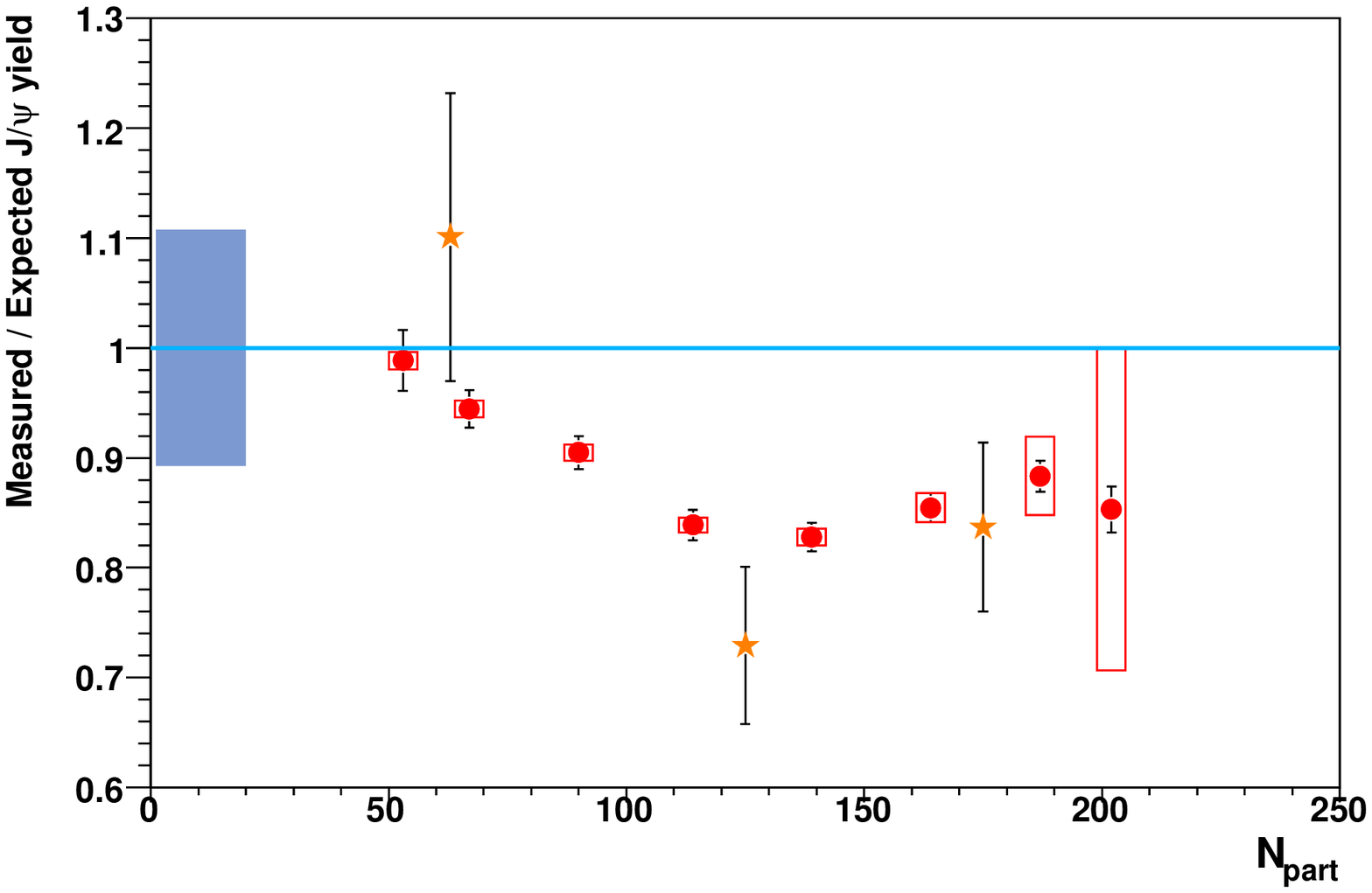}}
\parbox{7.8cm}{
\caption{Centrality dependence of the \jpsi\ suppression in 
\mbox{In-In} collisions for  
the ratio  $\sigma_{\rm J/\psi}/\sigma_{\rm DY}$ (star symbols) and for the 
\jpsi\ absolute yield (circle symbols). 
Also shown are the common global systematic error (left box) and the relative 
point to point uncertainties.}
\label{NA60fig:2}}
\end{minipage}
\hspace*{6pt}
\begin{minipage}[t]{8cm}
\resizebox{0.99\textwidth}{0.99\textwidth}{
\includegraphics*{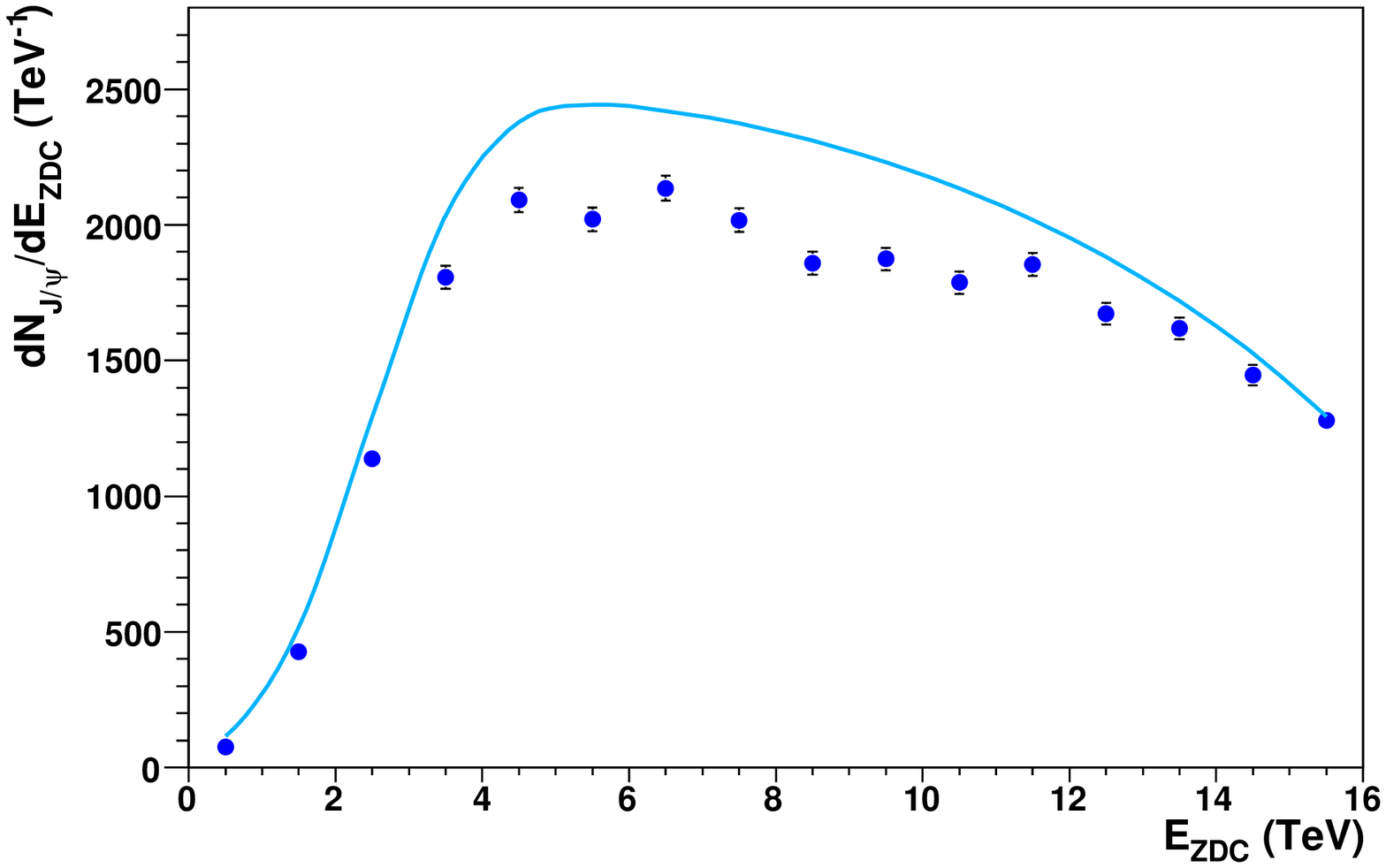}}
\hspace*{5pt}
\parbox{7.5cm}{
\caption{The \jpsi\ $E_{\rm ZDC}$ distribution (circles), compared with 
expectations from nuclear absorption (line).}
\label{NA60fig:3}}
\end{minipage}
\end{figure}

In a first analysis strategy, experiment NA60 
follows 
the approach adopted by experiments NA38/NA50 and makes use of the 
ratio \J/~Drell-Yan. This ratio is free from systematic uncertainties 
and provides its own absolute normalization, without any external input. 
It thus leads to the most robust estimate of the  \J yield normalized 
to the number of nucleon-nucleon collisions, through the experimentally 
measured  Drell-Yan events. Unfortunately the number of Drell-Yan 
events in NA60 is extremely small and allows, in a centrality dependent 
study, for only 3 bins in centrality. In order to identify any 
anomalous behavior,
the results are compared with normal nuclear absorption as determined 
experimentally by experiments NA38/NA50. 
The results can then be expressed in terms of the ratio "measured/expected",
where "expected" refers to normal nuclear absorption corresponding to an 
absorption cross-section of \mbox{4.18$\pm$0.35~mb}. The 3 measured ratios
are plotted in Fig.~\ref{NA60fig:2} as a function of $\npt$. 

\medskip

\noindent In a second approach, the measured "unnormalized" \J yield is directly 
compared to the analytically calculated centrality distribution of \J. 
The latter is derived in the frame of the Glauber model
using the same absorption cross-section of 4.18~mb. Relative normalization
between the data and the reference curve is obtained by requiring that the 
ratio of the centrality integrated distributions is equal to the centrality 
integrated ratio \J /~Drell-Yan. Fig.~\ref{NA60fig:3} 
shows both the measured and calculated distributions as a function of 
centrality, estimated here from the forward energy of the spectators nuclei. 
Their ratio is plotted in Fig.~\ref{NA60fig:2} as a function of $\npt$. 

\medskip
 
\noindent The method does not suffer any more from the limited number of Drell-Yan 
events, makes use of the statistical power of the \J sample  and reaches 
unprecedented small statistical uncertainties. It should
be noted here, nevertheless, that the same method could have been applied 
by experiment NA50 which, indeed and as compared to NA60,  had more than a 
factor 3 larger \J sample of events. The NA50 experiment 
made the choice of studying the \J suppression pattern in Pb-Pb collisions
through the robust, unbiased \J/~Drell-Yan ratio which is free, at first 
order, from all potential systematic experimental effects which cancel out 
in the ratio. It is also free from analytical model dependent descriptions 
of the detectors.      
It therefore minimizes systematic uncertainties, sometimes difficult to find 
and evaluate.
NA50 discarded this type of analysis from its final results which are thus 
dominated, by far, by pure statistical uncertainties in the Drell-Yan sample 
of events.  

\medskip

\noindent The results of experiment NA60 can be compared to those of NA50
shown in Fig.~\ref{npart} which superposes the results obtained, 
separately, through the three independent centrality estimators.
The suppression patterns in Pb-Pb and In-In are qualitatively similar
and compatible, thanks to the errors quoted by NA50. 
In some more detail, nevertheless, some differences can 
be noticed in this comparison. 

\medskip

\noindent The results on In-In do not exhibit a solid evidence of normal 
behaviour for peripheral collisions. There is one single centrality
point with a ratio "measured/expected" amounting to 0.99$\pm$0.06, 
which is somewhat insufficient to support a constant trend 
close to unity in the vicinity of  $\npt\simeq 50$.
On the other hand, the Pb-Pb results do exhibit a normal behaviour
extending over 2 or 3 centrality bins and up to about 
76 $\npt$. 
Taking into account the resolution of the centrality measurement, 
In-In collisions would depart from normal behaviour 
for $\npt\simeq 80~\npt$ corresponding to an energy density 
of \mbox{1.5~GeV/fm$^3$}.  For Pb-Pb collisions, the departure 
is detected at $\npt \simeq 100~\npt$ when using the same 
centrality detector and at $\npt \simeq 120~\npt$ when using the 
most precise centrality estimator, which corresponds to 
2.25 and \mbox{2.45~GeV/fm$^3$} respectively. 

\medskip

\noindent The In-In and Pb-Pb suppression patterns themselves show different
trends. 
Pb-Pb exhibits a steady decreasing trend, in particular 
for \mbox{$\npt > 150$}, as opposed to a kind of an unexpected rise, 
starting for \mbox{$\npt > 140$}, in the case of In-In. 
This rise could perhaps accommodate, within errors, some flat 
behaviour.\\   
 

\subsection{\J~Suppression in A-A Collisions at 
\mbox{$\sqs =200$~GeV}}

The PHENIX experiment at RHIC measures \J production 
at $\sqs = 200~\rm GeV$, for rapidities 
\mbox{$ |y| < [1.2,2.2]$} and also  \mbox{$ |y| < 0.35 $}. 
The study is made in terms of the "nuclear modification factor", 
$\Rab$, defined as:

\begin{displaymath}
\Rab(y) = {(d\sigma/dy)_{A-B} \over N_{coll}~ (d\sigma/dy)_{p-p}}
\label{R-AB}
\end{displaymath}

\noindent
when referring, for example, to the reaction of nucleus A with nucleus B.\\

\begin{figure}[h!]
\begin{minipage}[t]{8cm}
\resizebox{0.99\textwidth}{0.99\textwidth}{
\hspace*{-26pt}
\includegraphics{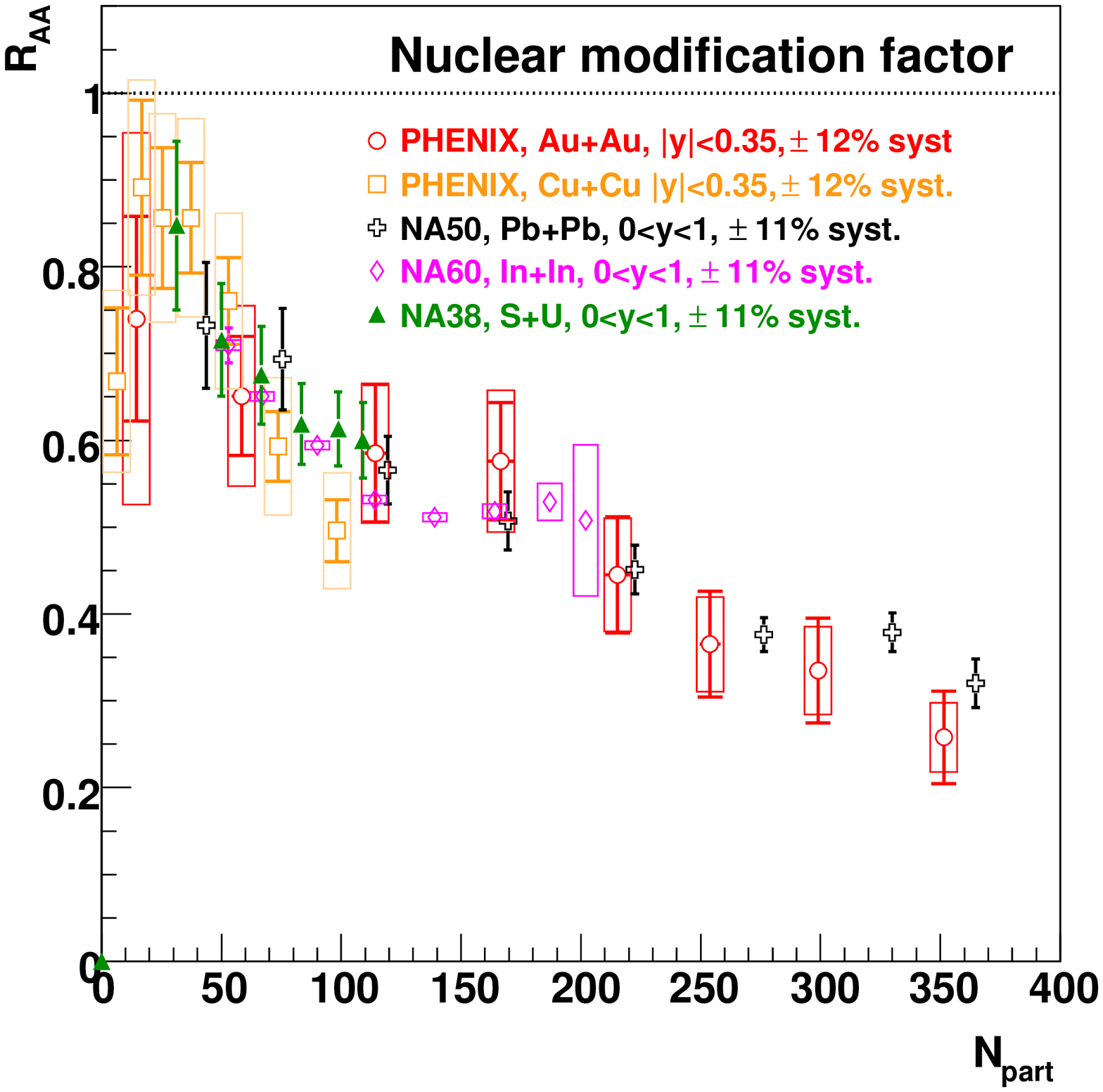}}
\vglue 3mm
\parbox{7.8cm}{
\caption{ \J nuclear modification factors for Au-Au, Pb-Pb, In-In and S-U 
colliding systems at their respective energies (200, 19 and 17.3~GeV  ) as 
a function of the number of participants, $\npt$. } 
\protect\label{Raphael05}}
\end{minipage}
\hspace*{6pt}
\begin{minipage}[t]{8cm}
\vspace*{-217pt}
\resizebox{0.99\textwidth}{0.99\textwidth}{%
\includegraphics*{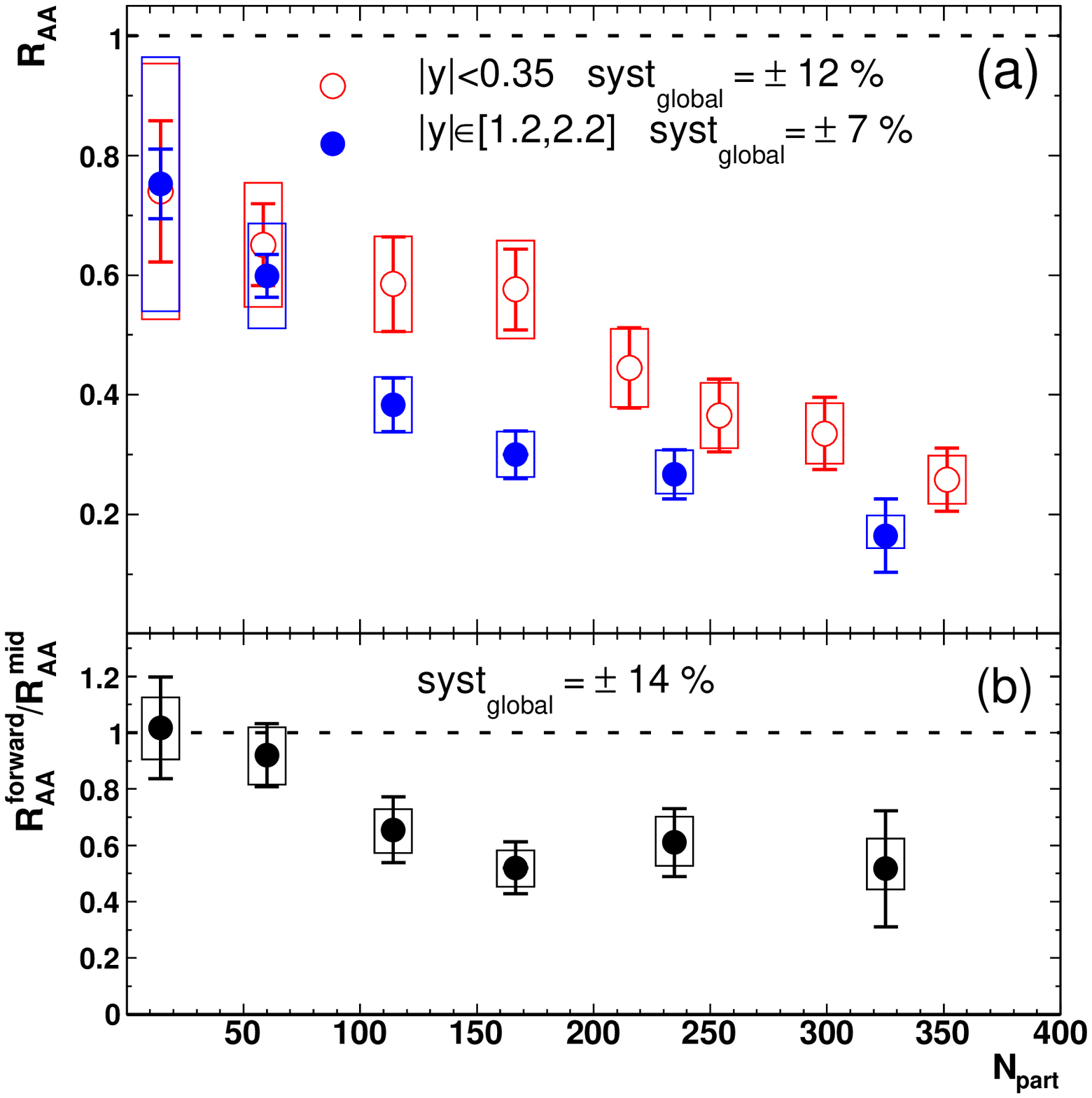}}
\vspace*{-3mm}
\parbox{7.8cm}{
\caption{ (a) The \J nuclear modification factor for Au-Au as a function of 
$\npt$, for central and forward rapidities. (b) Ratio of forward/mid rapidity
\J $\Raa$ vs. $\npt$. } 
\protect\label{Raphael06}}
\end{minipage}
\end{figure}

\noindent Fig.~\ref{Raphael05}  shows the ratio $\Raa$~ for \J 's 
of low rapidity in the collision c.m.s. obtained by the PHENIX
experiment for Au-Au and Cu-Cu collisions. 
The same plot shows the results obtained with other  
colliding systems at lower energies, derived from the measurements
performed at CERN.
The ratio $\Raa$ (or  $\Rab$) is plotted as a function of the 
$\npt$ value corresponding to the selected centrality bins made 
in the different analyses. 
It exhibits, within errors, an in-discriminable pattern for all 
the colliding systems, including S-U and Pb-Pb. 
On the other hand and from different type of studies, as detailed 
above, it is known that S-U behaves normally as well as the 2 more 
peripheral points of Pb-Pb which are plotted in the figure, whereas 
abnormality sets in for $\npt>100$\, in the case of Pb-Pb, at least. 
It therefore appears that this kind of representation, namely $\Raa$ 
as a function of $\npt$, is 
not the most sensitive 
in order to experimentally detect, without external theoretical 
inputs, any abnormal \J suppression.\\       
Fig.~\ref{Raphael06} shows that the nuclear modification factor is 
a function of the \J rapidity for  Au-Au collisions. Indeed, this 
feature was already visible in fixed target p-A experiments with 
200 and 800~GeV/c incident protons by experiments NA3~\cite{papNA3} 
and E866~\cite{E866paper}.
The understanding of such an observation on Au-Au at 200~GeV requires 
again complete experimental knowledge of the reference for the 
two different kinematic windows, under conditions excluding the 
formation of a new state of matter (see section "Discussion" below).

\begin{figure}[h!]
\begin{minipage}[t]{8cm}
\resizebox{0.99\textwidth}{0.75\textwidth}{
\hspace*{-26pt}
\includegraphics*{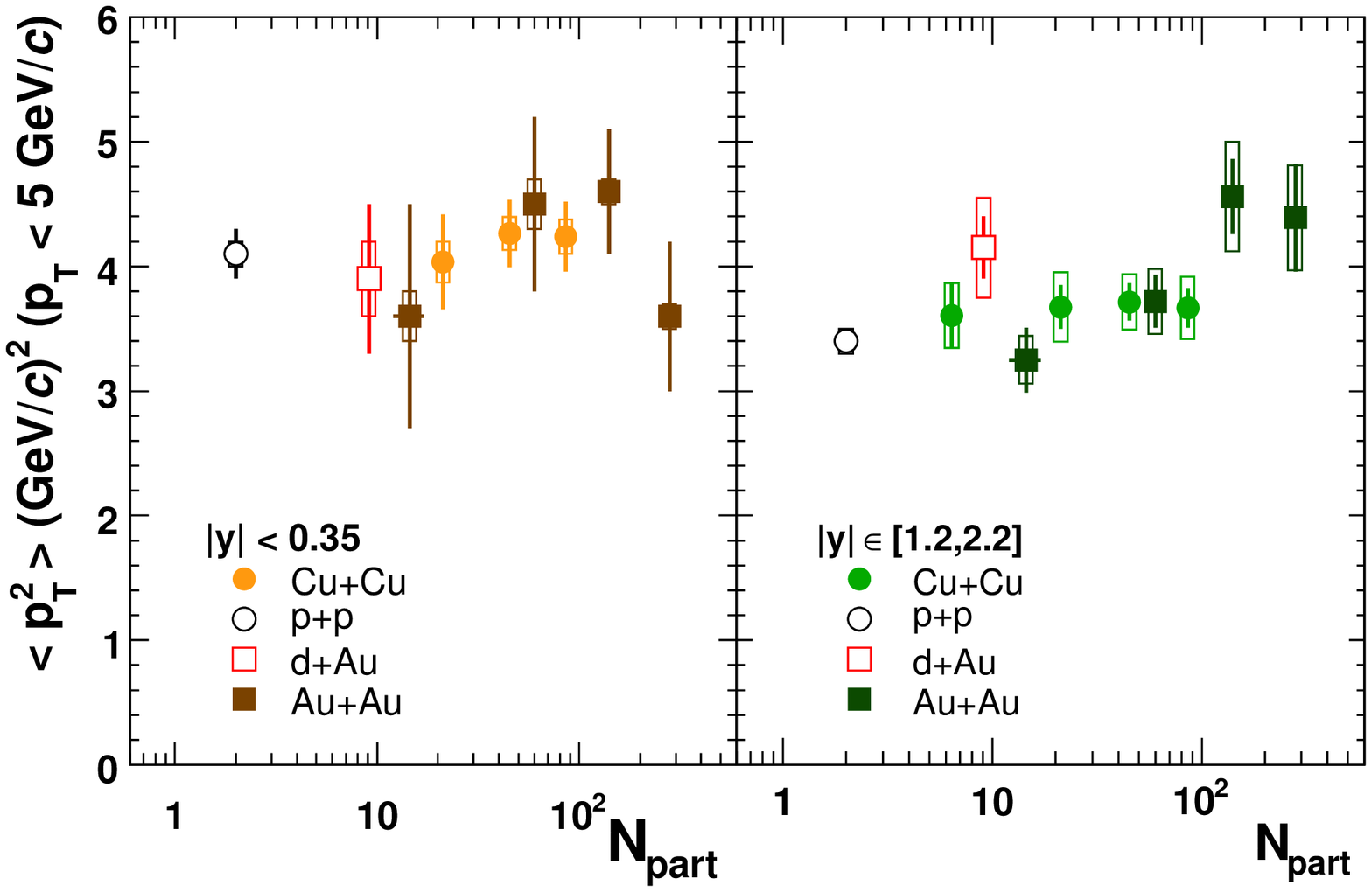}}
\parbox{7.8cm}{
\caption{The surviving \J average \mbox{$< \TR2 > $} as a function of 
$\npt$ for Au-Au, as measured by PHENIX at mid  (left) and forward 
(right) rapidity. }
\label{Raphaelpt2npart}}
\end{minipage}
\hspace*{6pt}
\begin{minipage}[t]{8cm}
\resizebox{0.99\textwidth}{0.75\textwidth}{%
\includegraphics{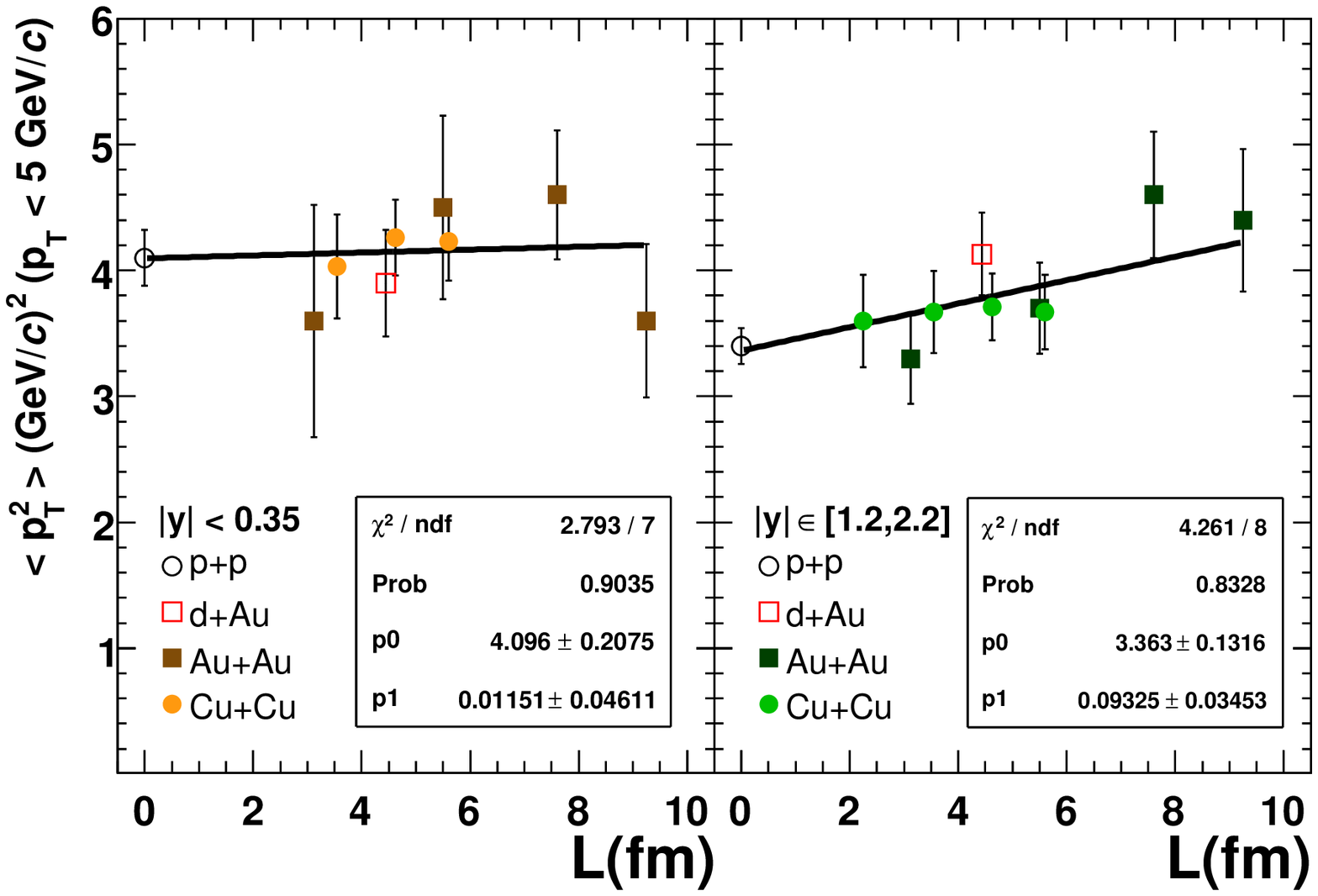}}
\vspace*{-3mm}
\parbox{7.8cm}{ 
\caption{Same as fig.~\ref{Raphaelpt2npart}, as a function \mbox{of L.}}
\protect\label{Raphaelpt2l}}
\end{minipage}
\end{figure}

~~\\ 
The \J suppression has also been studied by PHENIX as a function of the
transverse momentum. 
Fig.~\ref{Raphaelpt2npart} shows, as a function of $\npt$,
the values of $ \,<{p_{\rm T}^2}\,>$  for various colliding 
sytems~\cite{PhenixCuCu}.  
Fig.~\ref{Raphaelpt2l}~\cite{RaphaelQM08} uses instead the variable L
in order to allow a comparison with the results obtained at the SPS,
successfully parametrized as a function of the path traversed by the \J
in nuclear matter (see Figs.~\ref{transv-sep} and~\ref{transv-allfitted}).
The PHENIX measurements are affected by very large uncertainties. 
Within their limited accuracy, they also suggest a linear behaviour 
as a function of L, the same for all the systems which are taken here 
at same colliding energy. The numerical values of the corresponding fitted
slopes, namely $0.012 \pm 0.046$  and  $0.093 \pm 0.035$ for central and
forward rapidity respectively can be compared with the 
accurate value measured at the SPS by the 
NA38/NA51/NA50 experiments which amounts to $0.081 \pm 0.003$ in
significantly different kinematical conditions. 
Much more accurate measurements are obviously needed at RHIC in order 
to become a quantitative probe for the simplistic interpretation of the 
lower energy results.    

\section{Discussion and Evaluation}

Future experimental progress could consolidate what appears 
already today like well established experimental evidences. 
The results detailed above do raise, nevertheless, two important
points which are discussed hereafter.  


\begin{itemize} 
\item{Anomalous \J~suppression in Pb-Pb collisions at \mbox{
$\sqs = 17.2$ GeV}.\\
The anomalous character of the observed \J suppression
is the first point 
to be clarified if charmonium suppression
is the signature of 
some new physics specific to high energy nucleus-nucleus 
collisions. 
For the moment, an evidence for this abnormal behaviour has been 
deduced from the systematic studies carried at the SPS 
by experiments~NA38/51/50. 
In order to define the reference curve attached to "normality", 
the experiments have made use of a set of 
systematic measurements made on p-induced reactions at 450, 400 and 
200~GeV/c. It has been found that also \J's produced
in O-Cu, O-U and S-U~collisions at 200~GeV behave normally, i.e. 
like in p-A reactions, {\em for the same rapidity range in the lab 
(or rather target) reference system}. 
It is with respect to this experimental reference, appropriately 
rescaled to 158~GeV under the explicit assumption that nuclear 
absorption of \J is beam-energy independent from 450 down to 158~GeV, 
that the abnormal character of \J suppression has been established 
for mid-central and central Pb-Pb reactions. 
The validity of this crucial assumption is supported by two independent 
fixed target experiments.

{\begin{enumerate}
\item{ Experiment E866} \\
At Fermilab, experiment E866~\cite{E866paper} has provided 
cross-section ratios measurements on \mbox{p-Be} and \mbox{p-W} 
at 800~GeV/c~($\sqs$ = 39~GeV).    
The measured ratio of cross-sections remains constant within the 
range~ $-0.10 < x_f < 0.25$ or, equivalently~
\mbox{$ 3.13 < y_{\rm lab} < 4.95 $}.   
When parametrized with the Glauber model,
it leads to 
an absorption cross-section of
\mbox{2.83 $\pm$ 0.77}.
This numerical value is therefore rapidity independent 
within the explored $y_{\rm lab}$ range which largely overlaps 
with the range \mbox{$ 3.0<y_{\rm lab}<4.0$} explored by 
the NA38/NA51/NA50 experiments. 
The latter find  \mbox{4.18$\pm$0.35~mb}, show that 
nuclear absorption indeed accounts for the measurements 
and, furthermore, that the absorption cross-section is rapidity 
independent within the covered range~\cite{Rubenpap,Rubenthesis}.   
%
\item{Experiment NA3} \\    
The most precise result on the topic is found in CERN experiment 
NA3~\cite{papNA3, Charpentierthesis}. Their data lead to a  precise 
value of the ratio of \J production cross-sections measured directly 
in p-p and in p-Pt collisions in a single simultaneous beam exposure 
at \mbox{200~GeV} or $\sqs = 19.4~{\rm GeV}$.
The Glauber model description allows to 
derive the nuclear absorption cross-section, with appropriate 
treatment of errors, in the $y_{\rm lab}$ rapidity range
where NA3 shows it stays constant, namely 
\mbox{$ 3.0<y_{\rm lab}<4.0$}.
The resulting numerical value is 
\mbox{4.1$\pm$1.0~mb}. 
This value is surprisingly close 
to the NA50 determination at 400/450~GeV, namely 
\mbox{4.18$\pm$0.35~mb}. 
The results obtained for p-a reactions at 800, 450/400
and 200~GeV/c suggest  
that the \J absorption cross-section has no or little dependence 
on $\sqs$ within this $E_{\rm beam}$ range, when considered in 
the same $y_{\rm lab}$ rapidity interval in which it stays constant.
They thus support the assumption made by NA50 when rescaling the normal 
absorption reference curve down to 158~GeV.
\end{enumerate}

}

%
%
\begin{figure}[h!]
\begin{center}
\begin{minipage}[t!]{6.8cm}
\resizebox{0.9\textwidth}{0.9\textwidth}{
\includegraphics{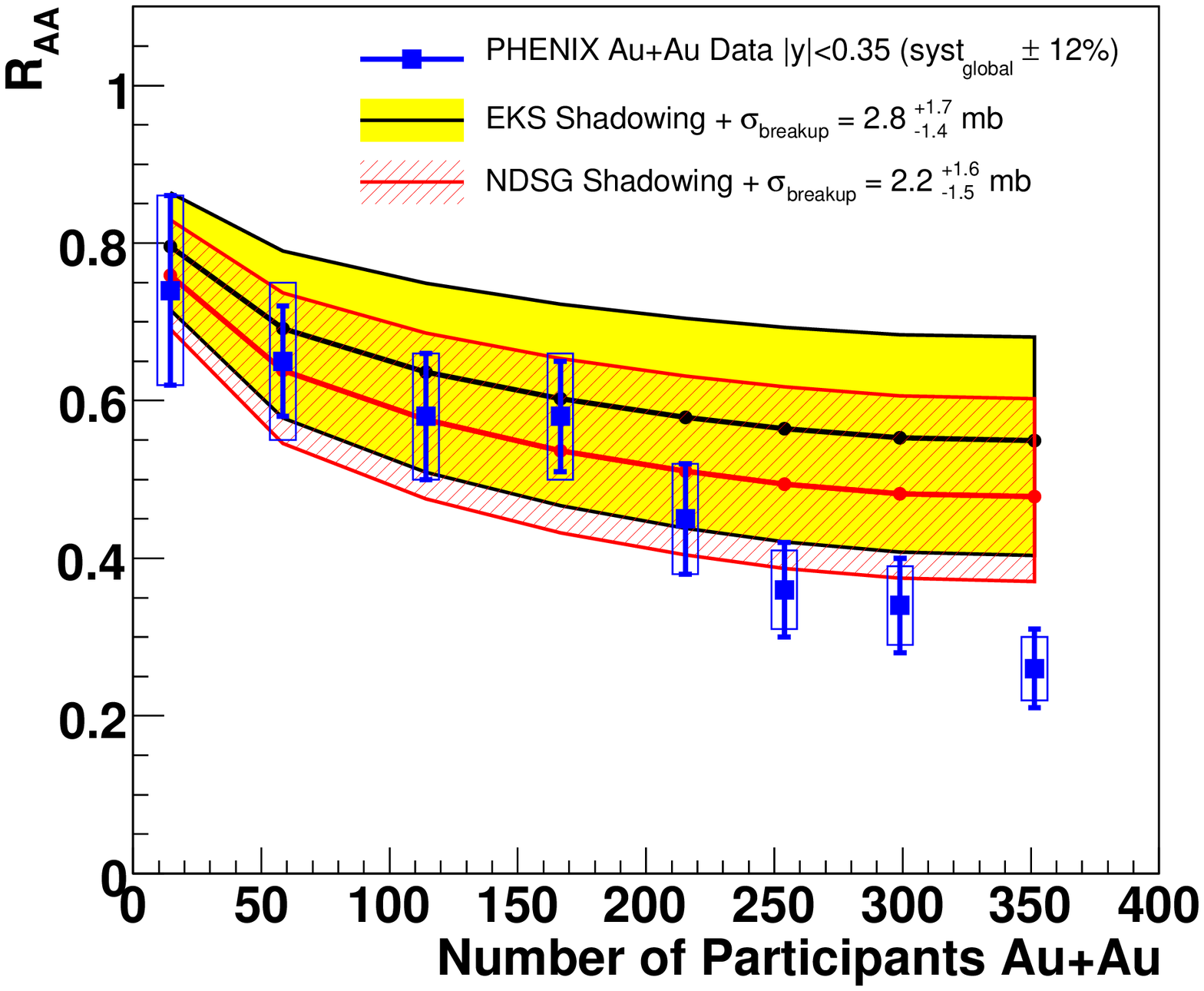}}
\end{minipage}
\begin{minipage}{6.8cm}
\resizebox{0.9\textwidth}{0.9\textwidth}{%
\includegraphics*{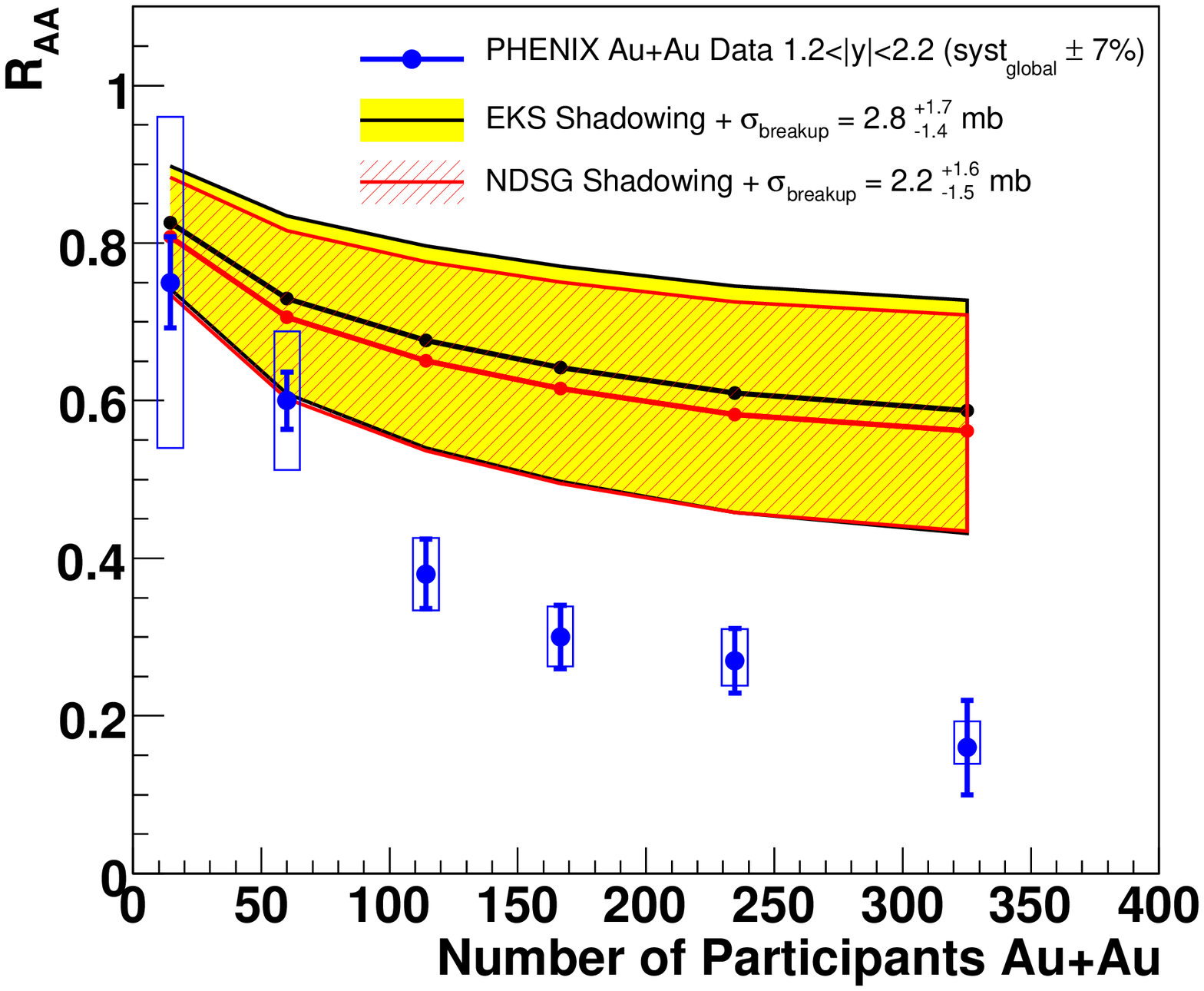}}
\end{minipage}
\parbox{0.99\textwidth}{
\caption{ $\Raa$ for Au-Au collisions compared to a band of theoretical curves for the \sgabs values found to be
consistent with the d-Au data.
EKS and NDSG shadowing are included for midrapidity (left). Same for forward rapidity  (right).}
\protect\label{rauauabnorm1}}\\[1mm] 
\end{center}
\end{figure}

\begin{figure}[h!]
\begin{center}
~~ 
\begin{minipage}[t]{6.8cm}
\resizebox{0.9\textwidth}{0.9\textwidth}{
\includegraphics{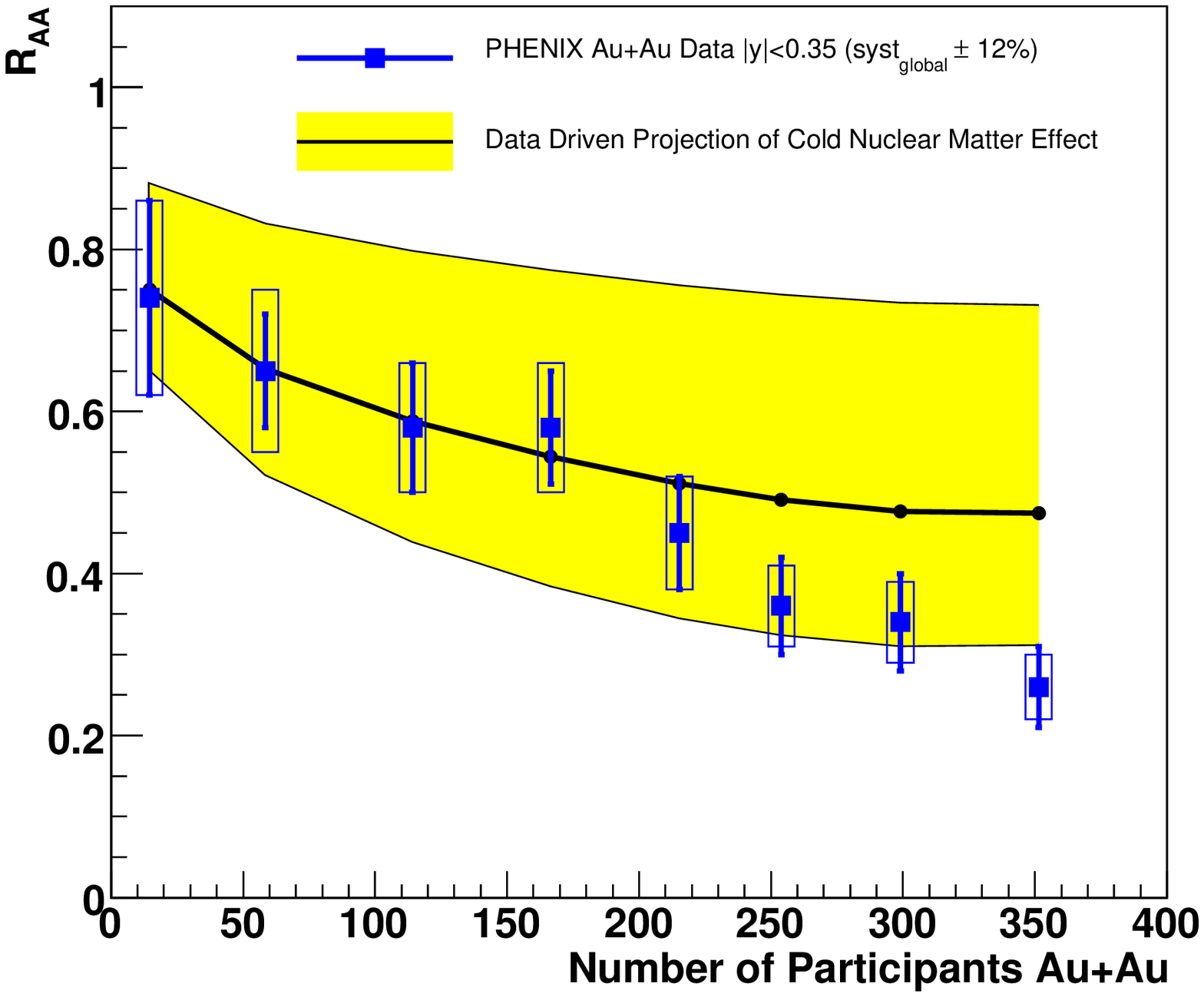}}
\end{minipage}
\vspace*{8pt} 
\begin{minipage}[t]{6.8cm}
\resizebox{0.9\textwidth}{0.9\textwidth}{%
\includegraphics*{./figures4stock/glauber_projectauaumid.ps}}
\end{minipage}
\parbox{0.99\textwidth}{
\caption{ $\Raa$ for Au-Au collisions compared to predictions of the data driven method constrained by the
$R_{d-Au}$ as a function of collision centrality, for midrapidity  (left) and forward rapidity (right).}
\protect\label{rauauabnorm2}}\\[1mm] 
\end{center}
\end{figure}

\item{Is there anomalous \J~suppression in Au-Au collisions at RHIC ?\\
In order to be able to detect an abnormal \J suppression
from an experimental point of view  and with minimal external 
input, appropriate experimental references are mandatory, measured 
under conditions such that no new physics can reasonably be expected.
The PHENIX experiment has made use of the presently available \mbox{d-Au}
data to establish the normal suppression or "cold nuclear 
matter effects" pattern expected for the Au-Au measurements 
at RHIC~\cite{normalPHENIX}. Figs.~\ref{rauauabnorm1} 
and~\ref{rauauabnorm2} show the values
of $\Raa$ as a function of centrality as measured in Au-Au collisions 
compared to the pattern extracted from \mbox{d-Au} measurements, taking  
also into account the shadowing effects expected at RHIC. PHENIX conclusion 
is that "{\sl Neither the predictions of cold nuclear 
effects in heavy ion collisions 
based on fitting of the \mbox{d-Au} data with theoretical curves nor those 
obtained directly from the \mbox{d-Au} data points are well enough constrained 
to permit quantitative conclusions about additional hot nuclear matter 
effects}". Much more precise reference data on \mbox{d-Au} collisions 
are needed to reach a solid conclusion.}



}
\end{itemize}

\section{Summary of the Experimental Status}

At the eve of the start of a new generation of experiments aimed at 
further studies of \mbox{quarkonium} abnormal suppression as a signature 
of QGP formation, the present experimental situation is summarized 
hereafter.

\begin{enumerate}
\vspace*{-0.2cm}
\item The production of \J suffers from cold matter nuclear effects also
called normal nuclear absorption. This absorption can be interpreted as 
resulting from interactions of a given $\C$ state while propagating in 
normal cold nuclear matter. Quantitatively, the corresponding absorption 
cross-section has been precisely measured in p-A interactions up to p-Pb, 
at $\sqs  = 29.0$ and 27.4~GeV. 
For a \J with rapidity in the range \mbox{$ 3.0<y_{\rm lab}<4.0$}.
with respect to the traversed nuclear matter, the best determination 
of the absorption cross-section leads to 
\mbox{${\Sgabs}^{\j} = 4.2 \pm 0.35~\rm mb$}. The values derived from data
collected at $\sqs = 39$ and 19.3~GeV in the same rapidity range are in 
in very good agreement with this devoted measurement.\\
Measurements performed on light nuclei interactions, namely O-S, O-U and S-U
at 19.3~GeV show excellent compatibility with the above quoted cross-section 
and, therefore, \J production in such reactions only suffers from normal 
nuclear effects.\\ 
There is thus compelling evidence that the value of $\Sgabs$ in the same 
rapidity window stays the same for $\sqs = 17.3$
\item The production of \P also suffers from cold matter nuclear effects. 
Systematic measurements performed on p-A reactions at 29.0 and 27.4~GeV
lead to a cross-section 
\mbox{${\Sgabs}^{\p} = 7.7 \pm 0.9~\rm mb$}, significantly higher than
\mbox{${\Sgabs}^{\j}$.}
 \vspace*{-0.2cm} 
\item From comparison with appropriate references deduced from the 
measurements on p and light ion-induced collisions, 
\J production in Pb-Pb interactions at 17.3~GeV exhibits only normal 
nuclear effects up to an energy density of 
\mbox{$\epsilon = 2.5~{\rm GeV}/{\rm fm^3}$}. 
An abnormal absorption appears above this value of $\epsilon$ and 
significantly increases with increasing energy density, up to an 
abnormal suppression factor of 0.6, for the most central Pb-Pb 
collisions.
\vspace*{-0.2cm} 
\item In contrast with the \J patterns, \P production exhibits an 
abnormal suppression starting with S-U reactions at 19~GeV. 
The observed suppression pattern is in very good agreement 
with the one exhibited by Pb-Pb interactions at 17.3~GeV. They both 
lead to an absorption cross-section of 
\mbox{${\Sgabs}^{\p} = 19.2\pm2.4 \rm mb$} as compared to 
\mbox{$7.3 \pm 1.6$} 
for normal \P absorption. Abnormal suppression for
\P starts above an energy density \mbox{$\epsilon = 1~{\rm GeV}/{\rm fm^3}$}.
\vspace*{-0.2cm}
\item Data obtained from In-In collisions at 17.3~Gev tend to qualitatively 
support the suppression pattern measured in Pb-Pb collisions.
\item For the moment being, there is no clear abnormality in Au-Au collisions 
at 200~GeV. Further measurements to precisely quantify cold nuclear effects 
are a must.  
\vspace*{-0.2cm}
\item \J absorption, as observed in Pb-Pb collisions, affects only 
low \pt \J's for peripheral collisions. Higher \pt \J's are more and more 
suppressed with increasing centrality of the collision. 
This feature requires a robust experimental reference 
in order to allow model-independent conclusions.
\vspace*{-0.2cm}
\item The surviving \J's in Pb-Pb collisions exhibit an average $<\TR2>$, 
or equivalently, a temperature $T_{\rm eff}$ which starts by increasing 
with increasing collision centrality. It saturates for energy densities 
above {$\simeq 2.8~{\rm GeV/fm^3}$} and up to {$3.9~{\rm GeV/fm^3}$}, 
the average value of
the most central bin.      
\vspace*{-0.2cm}
\end{enumerate}


%% file: out.tex
\oddsidemargin 15pt
\topmargin 0pt
\headheight 00pt
\headsep 00pt
\textheight 235mm
\textwidth 161mm
\hoffset=-0.5cm
\parindent=0pt

\def\J{$J/\psi$}

~~~
\vskip1cm

{\LARGE \bf Outlook}

\vskip0.5cm

The theoretical analysis of the in-medium behaviour of quarkonia has 
greatly advanced in the past decade. Potential model studies based on
lattice results for the colour-singlet free energy as well as direct
lattice calculations appear to be converging, and within a few more
years, the dissociation temperatures for the different quarkonium
states will be calculated precisely. Through corresponding calculations
of the QCD equation of state, these temperatures provide the energy
density values at which the dissociation occurs. In statistical QCD,
quarkonia thus allow a spectral analysis of the quark-gluon plasma.

\medskip

The application of this analysis in high energy nuclear collisions is
so far less conclusive. There exists a wealth of data from different
interactions at the SPS, and first RHIC results are now also available.
These results indicate the production of a hot, dense, strongly 
interacting medium. It is not yet clear, however, to what extent this
medium is indeed the quark-gluon plasma studied in statistical QCD. Is
it possible to identify dissociation onsets for different charmonium
states? Does the initial (non-thermal) overabundance of charm survive a
subsequent thermalization? Is the apparent \J~suppression saturation
from SPS up to fairly central RHIC collisions the survival of the
directly produced \J's as seen in lattice QCD studies, or is it the
production of additional \J's through secondary $\C$ pairings?
These questions can only be answered by further experiments, ideally
in all the possible energy ranges.

\medskip

Future nucleus-nucleus experiments at the CERN-LHC will  
extend the field to much higher energy densities; but the precise 
experimental measurement of cold matter nuclear effects 
in this new energy domain will still remain the prime condition 
for the understanding and interpretion of the results.

%% file: ref.tex
\oddsidemargin 15pt
\topmargin 0pt
\headheight 00pt
\headsep 00pt
\textheight 235mm
\textwidth 160mm
\voffset=0.5cm
\hoffset=-0.5cm
\parindent=0pt

\def\CMP{{ Comm.\ Math.\ Phys.\ }}
\def\NP{{ Nucl.\ Phys.\ }}
\def\PL{{ Phys.\ Lett.\ }}
\def\PR{{ Phys.\ Rev.\ }}
\def\PRep{{ Phys.\ Rep.\ }}
\def\PRL{{ Phys.\ Rev.\ Lett.\ }}
\def\RMP{{ Rev.\ Mod.\ Phys.\ }}
\def\ZP{{ Z.\ Phys.\ }}
\def\EPJ{{Eur.\ Phys.\ J.\ }}
\def\B{\boldmath}

~~~